\documentclass[%
twocolumn,
superscriptaddress,
 amsmath,amssymb,
 prc
 aps,
]{revtex4-1}

\def\am{$^{241}$Am~}
\def\na{$^{22}$Na~}

\def\cenns{CE$\nu$NS~}
\def\cennsp{CE$\nu$NS}
\usepackage{romanbar}
\usepackage{xcolor}
\usepackage{graphicx}% Include figure files
\usepackage{dcolumn}% Align table columns on decimal point
\usepackage{bm}% bold math
\usepackage{SIunits}
\usepackage{array}
\usepackage{makecell}
\usepackage[%  
    colorlinks=true,
    linkcolor=red
]{hyperref}

\begin{document}

\title{ Exploring coherent elastic neutrino-nucleus scattering using reactor electron antineutrinos in the NEON experiment 
} 
\author{J.J.~Choi}
\affiliation{Department of Physics and Astronomy, Seoul National University, Seoul 08826, Republic of Korea} 
\affiliation{Center for Underground Physics, Institute for Basic Science (IBS), Daejeon 34126, Republic of Korea}
\author{E.J.~Jeon}
\affiliation{Center for Underground Physics, Institute for Basic Science (IBS), Daejeon 34126, Republic of Korea}
\affiliation{IBS School, University of Science and Technology (UST), Deajeon 34113, Republic of Korea}
\author{J.Y.~Kim}
\affiliation{Department of Physics, Sejong University, Seoul 05006, Republic of Korea}
\author{K.W.~Kim}
\affiliation{Center for Underground Physics, Institute for Basic Science (IBS), Daejeon 34126, Republic of Korea}
\author{S.H.~Kim}
\affiliation{Center for Underground Physics, Institute for Basic Science (IBS), Daejeon 34126, Republic of Korea}
\author{S.K.~Kim}
\affiliation{Department of Physics and Astronomy, Seoul National University, Seoul 08826, Republic of Korea} 
\author{Y.D.~Kim}
\affiliation{Center for Underground Physics, Institute for Basic Science (IBS), Daejeon 34126, Republic of Korea}
\affiliation{IBS School, University of Science and Technology (UST), Deajeon 34113, Republic of Korea}
\author{Y.J.~Ko}
\affiliation{Center for Underground Physics, Institute for Basic Science (IBS), Daejeon 34126, Republic of Korea}
\author{B.C.~Koh}
\affiliation{Department of Physics, Chung-Ang University, Seoul 06973, Republic of Korea}
\author{C.~Ha}
\affiliation{Department of Physics, Chung-Ang University, Seoul 06973, Republic of Korea}
\author{B.J.~Park}
\affiliation{IBS School, University of Science and Technology (UST), Deajeon 34113, Republic of Korea}
\affiliation{Center for Underground Physics, Institute for Basic Science (IBS), Daejeon 34126, Republic of Korea}
\author{S.H.~Lee}
\affiliation{IBS School, University of Science and Technology (UST), Deajeon 34113, Republic of Korea}
\affiliation{Center for Underground Physics, Institute for Basic Science (IBS), Daejeon 34126, Republic of Korea}
\author{I.S.~Lee}
\affiliation{Center for Underground Physics, Institute for Basic Science (IBS), Daejeon 34126, Republic of Korea}
\author{H.~Lee}
\affiliation{IBS School, University of Science and Technology (UST), Deajeon 34113, Republic of Korea}
\affiliation{Center for Underground Physics, Institute for Basic Science (IBS), Daejeon 34126, Republic of Korea}
\author{H.S.~Lee}
\affiliation{Center for Underground Physics, Institute for Basic Science (IBS), Daejeon 34126, Republic of Korea}
\affiliation{IBS School, University of Science and Technology (UST), Deajeon 34113, Republic of Korea}
\author{J.~Lee}
\affiliation{Center for Underground Physics, Institute for Basic Science (IBS), Daejeon 34126, Republic of Korea}
\author{Y.M.~Oh}
\affiliation{Center for Underground Physics, Institute for Basic Science (IBS), Daejeon 34126, Republic of Korea}
\collaboration{NEON Collaboration}
\date{\today}

\begin{abstract}
Neutrino elastic scattering observation with NaI (NEON) is an experiment designed to detect neutrino-nucleus coherent scattering using reactor electron antineutrinos. NEON is based on an array of six NaI(Tl) crystals with a total mass of 13.3\,kg, located at the tendon gallery that is 23.7\,m away from a reactor core with a thermal power of 2.8\,GW in the Hanbit nuclear power complex.
 The installation of the NEON detector was completed in December 2020, and since May 2021, the detector has acquired data at full reactor power. 
 Based on the observed light yields of the NaI crystals of approximately 22, number of photoelectrons per unit keV electron-equivalent energy (keVee), and 6 counts/kg/keV/day background level at 2--6\,keVee energy, coherent elastic neutrino-nucleus scattering~(\cennsp) observation sensitivity is evaluated as more than 3$\sigma$ assuming one-year reactor-on and 100\,days reactor-off data, 0.2\,keVee energy threshold, and 7\,counts/keV/kg/day background in the signal region of 0.2--0.5\,keVee. 
 This paper describes the design of the NEON detector, including the shielding arrangement, configuration of NaI(Tl) crystals, and associated operating systems. The initial performance and associated sensitivity of the experiment are also presented. 
\end{abstract}
\maketitle

\section{Introduction}
Since it was predicted in 1974~\cite{PhysRevD.9.1389,Kopeliovich:1974mv}, coherent elastic neutrino-nucleus scattering~(\cennsp) has garnered the attention of particle physicists owing to its potential use in completing the standard model picture~\cite{Drukier19842295,Krauss:1991ba,Patton:2012jr} and searching for novel physics phenomena~\cite{Dutta:2015nlo,Kosmas:2017zbh,deNiverville:2015mwa}. Several experiments using advanced detector technologies had been proposed and some were conducted~\cite{Cabrera:1984rr,Formaggio:2011jt,Akimov:2012aya,Belov:2015ufh,Agnolet:2016zir,Aguilar-Arevalo:2016khx,Kerman:2016jqp,Billard:2016giu,Hakenmuller:2019ecb,Angloher:2019flc,CONUS:2020skt}, but they had a difficulty due to the tiny deposited energy ($\sim$ keV) into the nucleus.  In 2017, the COHERENT collaboration announced the detection of \cenns with a conventional CsI(Na) detector using a stopped-pion source~\cite{Akimov:2017ade}. Subsequently, the same group confirmed the process using a liquid argon detector~\cite{COHERENT:2020iec}. 

The COHERENT experiment used neutrinos from a spallation neutron source (prompt muon neutrino followed by delayed electron neutrino and muon antineutrino) with energies of approximately 30\,MeV~\cite{PhysRevD.73.033005,COLLAR201556}. Relatively high-energy neutrinos with significant background reduction using the timing information of the pulsed beam had allowed the capture of a low-energy signal induced by the \cenns process~\cite{Akimov:2017ade,COHERENT:2020iec}. However, such success has not been achieved using other neutrino sources, such as reactors~\cite{CONUS:2020skt,Colaresi:2021kus,CONNIE:2021ggh,nGeN:2022uje} or solar neutrinos~\cite{XENON:2020gfr}, although extensive efforts have been made. In particular, reactor neutrinos with energy reaching a few MeV produce visible recoils with energy less than 1\,keV, which is significantly lower than the typical energy threshold of a kg-size particle detector. 

Measurement of \cenns using reactor electron antineutrinos provides valuable information for understanding neutrinos. In astrophysics, understanding neutrino interactions at the MeV-scale is important for modeling supernova energy transport~\cite{Janka:2017vlw}. The monitoring of nuclear reactors can be accomplished via \cenns using detectors weighing several tens of kilogram~\cite{Cogswell2016,RevModPhys.92.011003}. Possible investigations of novel physics beyond the standard model with reactor neutrinos have been proposed~\cite{Formaggio:2011jt,Liao:2017uzy,Dev:2019anc,Aguilar-Arevalo:2019jlr}. 

Neutrino elastic scattering observation with NaI (NEON) is an experiment that aims to observe \cenns using reactor antineutrinos. 
Exploiting the expertise of the short baseline reactor neutrino experiment NEOS~\cite{YJKO:2017NEOS} and developing a high-quality NaI(Tl) detector for the COSINE experiment~\cite{Adhikari:2018ljm,COSINE-100:2019lgn}, NEON proceeded smoothly from the initial development of the high-light-yield NaI(Tl) detector~\cite{Choi:2020qcj} to the installation of the NEON detector at the reactor site that was completed in December 2020. The detector has been acquiring data at full reactor unit operational power since May 2021. 
This study describes the detector configuration of the NEON experiment and its performance in the detection of \cenns from the reactor electron antineutrinos. 

The remainder of this paper is organized as follows: Section 2 describes the NaI(Tl) crystals for \cenns searches. Section 3 discusses the NEON experiment from  the experimental site to the shielding arrangement. Section 4 gives details about the internal radioactivity levels of individual crystals and describes how they are assembled into the detector array. Section 5 provides details about the liquid scintillator veto system.  Section 6 provides a brief overview of the data acquisition system, and Section 7 describes the environmental monitoring system. Section 8 reports the detection sensitivity of \cenns in the NEON experiment.  Finally, Section 9 presents a summary.

\section{NaI(Tl) as a \cenns detector}
The detection of light signals from scintillation crystals is a well-established technology used to search for extremely rare events, such as weakly interacting massive particles (WIMPs)~\cite{Kim_2010,Pandey_2018} and \cenns~\cite{Akimov:2017ade}.
Among various scintillation crystals, NaI(Tl) has drawn particular attention because the DAMA/LIBRA collaboration reported a positive signal that manifested as an annual modulation in the rate of low-energy events in an array of NaI(Tl) crystals~\cite{Bernabei:2014jba,Bernabei:2018jrt}. 
This claim has triggered worldwide independent efforts to reproduce the DAMA/LIBRA observations with the same NaI(Tl) crystals~\cite{Kim:2014toa,sabre,Adhikari:2017esn,Fushimi:2018qzk,Coarasa:2018qzs,Amare:2018sxx}. Eventually, these efforts  become realized as high-light yield~\cite{Olivan:2017akd,Choi:2020qcj} and low-background NaI(Tl) detectors~\cite{Suerfu:2019snq,Park:2020fsq,Adhikari:2021rdm,Fushimi:2021mez}, which are essential for both WIMP dark matter searches and \cenns observations.

The COSINE-100 experiment is one such experiment that is currently operating with 106\,kg of low-background NaI(Tl) crystals~\cite{Adhikari:2017esn}. The average level of background in the energy region of 1--6\,keVee (keVee denotes an electron equivalent energy in keV) is obtained as 2.73$\pm$0.14\,counts/kg/keV/day with dominant background sources from $^{210}$Pb and $^{3}$H~\cite{Adhikari:2021rdm}.  
As the NEON experiment operates at sea level with commercial-grade crystals, we expect higher backgrounds from both internal and external radiations. 
For internal sources, $^{3}$H will be similar because it originates from cosmogenic activation. In the case of $^{210}$Pb, commercial crystals contain similar or slightly larger amounts than those of COSINE-100 crystals. 
Considering the similarity between the COSINE-100 shield~\cite{Adhikari:2017esn} and the NEON shield, no significant increase of the background level due to external radioactive elements is expected~\cite{cosinebg,Adhikari:2021rdm}. 
The initial background level target in the signal region is 10\,counts/kg/keV/day, which is similar to the measured background levels from the CONUS experiment in the 0.5--1\,keVee energy region~\cite{CONUS:2020skt}.

The NEON experiment uses a 13.3\,kg NaI(Tl) array, which is one of the largest mass detectors operated in the reactor for \cenns observation, except for the RED-100 experiment~\cite{Abdullah:2022zue}. 
The maximum recoil energy from \cenns for a given target species with nuclear mass $m_A$ and neutrino energy $E_\nu$ is approximately $2E_\nu^2/m_A$~\cite{Abdullah:2022zue}. Owing to the low atomic mass number of sodium, NaI(Tl) crystals have the advantage of a larger recoil energy. 
In addition, low-background and high-light yield NaI(Tl) detectors have been developed for the next phase of NaI-based dark matter search experiments~\cite{Suerfu:2019snq,Park:2020fsq,Adhikari:2021rdm,Fushimi:2021mez} which makes it easy to upgrade the current NEON experiment for the next phase 100\,kg or 1\,tonne-scale experiments to achieve precision measurement of \cenns as well as search for new physics phenomena.

The light output of NaI(Tl) crystals has continuously improved in recent decades.
In the DAMA/LIBRA-phase 2 experiment, NaI(Tl) crystals with light yields of 5--10\,NPE/keVee (NPE denotes the number of photoelectrons), were operated at a 1\,keVee energy threshold~\cite{Bernabei:2018jrt}. Similarly, ANAIS-112 operated their crystals with approximately 15\,NPE/keVee and reported physics results with a 1\,keVee energy threshold~\cite{Amare:2019jul,Amare:2021yyu}. Furthermore, in the COSINE-100 experiment run at the Yangyang Underground Laboratory, NaI(Tl) crystals with light yields of approximately 15\,NPE/keVee were used~\cite{Adhikari:2017esn}.
The NEON detector applied a novel technique of crystal encapsulation that significantly increased the light collection efficiency and obtained approximately 22\,NPE/keVee~\cite{Choi:2020qcj}, which was approximately 50\,\% larger than the light yield of COSINE-100 crystals~\cite{Adhikari:2017esn}. 

In the COSINE-100 data, the trigger of an event was satisfied with coincident photoelectrons in two PMTs attached to each side of the crystal, resulting in an approximately 0.13\,keVee threshold. However, in the low-energy signal region below 10\,keVee, PMT-induced noise events were predominantly triggered. A multivariable analysis technique using a boosted decision tree (BDT)~\cite{BDT} achieved 1\,keVee analysis threshold of less than 0.1\,\% noise contamination and above 80\,\% selection efficiency~\cite{adhikari2020lowering,COSINE-100:2021xqn}. 
Further improvement of low-energy event selection is ongoing based on the development of new parameters for the BDT as well as the use of a machine learning technique that uses raw waveforms directly. 
NEON targets an analysis threshold of 5\,NPE (0.2\,keVee), which is similar to the energy threshold that has already been achieved by the COHERENT experiment with a CsI(Na) crystal~\cite{Akimov:2017ade} and has the same target threshold as the next phase COSINE-200 experiment~\cite{COSINE-100:2021poy}.
With this detector performance, NaI(Tl) detectors are suitable for searching for \cenns from the reactor electron antineutrinos.

\section{Experiment}

\subsection{Hanbit nuclear power complex}
\begin{figure}[!htb]
  \centering
  \begin{center}
    \includegraphics[width=0.45\textwidth]{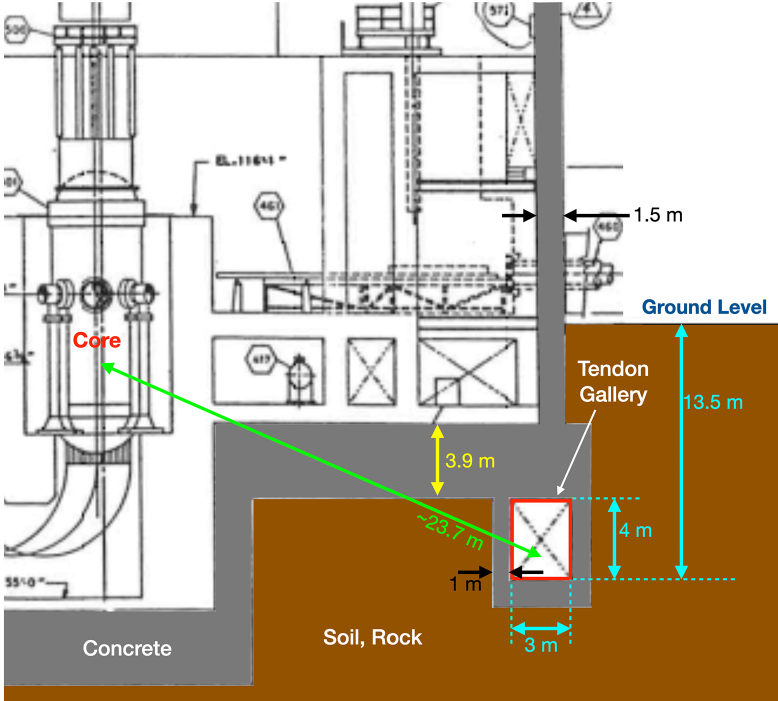}
  \end{center}
	\caption{Schematic view of Hanbit nuclear reactor unit-6. 
			The tendon gallery is 23.7\,m away from the reactor core that has  more than 10\,m shields of concrete and rock. 
	}
  \label{fig:reactor}
\end{figure}

The NEON detector was installed in November 2020 at the tendon gallery of reactor unit-6 of the Hanbit nuclear power complex in Yeonggwang, Korea. 
The location and distance from the reactor core are similar to those in the NEOS experiment, which was installed in reactor unit-5 in the same reactor complex~\cite{YJKO:2017NEOS}. 
In addition, this is the same reactor complex used for the RENO experiment~\cite{RENO:2019otc}. 
The active core size of unit-6 has a diameter of 3.1\,m, height of 3.8\,m, is cylindrical in shape, and contains 177 low-enrichment uranium fuel assemblies. 
The detector is located 23.7$\pm$0.3\,m away from the center of the reactor core, as shown in Fig.~\ref{fig:reactor}, whereas the distance to the closest neighboring reactor core is 256\,m.
The expected neutrino flux at the tendon gallery is $8.09\times10^{12}~/\mathrm{cm}^2/\mathrm{sec}$ based on the reactor neutrino flux model in Ref.~\cite{Kopeikin:2012zz}.
Between the reactor core and tendon gallery, over 10\,m of heavy concrete shielding mitigates radiation from the reactor operation, as shown in Fig.~\ref{fig:reactor}. As the tendon gallery is not a radiation-restricted area, the experimental site can be accessed without a dosimetry badge. 
Furthermore, the tendon gallery is located 10\,m below ground level under the wall of the concrete building. The experimental site has an approximately 20\,m water-equivalent overburden, which has six times lower muon flux than that at sea level. 

The NEON detector is located inside a temporary housing structure for temperature control and dust protection.
Owing to the maximum electricity usage of 3\,kW, an air control system with a low power consumption (1\,kW maximum) has been installed, which maintains a temperature of 23$\pm$2\,$^{\circ}$C in the detector room.

\subsection{Shielding design}
To observe the \cenns signal, excellent background suppression is crucial.
Background originating from environmental radioisotopes, cosmic muon decays, and reactor cores must be effectively reduced by shielding materials. 
As the NEOS experiment in the same tendon gallery of reactor unit-5 reported no significant reactor-correlated backgrounds for both $\gamma$~\cite{Ko:2019cip} and neutrons~\cite{Ko:2016neu}, the NEON shield follows a design similar to that used for the COSINE-100 experiment~\cite{Adhikari:2017esn}. 
The detailed design considers the limited space of the tendon gallery (3\,m in width and 4\,m in height)~\cite{KIM2016285}, background measurements of the NaI(Tl) crystals from the COSINE-100 experiment~\cite{cosinebg,Adhikari:2021rdm}, and neutrons in the shallow-depth tendon gallery, such as muon-induced and reactor-related ones. 
This reduces the thickness of the lead-shielding layer and additional neutron-shielding layers using polyethylene and borated polyethylene blocks. 

\begin{figure*}[!htb]
  \begin{center}
    \includegraphics[width=0.95\textwidth]{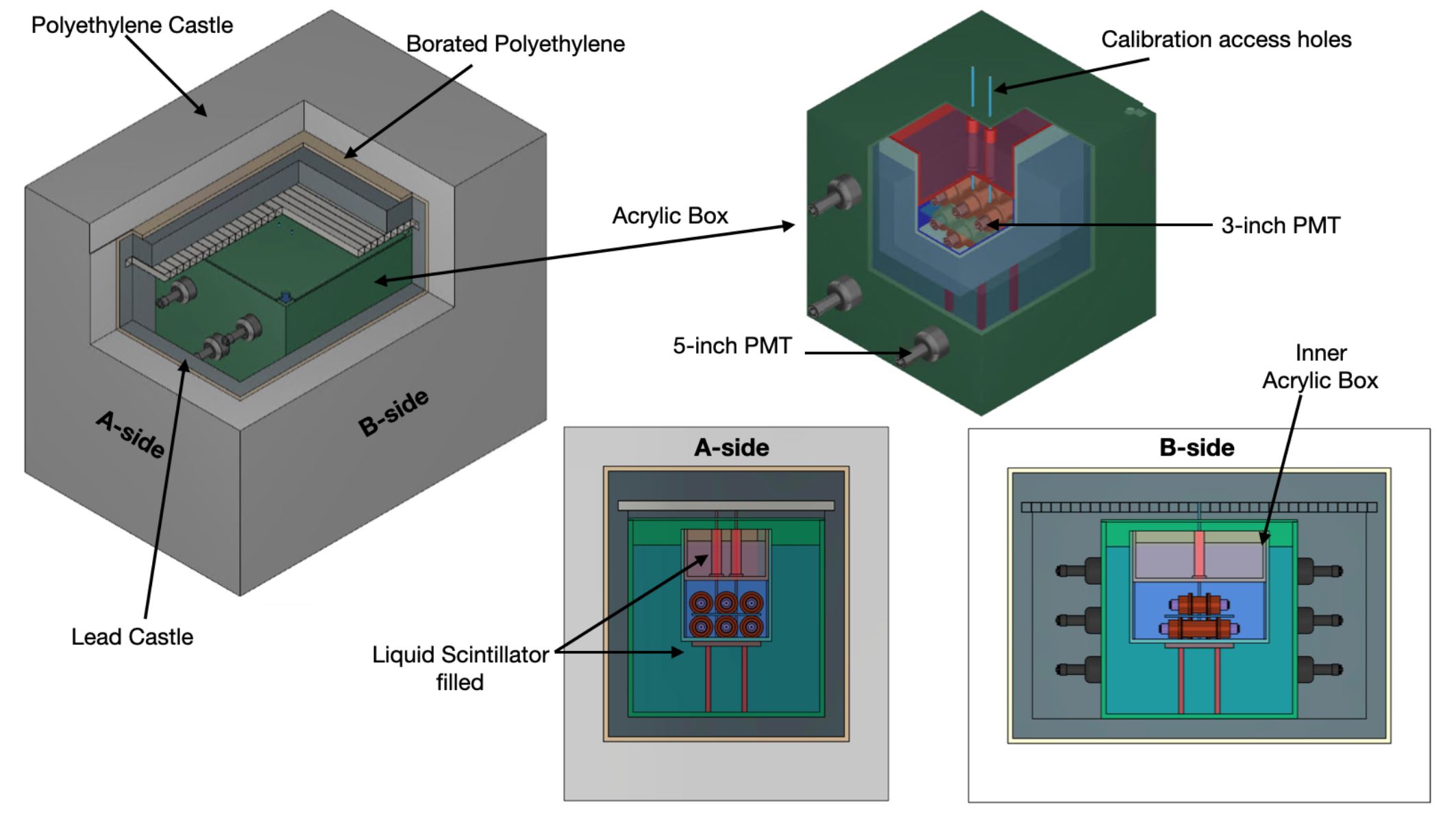}
  \end{center}
  \caption{NEON shielding overview. From outside inward, a polyethylene castle and borated polyethylene boards, a lead brick castle, stainless-steel pipes, and acrylic box ($2.5~\rm{cm}$ thick) with ten 5-inch PMTs and six encapsulated crystal detectors immersed in the liquid scintillator are indicated. In addition, the locations of the calibration holes and size of the PMTs are indicated. Projections of the NEON shielding on the A-side and B-side are presented in the bottom plots.}
  \label{neondesign}
\end{figure*}

\begin{figure*}[!htb]
  \begin{center}
    \includegraphics[width=0.95\textwidth]{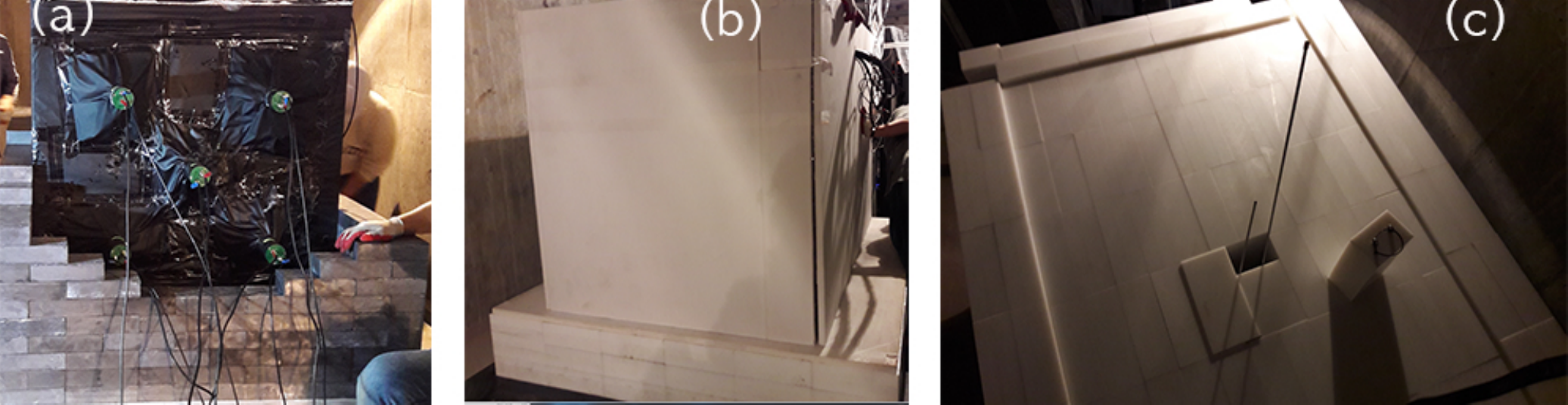}
  \end{center}
  \caption{Pictures of the NEON shield during installation. 
(a) The LS container and readout PMTs are surrounded by 10~cm lead bricks. (b) The lead castle is surrounded by 2.5\,cm thick borated PE and (c) 20\,cm high density PE. Two calibration tubes installed with the calibration rods (sources) are shown. 
	}
  \label{neon_pic}
\end{figure*}

The NEON detector is contained within a 4-layer nested arrangement of shielding components, as shown in Fig~\ref{neondesign}. 
It provides $4\pi$ coverage to shield external radiation from various sources as well as an active veto for internal or external sources. 
The shield is placed on a $250~\rm{cm} \times 200~\rm{cm} \times 20~\rm{cm}$ steel palette. 
From the outside inward, the four shielding layers are a polyethylene castle, borated polyethylene board, lead castle, and linear alkylbenzene (LAB)-based liquid scintillator (LS), as described below. 
The six NaI(Tl) crystal assemblies are placed in an acrylic box to avoid direct contact with LS. 
Noise generation was observed in the PMTs when the PMT bases were exposed to LS. 
A new crystal encapsulation design to embed crystals in an LS without an acrylic box is an ongoing research area aimed at improving veto efficiency. 
This acrylic box and its support acrylic table are immersed in the LAB-LS. 
A few pictures of the NEON shielding obtained during detector installation are shown in Fig.~\ref{neon_pic}. 

\subsubsection{Polyethylene castle and borated polyethylene}
Two types of polyethylene are used to prevent external neutrons: $2.5~\rm{cm}$-thick polyethylene boards with $5~\%$ boron loading tightly cover the lead castle, whereas  a $30~\rm{cm}$- (top and bottom) and $20~\rm{cm}$-thick (side) high-density polyethylene castle cover the borated polyethylene.
Owing to the limited space of the tendon gallery, the total width of the NEON detector must be less than 2\,m. This results in a slightly narrower thickness of the polyethylene castle on the sides.  
\subsubsection{Lead castle}
A $15~\rm{cm}$- (top and bottom) and $10~\rm{cm}$-thick (side) low-activity lead castle surround the acrylic box filled with LS.
To reinforce the top, a $5~\rm{cm}$-thick square and 120\,cm-long stainless-steel pipes support the lead bricks. 
There is a $10~\rm{cm}$ space between the stainless-steel pipes and the LS-containing acrylic box.

\subsubsection{Liquid scintillator}
The innermost shield is provided by 800\,L of LAB-LS contained in a $2.5~\rm{cm}$-thick $100~\rm{cm} \times 100~\rm{cm} \times 100~\rm{cm}$ acrylic box. The outer wall of the box is wrapped with teflon sheets to increase the light collection efficiency of the LS, which is then covered by a black polyvinyl chloride sheet to prevent light leakage. 
The LS-produced scintillation signals are detected via ten 5-inch Hamamatsu PMTs (R877) that are attached to two sides of the box. 

A variety of backgrounds produced by radiogenic particles from components in and near the NaI(Tl) crystals are efficiently rejected owing to the anticoincidence requirement of the PMT signals from the LS~\cite{Adhikari:2017esn}. 
In addition, the LS shield provides effective shielding of external neutrons. 

\subsubsection{Simulation study}
\begin{figure*}[!htb]
  \begin{center}
    \begin{tabular}{ccc}
    \includegraphics[width=0.33\textwidth]{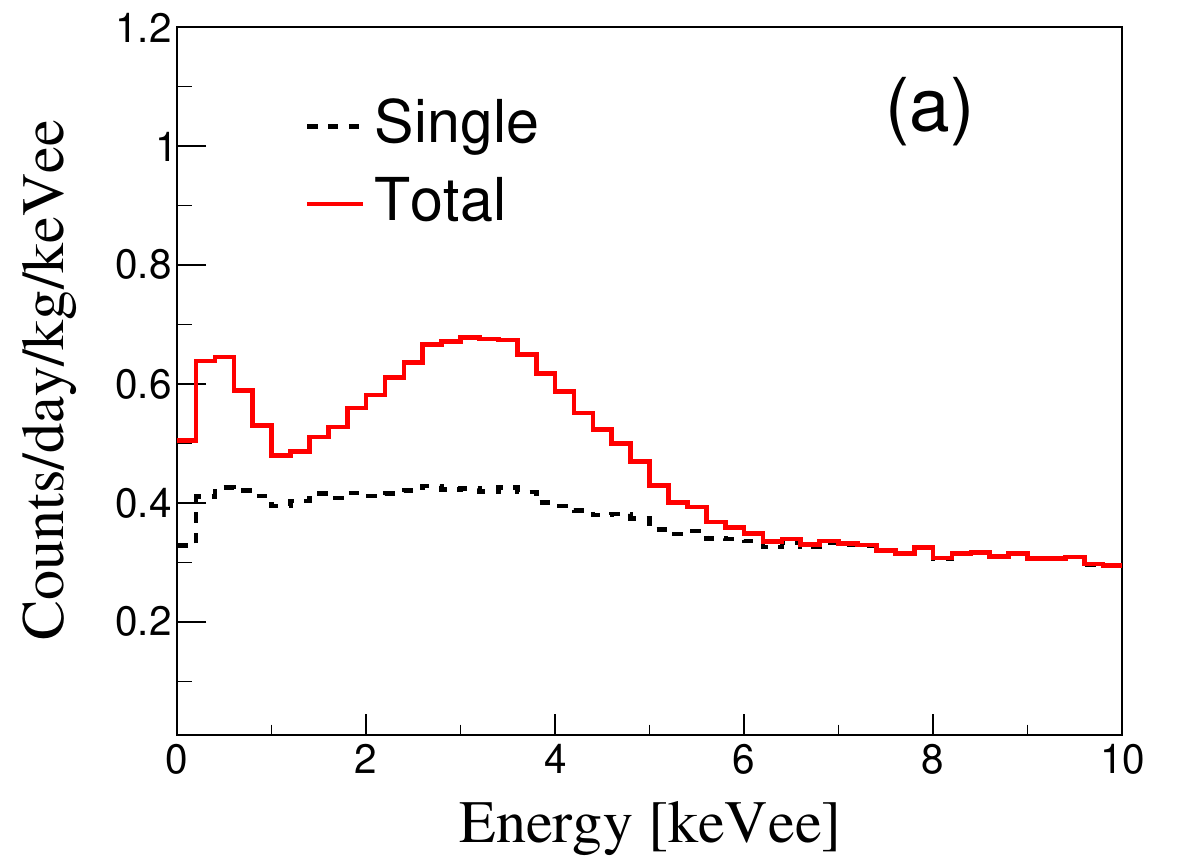} &
    \includegraphics[width=0.33\textwidth]{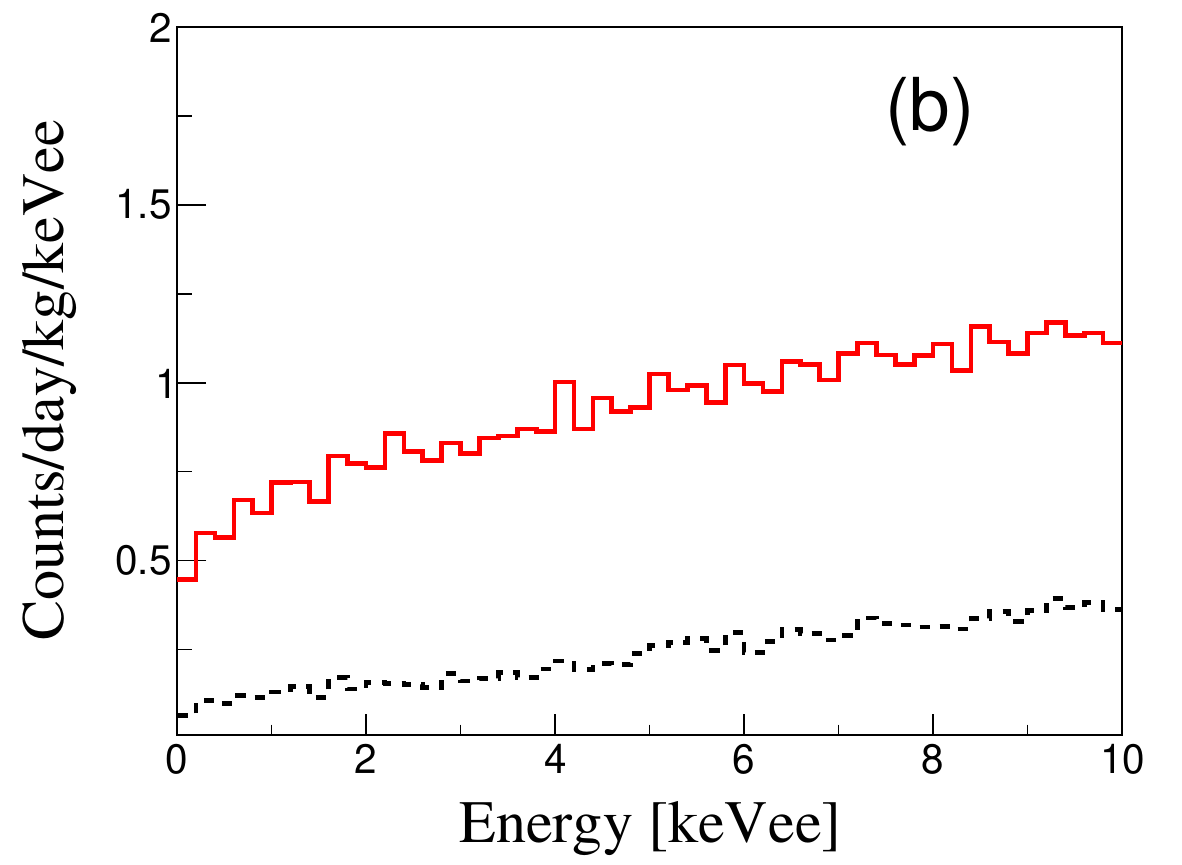} &
    \includegraphics[width=0.33\textwidth]{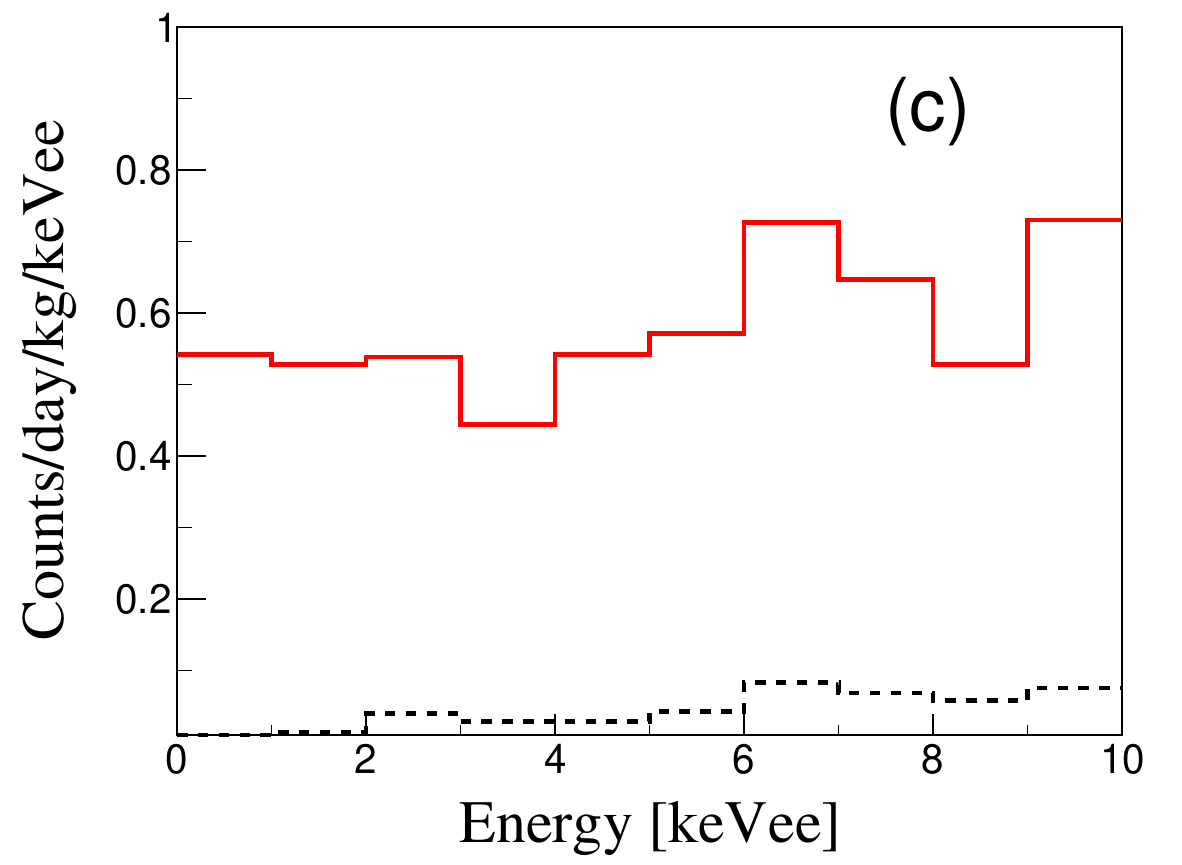} \\
		(a) Internal $^{210}$Pb and $^{40}$K & (b) PMT radioactivities & (c) External neutrons\\
    \end{tabular}
  \end{center}
	\caption{Simulated energy spectra of the NEON crystal in 0--10\,keV energy region from different background sources for the total events (solid-red lines) and single crystal hit events (black-dashed lines) that have no hit in the LS and other crystals. (a) Internal $^{210}$Pb and $^{40}$K assuming the NEO-5 contaminations in Table~\ref{table:LightYield} are generated. (b) The PMT radioactivities from the COSINE-100 data modeling are simulated with the NEON detector geometry. (c) External neutrons assuming the same flux of the CONUS site~\cite{Hakenmuller:2019ecb} without the muon-induced neutrons are generated. In this case, dominant interactions are neutron-induced X-rays and $\gamma$-rays in the shielding materials. }
  \label{simulation}
\end{figure*}

We construct a simplified geometry for the NEON detector and  generated background events using the Geant4 simulation toolkit~\cite{Agostinelli:2002hh}.
To understand the effectiveness of the active veto of the LS detector depending on the source locations, we study the simulated background events from three different locations: internal crystals, crystal PMTs, and outside the shield. 
Figure~\ref{simulation} shows the simulated energy spectra of the 0--10 keV energy regions in the NEON crystal for three different cases. 
Internal sources of $^{210}$Pb and $^{40}$K assuming the NEO-5 contamination in Table~\ref{table:LightYield}, PMT radioactivities assuming the same contaminations of the COSINE-100 detector~\cite{Adhikari:2021rdm}, and external neutrons without muon induced neutrons assuming neutron flux from the CONUS experiment~\cite{Hakenmuller:2019ecb} are simulated, and their energy spectra are presented in Figs.~\ref{simulation} (a), (b), and (c). 
The single hit ratios to the total events in the 0.2--1\,keV regions are 70$\pm$6\,\%, 21$\pm$5\,\%, and less than 1\,\% for internal sources, PMT radioactivity, and outside neutron sources, respectively. 
The veto efficiencies of the LS detector are highly dependent on the locations of the background sources, which makes it easy to understand their origins. For instance, unexpected backgrounds from the reactor operation can be easily identified by enhanced backgrounds in the reactor from data on the total event rates rather than that on single-crystal hit rates, similar to Fig.~\ref{simulation} (c).

Various background sources that can contribute to the NEON detector are under studied with known elements from the COSINE-100 experiments~\cite{cosinebg,Adhikari:2021rdm}, CONUS experiment~\cite{Hakenmuller:2019ecb}, and NEOS experiment~\cite{YJKO:2017NEOS}. 
By incorporating all known background sources and modeling the NEON data for both reactor-on and reactor-off periods, a precise understanding of the background contributions to the NEON detectors can be achieved. 
By taking advantage of the active LS veto detector shown in Fig.~\ref{simulation}, simultaneous modeling of single crystal hit events and multiple crystals, or LS hit events can be performed, making it easy to disentangle and understand the contributions from various background sources. Although this is under development with the NEON data, similar studies with the COSINE-100 data were successfully done~\cite{cosinebg,Adhikari:2021rdm}.

\subsection{Calibration sources}
\begin{figure*}[!htb]
  \begin{center}
    \begin{tabular}{cc}
    \includegraphics[height=0.3\textwidth]{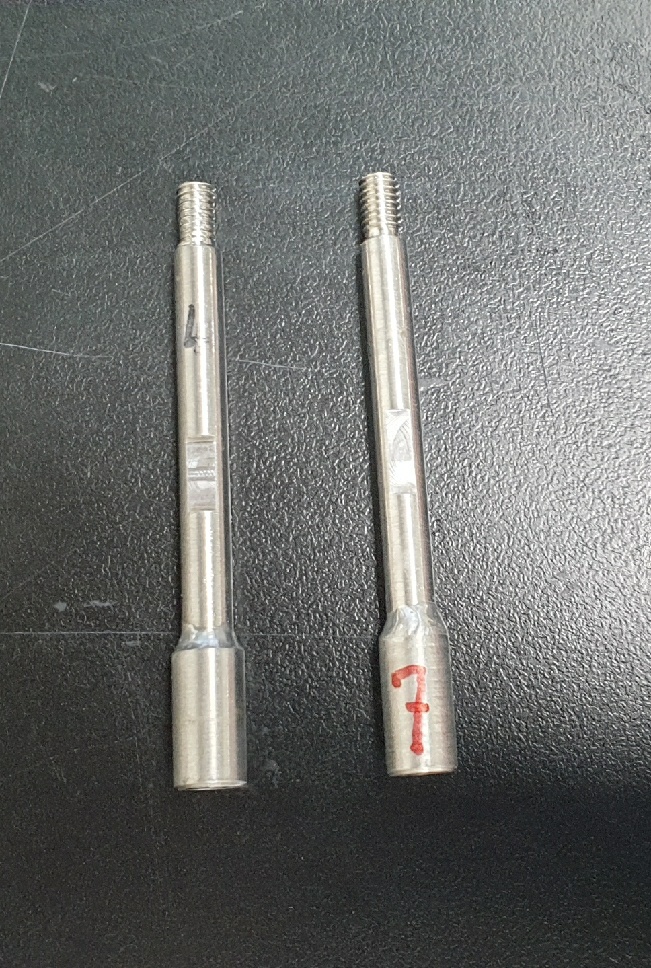} &
    \includegraphics[height=0.3\textwidth]{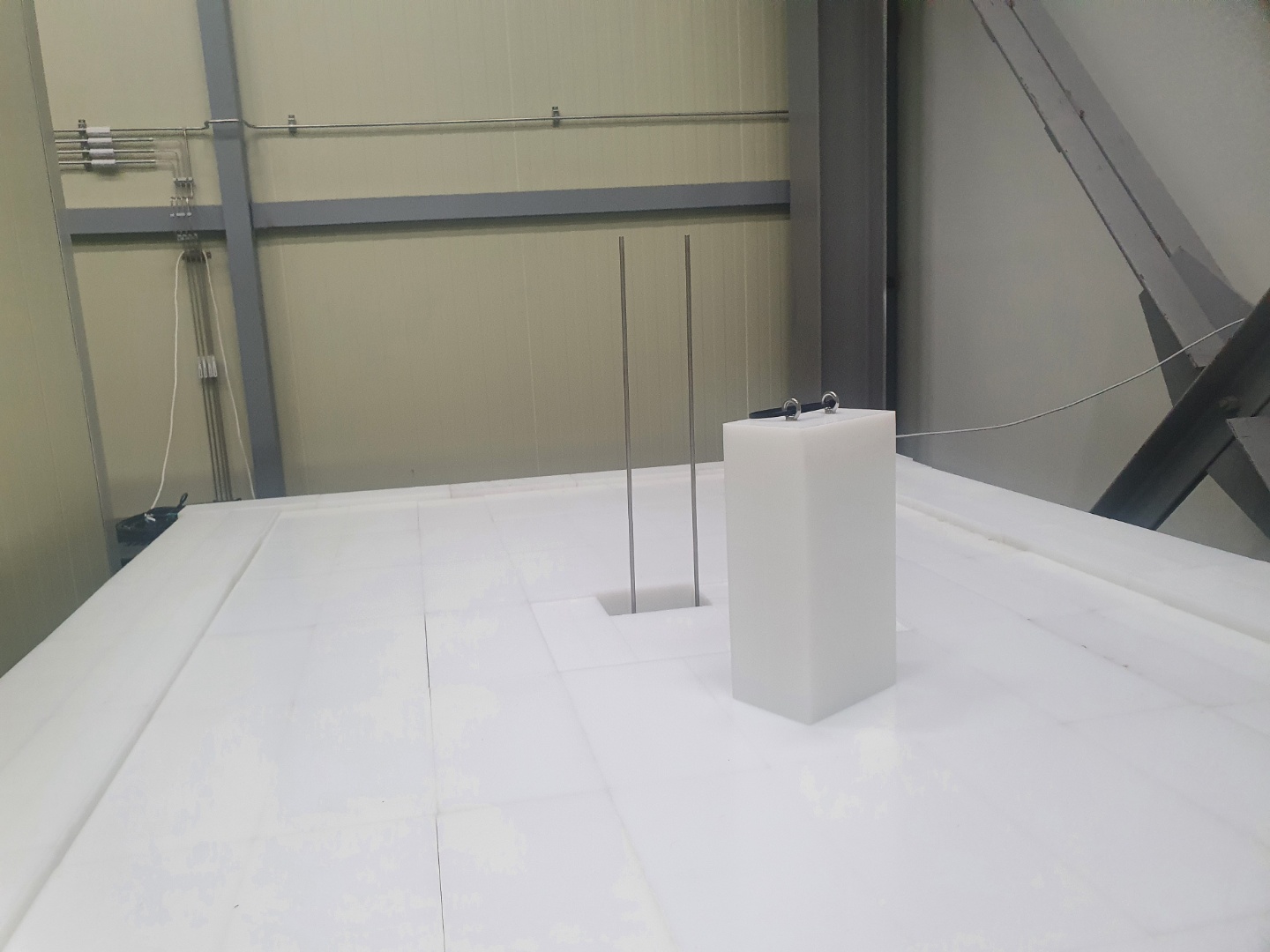} \\
      (a) & (b)\\
    \end{tabular}
  \end{center}
  \caption{(a) $^{22}{\rm Na}$ and $^{241}{\rm Am}$ calibration sources are contained in cases of 6 mm diameter and 12 mm height. (b) The encapsulated sources are connected to stainless-steel rods and located in the calibration pipe.}
  \label{CalibrationSource}
\end{figure*}

The calibration sources are prepared by sealing each isotope in a stainless-steel case suitable for the calibration tube.
Further, using standard isotope solutions \footnote{procured from Eckert and Ziegler Isotope Products}, calibration sources are produced to yield approximately 100\,Bq activities. 
During the calibration data acquisition, these calibration sources are connected to a 1.5\,m stainless-steel rod and installed on the calibration tubes to reach near the crystal detectors. 
Figure~\ref{CalibrationSource} shows the encapsulated $^{22}{\rm Na}$ and $^{241}$Am calibration sources and stainless-steel rods. 

\section{NaI crystal detectors}
\label{sec:NaI}
\subsection{Crystal assembly}
Six commercial-grade crystals with two different dimensions are manufactured by Alpha Spectra Inc.\,(AS): four with 3-inch diameter and 4-inch length, and two with 3-inch diameter and 8-inch length. 
These crystals are labeled NEO-1 to NEO-6.
Further, the lateral surfaces of each crystal are wrapped in approximately 10 layers of 250-$\mu$m-thick teflon reflective sheets, inserted into the copper tubes in a nitrogen gas environment, and sealed to render them airtight.
A novel technique for crystal encapsulation is developed to collect scintillation photons efficiently and maximize the measured light yield, as described in Ref.~\cite{Choi:2020qcj}. First, the size of the crystal end-face is matched to that of the PMT photocathode. In addition, only a single optical pad is used between the PMT window and the NaI(Tl) end face, although the typical encasement of the NaI(Tl) crystal requires three layers of optical interfaces owing to quartz windows. This detector-sensor combined assembly reduces light losses due to reflections at each optical interface. Consequently, by applying this design to the NEON crystals, an approximately 50\,\% increased light yield is achieved compared to that of the COSINE-100 crystals~\cite{Choi:2020qcj}.

\begin{figure*}[!htb]
  \begin{center}
    \begin{tabular}{cc}
      \includegraphics[width=0.36\textwidth]{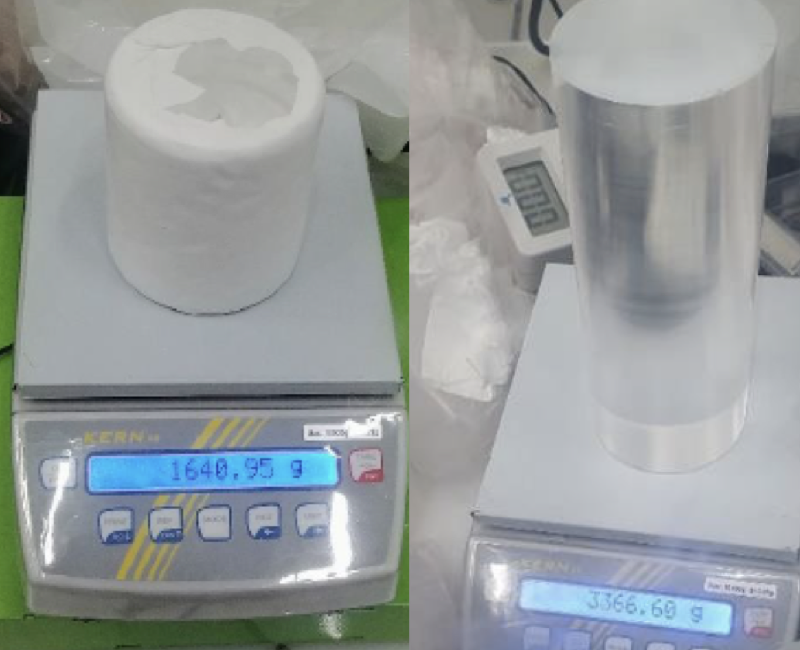} &
      \includegraphics[width=0.54\textwidth]{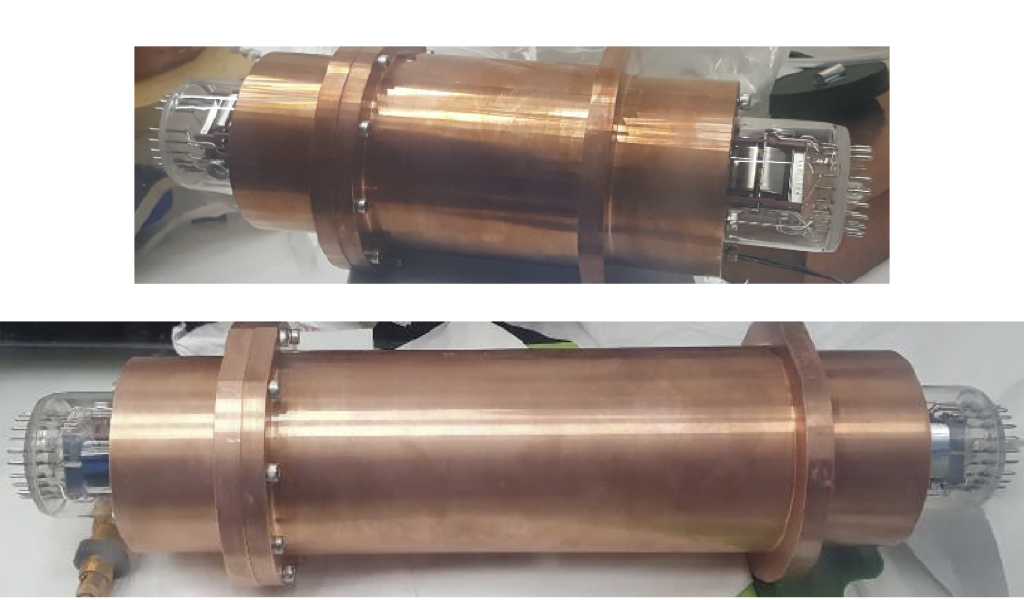} \\
      (a) & (b)\\
    \end{tabular}
  \end{center}
  \caption{NaI(Tl) crystals are polished manually
    with lapping films and encapsulated together with PMTs
    inside a copper encapsulation.
    (a) Bare crystals of 1.6~kg and 3.7~kg types 
    are shown. (b) Completed detector assemblies are presented. 
  }
  \label{ref:crystal_det}
\end{figure*}

The bare crystals and completed assemblies are shown in Fig.~\ref{ref:crystal_det} for the two different sizes. The measured light yields for the NEON crystals are summarized in Table~\ref{table:LightYield}.
Further details regarding encapsulation and detector assembly can be found in~\cite{Choi:2020qcj,Park:2020fsq}.

\begin{figure*}[!htb]
  \begin{center}
    \begin{tabular}{cc}
      \includegraphics[width=0.55\textwidth]{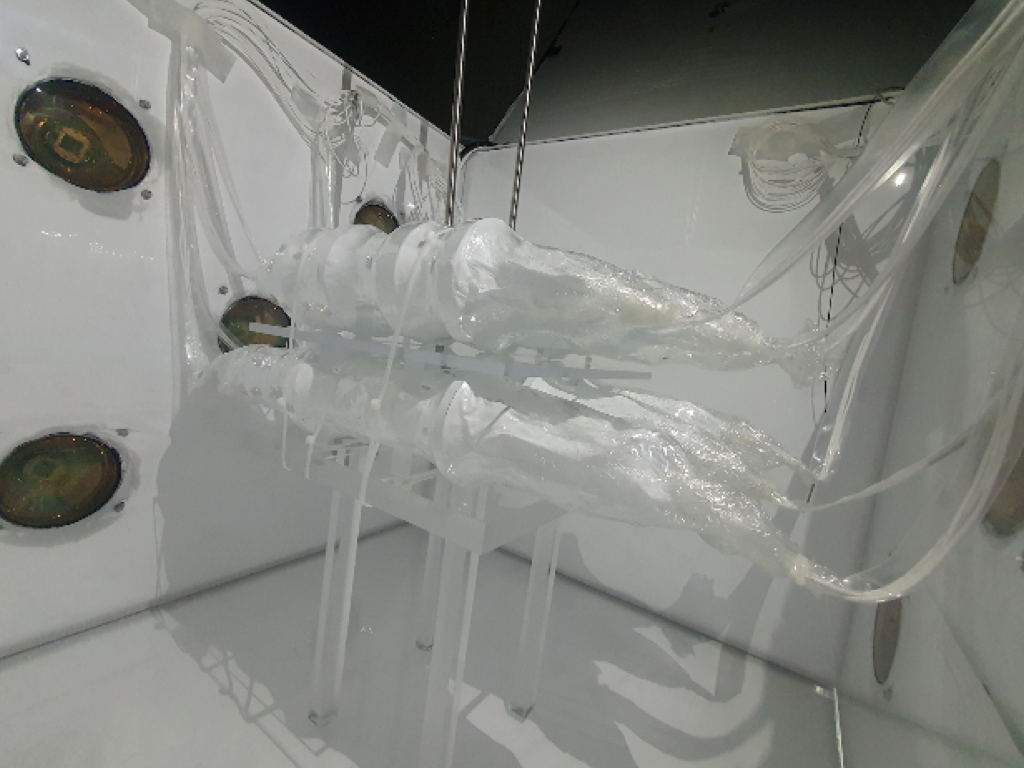} &
      \includegraphics[width=0.45\textwidth]{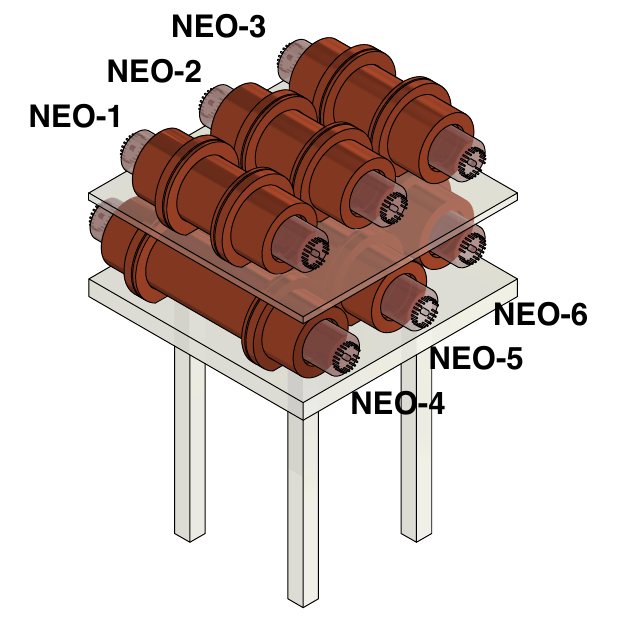} \\
      (a) & (b)\\
    \end{tabular}
  \end{center}
  \caption{$3\times2$ NaI(Tl) crystal array, with each crystal coupled with two PMTs and encased in the copper housing, are presented. 
    The crystal arrangement photon (a) and its diagram (b) show that the crystals stand on the acrylic supporting table. 
    The six crystals are labeled as NEO-1 through NEO-6.
    Two vertical stainless-steel pipes are used for calibration. 
  }
  \label{ref:crystals}
\end{figure*}

\subsection{Crystal placement and detector calibration}
The six NaI(Tl) crystals are arranged in a 3$\times$2 array supported by a two-layer acrylic table located in the central region of the LS. The crystal arrangement and numbering scheme are shown in Fig.~\ref{ref:crystals}.

\begin{figure*}[!htb]
  \begin{center}
    \begin{tabular}{cc}
      \includegraphics[width=0.5\textwidth]{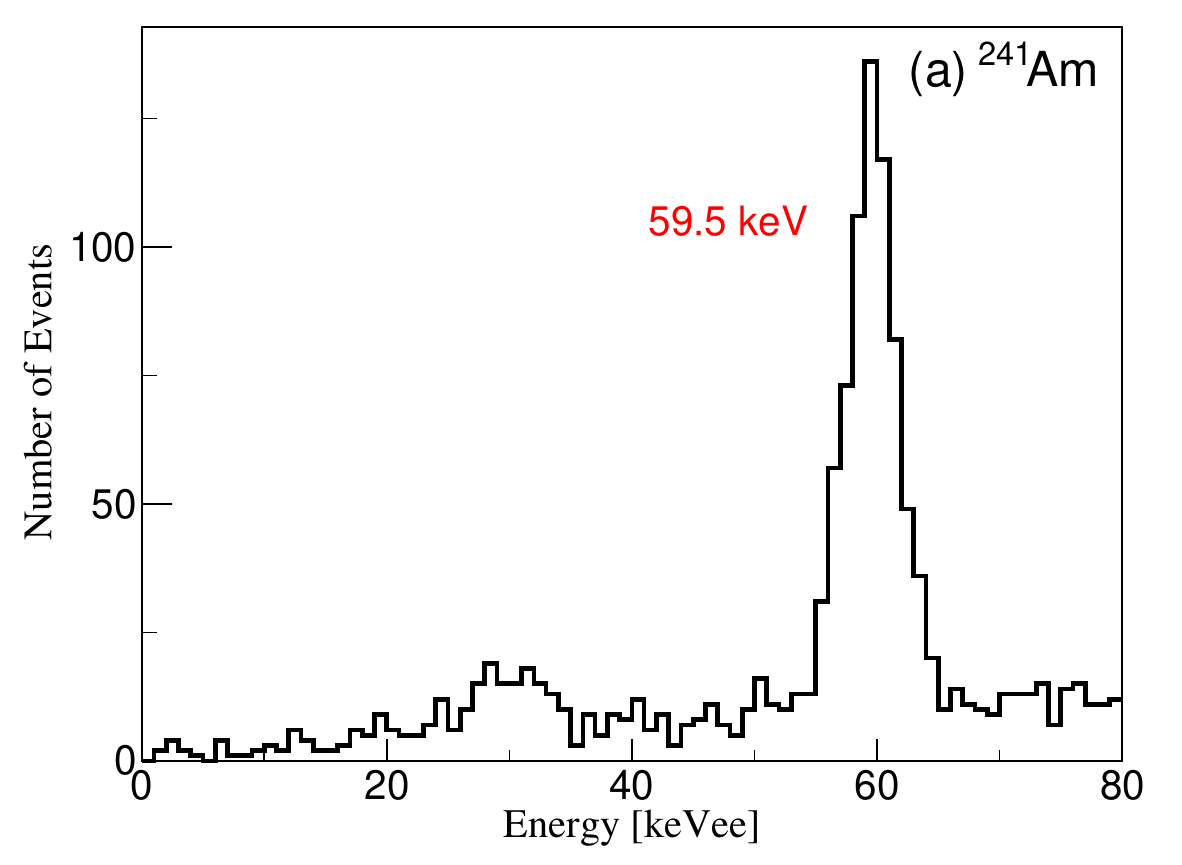} &
      \includegraphics[width=0.5\textwidth]{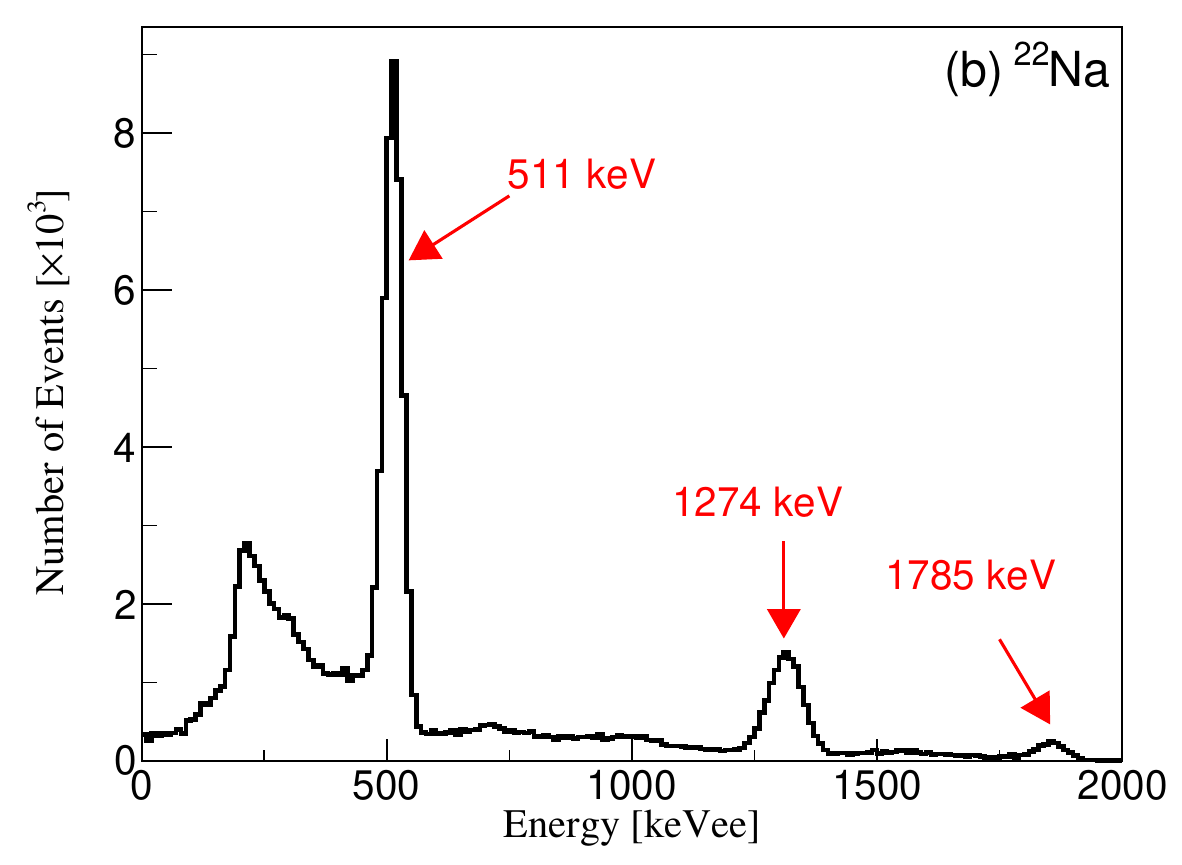} \\
    \end{tabular}
  \end{center}
  \caption{Energy spectra of \am{} (a) and \na{} (b) calibration data of the NEO-5 crystal.}
  \label{ref:energy_cal}
\end{figure*}

The energy scales of the NaI(Tl) crystals and LS are measured with two $\gamma$-ray sources, $^{241}$Am and $^{22}$Na. 
Figure~\ref{ref:energy_cal} shows the energy spectra of the NEO-5 crystal obtained using the \am{} (a) and \na{} (b) sources. 

\subsection{Internal contamination of radioactive elements in the NaI(Tl) crystals} 
Dominant background contributions in the low-energy signal region for NaI(Tl) crystals are due to the internal contaminants, such as $^{40}$K and $^{210}$Pb, of the radioactive materials~\cite{cosinebg,Adhikari:2021rdm}. Table~\ref{table:LightYield} presents the measured results for the internal background of the six crystals.

\begin{table*}[htb]
  \begin{center}
    \begin{tabular}{c c c c c c c c c} 
      \hline
      Crystal & Mass (kg)    & Size      &  $^{nat}$K    & $\alpha$ Rate  & $^{210}$Pb & $^{216}{\rm Po}$ & $^{218}{\rm Po}$  & Light yield   \\
              &              & (inch, D $\times$ L) & (ppb)& (mBq/kg)&(mBq/kg) &($\micro$Bq/kg)&($\micro$Bq/kg)& (NPE/keV) \\
      \Xhline{4\arrayrulewidth}
      NEO-1 & 1.62  &  3 $\times$ 4  & 50$\pm$20& 2.16$\pm$0.02& 1.89$\pm$0.26 & 1.6$\pm$0.7 & 10.6$\pm$4.2 & 20.5$\pm$0.9\\
      NEO-2 & 1.67  &  3 $\times$ 4  & 137$\pm$28&7.78$\pm$0.03& 7.46$\pm$0.73 &    $<$59.8         &    $<$57.2          & 19.3$\pm$0.9\\
      NEO-3 & 1.67  &  3 $\times$ 4  & 46$\pm$20&0.56$\pm$0.01& 0.53$\pm$0.13 &  $<$3.6       &      $<$11.2      & 21.8$\pm$0.9\\
      NEO-4 & 3.35  &  3 $\times$ 8  & 22$\pm$11&0.76$\pm$0.01& 0.69$\pm$0.18 & 1.6$\pm$0.8 &      $<$3.3    & 22.4$\pm$1.0\\
      NEO-5 & 3.35  &  3 $\times$ 8  & $<$29&0.76$\pm$0.01& 0.68$\pm$0.17 & 1.6$\pm$0.5 & 2.9$\pm$1.6  & 21.8$\pm$0.9\\
      NEO-6 & 1.65  &  3 $\times$ 4  & $<$38&0.94$\pm$0.01& 0.88$\pm$0.21 & 5.8$\pm$1.3 & 11.0$\pm$3.3 & 21.7$\pm$1.0\\
      \hline
      COSINE-100(C6) & 12.5   &  4.8 $\times$ 11.8  &17$\pm$3& 1.52$\pm$0.04  & 1.46$\pm$0.07  & 2.5$\pm$0.8 &  $<$ 
      0.25 & 14.6$\pm$1.5\\
      \hline
    \end{tabular}
  \end{center}
  \caption{Measured radioactivity levels in the NEON crystals with their specifications are compared with one of the COSINE-100 crystals~\cite{Adhikari:2017esn}. 
	The light yields are measured with the 59.6\,keV $\gamma$ peak from a $^{241}$Am source. The levels of $^{216}$Po and $^{218}$Po are measured by the time coincidence measurements of the $\alpha$ particles that are elements of decay chains from $^{232}$Th and $^{238}$U chain, respectively. 
A novel detector encapsulation technique used for the NEON crystals enhances the light yields such that they are 53\,\% higher than that of the COSINE-100 crystal. 
Although commercial quality crystals are used, contaminations of radioactive elements in the NEON crystals are similar with those of the COSINE-100 crystal without NEO-2. 
  }\label{table:LightYield}
\end{table*}

\subsubsection{$^{40}$K background}

\begin{figure}[!htb]
      \includegraphics[width=0.5\textwidth]{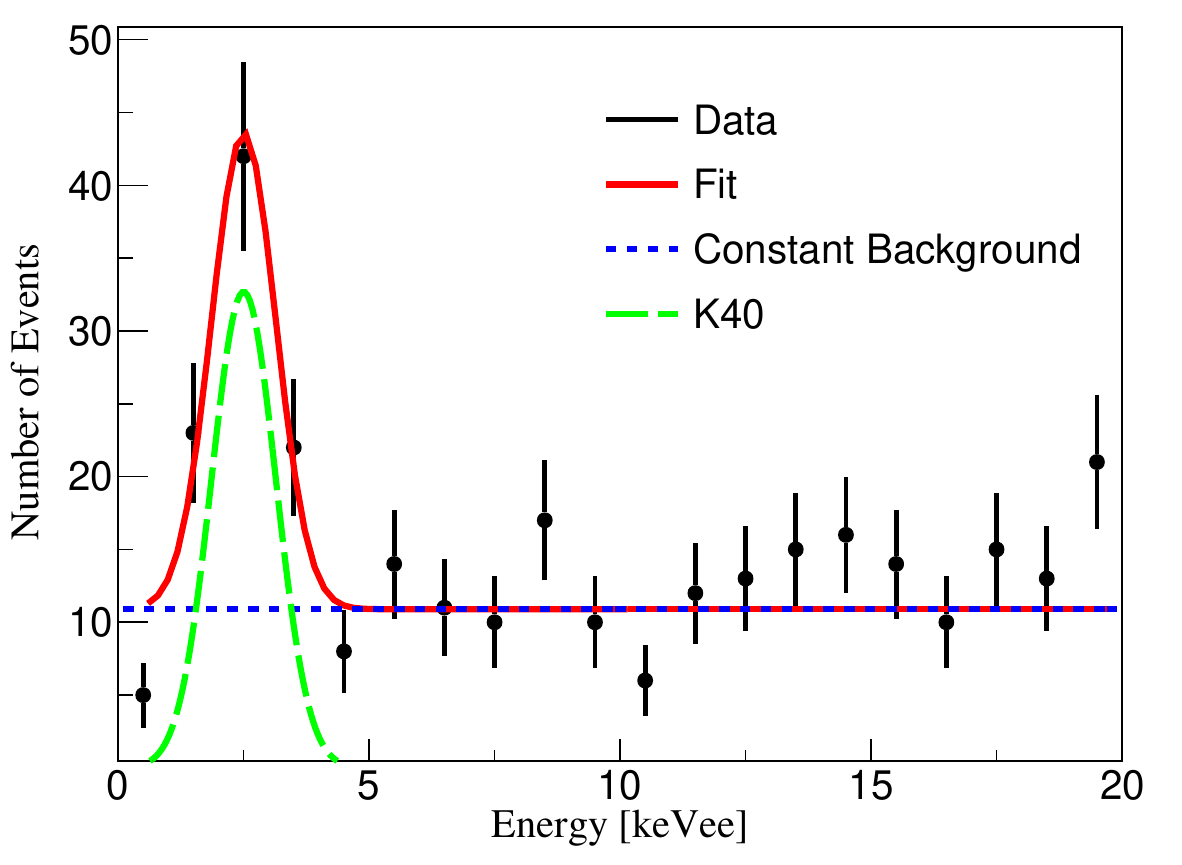} 
  \caption{$^{40}$K 3.2~keV spectrum in the NEO-2 crystal tagged by the 1460~keV $\gamma$s in the surrounding NaI(Tl) crystals.
  }
  \label{ref:K40}
\end{figure}

The $^{40}$K contamination is evaluated by studying the coincidence signals of approximately 3.2\,keV X-rays and 1460\,keV $\gamma$-rays tagged using surrounding crystals. 
Figure~\ref{ref:K40} shows the low-energy background spectrum of NEO-2 when 1460\,keV $\gamma$ events are tagged by surrounding the other crystals (NEO-1, NEO-3, NEO-4, NEO-5, and NEO-6). 
The $^{40}$K background level in each crystal is determined by comparing the measured coincidence rate with a GEANT4-simulated rate~\cite{Kim:2014toa,adhikari16} and is summarized in Table~\ref{table:LightYield}. 

\subsubsection{$\alpha$ analysis}
\begin{figure}[!htb]
      \includegraphics[width=0.5\textwidth]{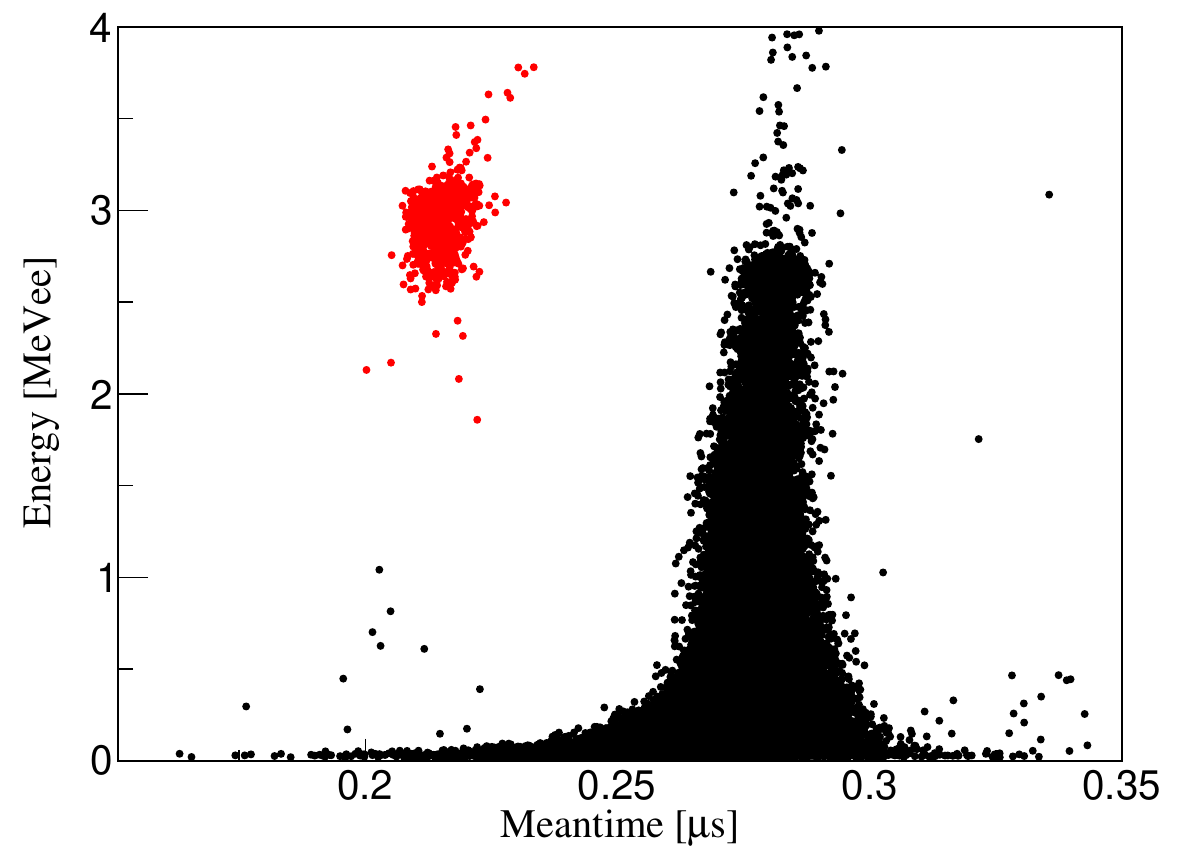} 
			\caption{Mean time {\sc versus} energy of the NEO-6 crystal. The mean time parameter separates the $\alpha$ events from the $\beta$/$\gamma$ events as indicated by the red dots in a short mean time.  
  }
  \label{ref:alpha_mt}
\end{figure}

Alpha-induced events inside the crystal can be identified based on the mean time of the signal, which is defined as
\begin{eqnarray}
\left<t\right> = \frac{\Sigma_i A_i t_i}{\Sigma_i A_i}-t_0,
 \label{meantime}
\end{eqnarray}
where $A_i$ and $t_i$ denote the charge and time of each time bin, respectively, and $t_0$ denotes the start time of an event evaluated from the rising edge near the trigger position. Here, a 1.5\,$\mu$s time window from $t_0$ is used for the mean time calculation. 
Figure~\ref{ref:alpha_mt} shows a scatter plot of the energy {\sc versus} mean time for the event signals from the NEO-6 crystal. Alpha-induced events are clearly separated from $\gamma$-induced events owing to the faster decay times of the $\alpha$-induced events. Alpha rates are summarized in Table~\ref{table:LightYield} that are the selected alpha events using the mean time parameter for the measured energy greater than 1\,MeV as shown in Fig.~\ref{ref:alpha_mt}.

\subsubsection{$^{232}$Th chain}
\begin{figure*}[!htb]
  \begin{center}
    \begin{tabular}{cc}
      \includegraphics[width=0.5\textwidth]{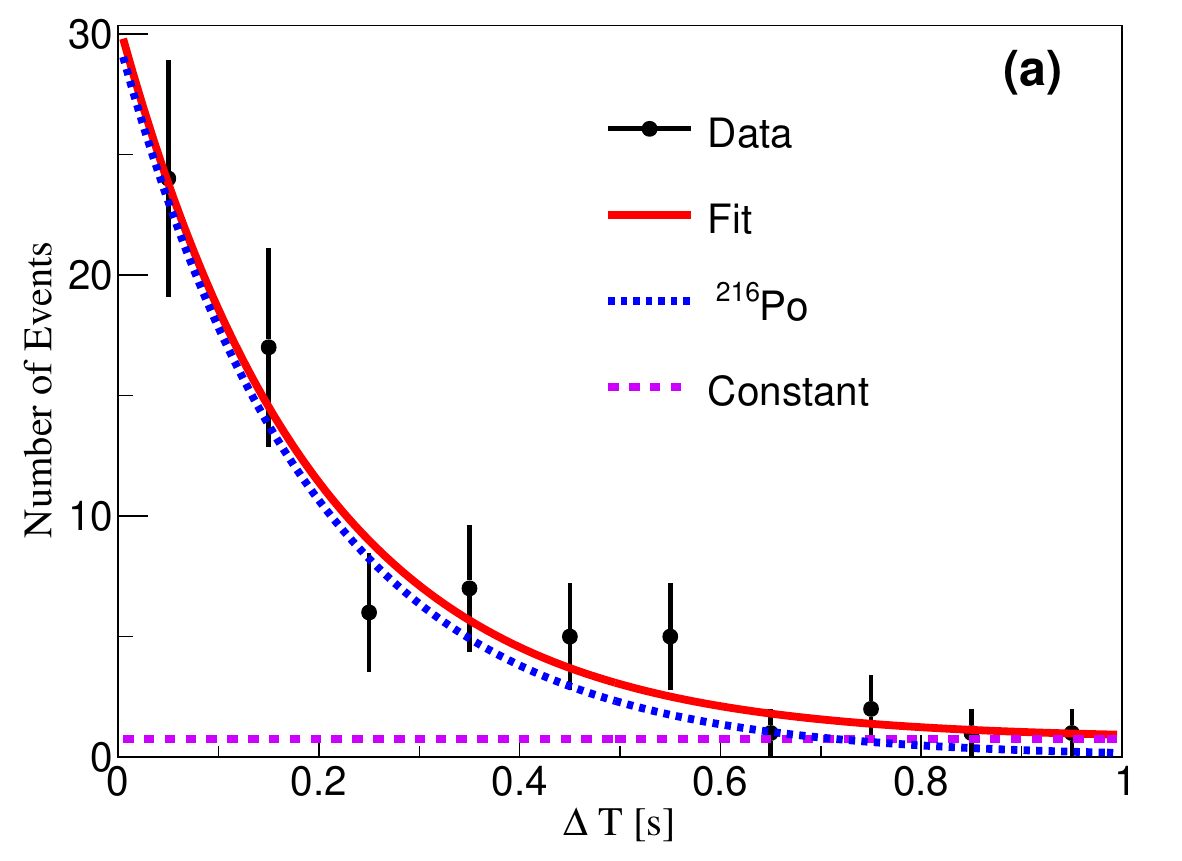} &
      \includegraphics[width=0.5\textwidth]{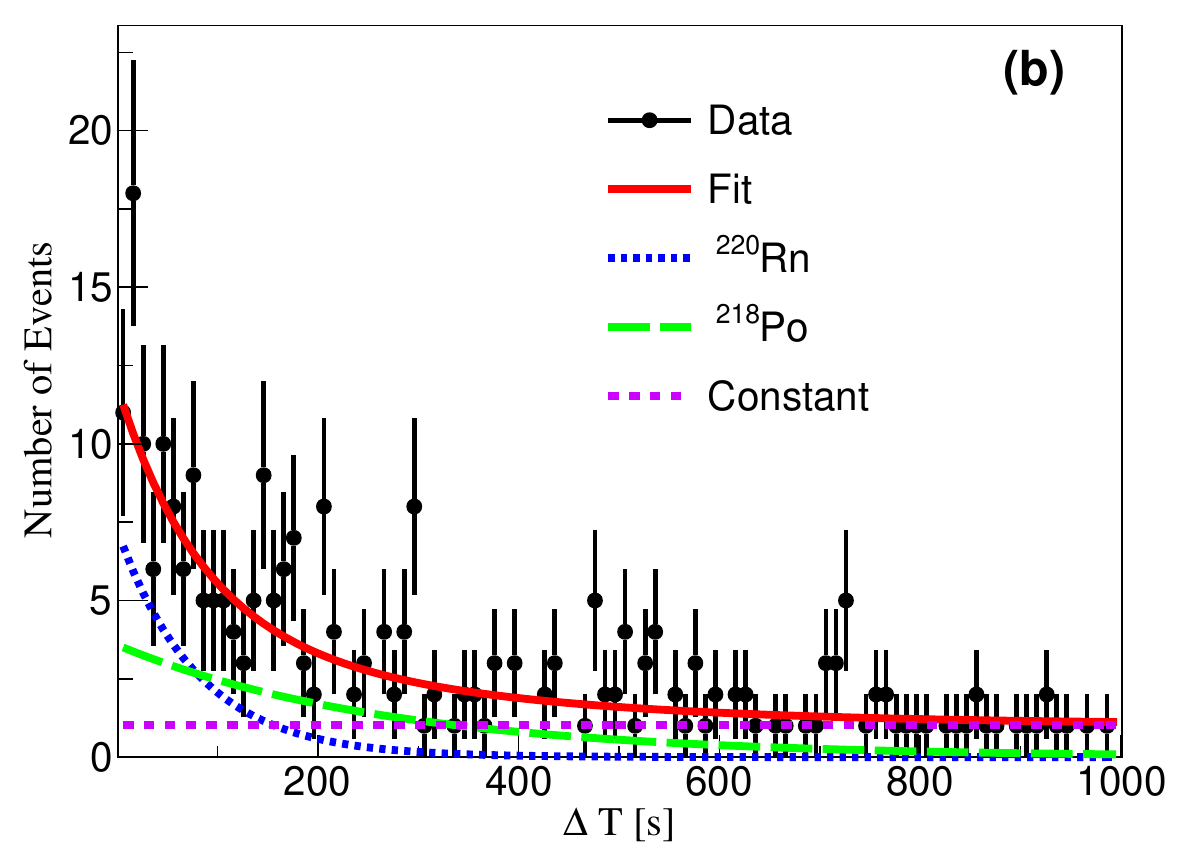} \\
    \end{tabular}
  \end{center}
	\caption{Time difference ($\Delta$T) distributions of data (points) and exponential fits (red solid line) between two successive $\alpha$-induced events are presented. Here, decay time used in the fit is fixed to the known lifetime of each radioisotope. (a) $^{216}$Po$\rightarrow ^{212}$Pb (half-life of 0.145\,s) events are extracted from the fit including an exponential component of the $^{216}$Po (blue dotted line) and random coincidence (purple dashed line) events. (b) $^{218}$Po$\rightarrow ^{214}$Pb (half-life of 186\,s) events are obtained from two exponential components of $^{218}$Po (green long dashed line) and $^{220}$Rn (half-life 55.6\,s), whereas the activity of $^{220}$Rn is constrained from $^{216}$Po measurement in (a), together with random coincidence events (purple dashed line) in the NEO-6 crystal. }
  \label{ref:alpha_alpha}
\end{figure*}

Contamination from the $^{232}$Th chain can be studied through $\alpha - \alpha$ time-interval measurements in the crystals. A $^{216}$Po $\alpha$ decay has a half-life of 145\,ms following its production via $^{220}$Rn$\rightarrow ^{216}$Po $\alpha$ decay. Figure~\ref{ref:alpha_alpha} (a) shows the distribution of the time difference between two $\alpha$-induced events of NEO-6, wherein an exponential component of the $^{216}$Po decay time can be observed. The $^{216}$Po contamination levels of all the crystals are listed in Table~\ref{table:LightYield} indicating $^{232}$Th contamination if we assume the chain equilibrium.  The chain equilibrium of $^{232}$Th decay is not verified.

\subsubsection{$^{238}$U chain}
Delayed coincidence $\alpha - \alpha$ events with a decay time  of 3.10\,min from $^{218}$Po$\rightarrow ^{214}$Pb can be used to infer the $^{238}$U contamination levels, as shown in Fig.~\ref{ref:alpha_alpha} (b). Furthermore, the measured rate for $^{220}$Rn$\rightarrow ^{216}$Po ( half-life of 55.6\,s) is extracted from the $^{232}$Th level, which is determined from $^{216}$Po$\rightarrow ^{212}$Pb decays. 
The $^{218}$Po contamination levels of all the crystals are listed in Table~\ref{table:LightYield} indicating $^{238}$U contamination with the assumption of the chain equilibrium. 

\subsubsection{$^{210}$Pb background}
The $^{238}$U and $^{232}$Th contamination levels measured by $\alpha - \alpha$ time correlation methods summarized in Table~\ref{table:LightYield} are too low to account for the total observed $\alpha$ rates although we assume the chain equilibrium. This suggests that the $\alpha$ rate is dominated by the decay of $^{210}$Po ($E_{\alpha}$ = 5.3\,MeV) nuclei. Considering $\alpha$ quenching in the NaI(Tl) crystals, the electron-equivalent measured energy of approximately 3\,MeV in Fig.~\ref{ref:alpha_mt} matches well with the $^{210}$Po $\alpha$ energy indicated in the literatures~\cite{Bernabei:2008yh,Adhikari:2017gbj}.  

\begin{figure}[!htb]
      \includegraphics[width=0.49\textwidth]{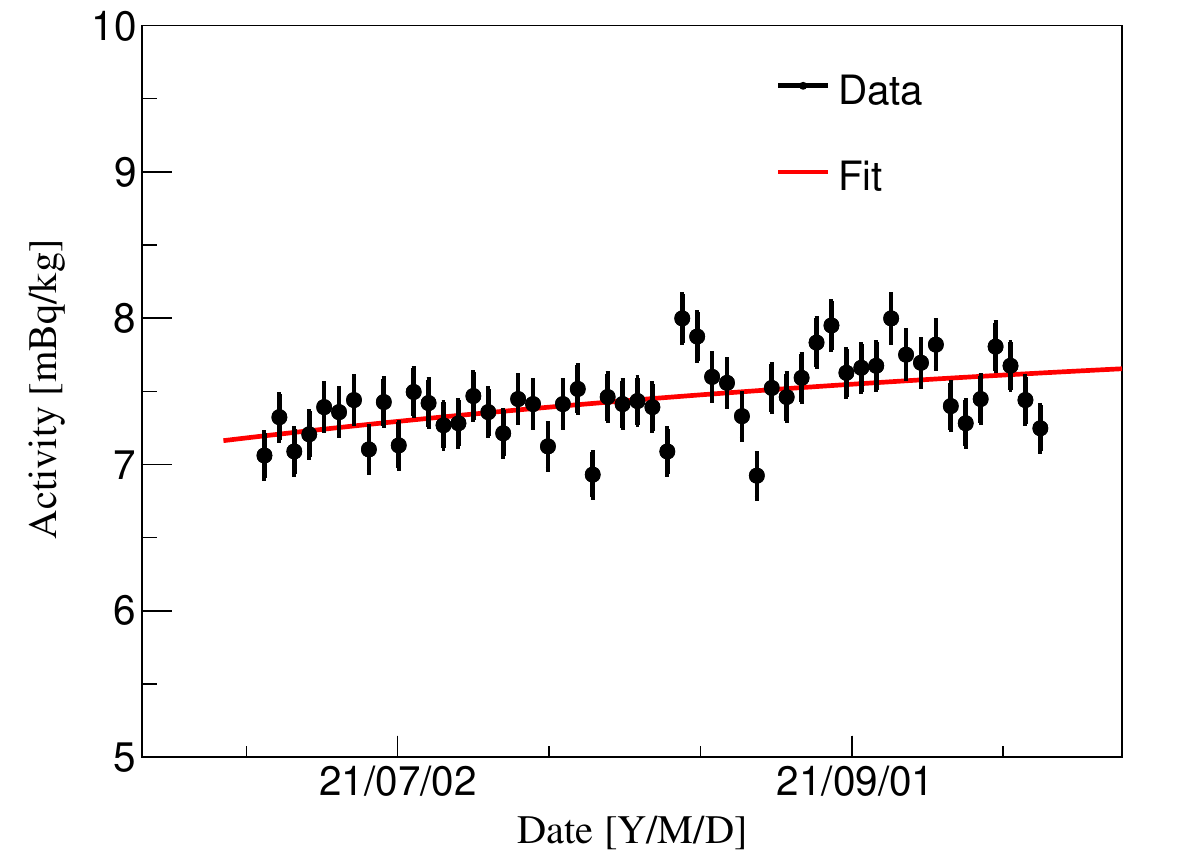} 
			\caption{ Total alpha rates in the NEO-2 crystal as a function of time are modeled with $^{210}$Po assuming a contamination of $^{222}$Rn (and/or $^{210}$Pb). }
	  \label{ref:Pb210rate}
	\end{figure}

In NaI(Tl) crystals, internal contamination of $^{210}$Pb was the dominant background in the low-energy signal region~\cite{Adhikari:2017gbj,Amare:2018ndh,Antonello2021,Fushimi:2021mez,Adhikari:2021rdm}.
$^{210}$Pb amounts can be studied with the alpha events owing to the decay of $^{210}$Po that originates from $\beta$-decay of the $^{210}$Pb nuclei as well as a 46.5\,keV $\gamma$ peak. 
As typical contamination of $^{210}$Pb occurred during crystallization by exposure to $^{222}$Rn, the $^{210}$Po decay grows with a lifetime of $^{210}$Po, $\tau_{^{210}\text{Po}}$=200\,days as an example shown in Fig.~\ref{ref:Pb210rate}. From this fit, the $^{210}$Pb amount can be extracted~\cite{Adhikari:2017esn,Park:2020fsq}. 
The measured $^{210}$Pb levels are summarized in Table~\ref{table:LightYield}.

\subsection{Light Yield}
\begin{figure}[!htb]
      \includegraphics[width=0.5\textwidth]{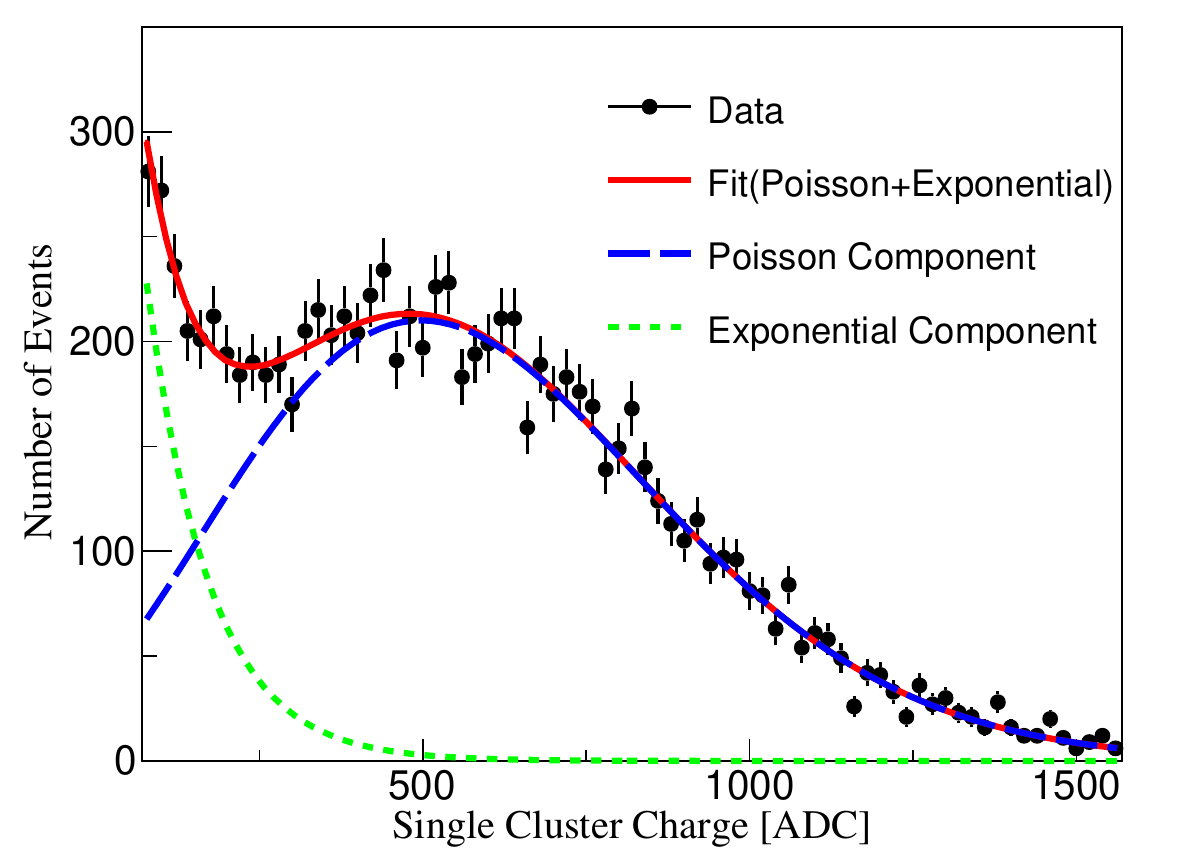} 
  \caption{Single cluster charge spectrum of one PMT attached to the NEO-5 crystal is modeled with exponential backgrounds (pedestal and low-gain dark current, green dashed line) and single photoelectron (Poisson component, blue long dashed line). 
  }
  \label{ref:LYspe}
\end{figure}

$\gamma$ rays from the $^{241}$Am source with an energy of 59.54\,keV are used to evaluate the light yields of the NaI(Tl) crystals. 
The charge distribution of single photoelectrons (SPEs) is obtained by identifying isolated clusters at the decay tails of the 59.54\,keV signal (2--5\,$\mu$s after the signal started) to suppress multiple photoelectron clusters, as shown in Fig.~\ref{ref:LYspe}. Furthermore, the total charges from 59.54\,keV $\gamma$ are divided by the measured single photoelectron charge to obtain the light yield per unit keVee. As shown in Table~\ref{table:LightYield}, approximately 22\,NPE/keVee light yields are achieved. Two crystals, NEO-1 and NEO-2, exhibit relatively small light yields of approximately 20\,NPE/keVee owing to the development of cracks because of the initial polishing process. 

\subsection{Background levels of the NaI(Tl) crystals}
\begin{figure*}[!htb]
  \begin{center}
    \begin{tabular}{ccc}
    \includegraphics[width=0.33\textwidth]{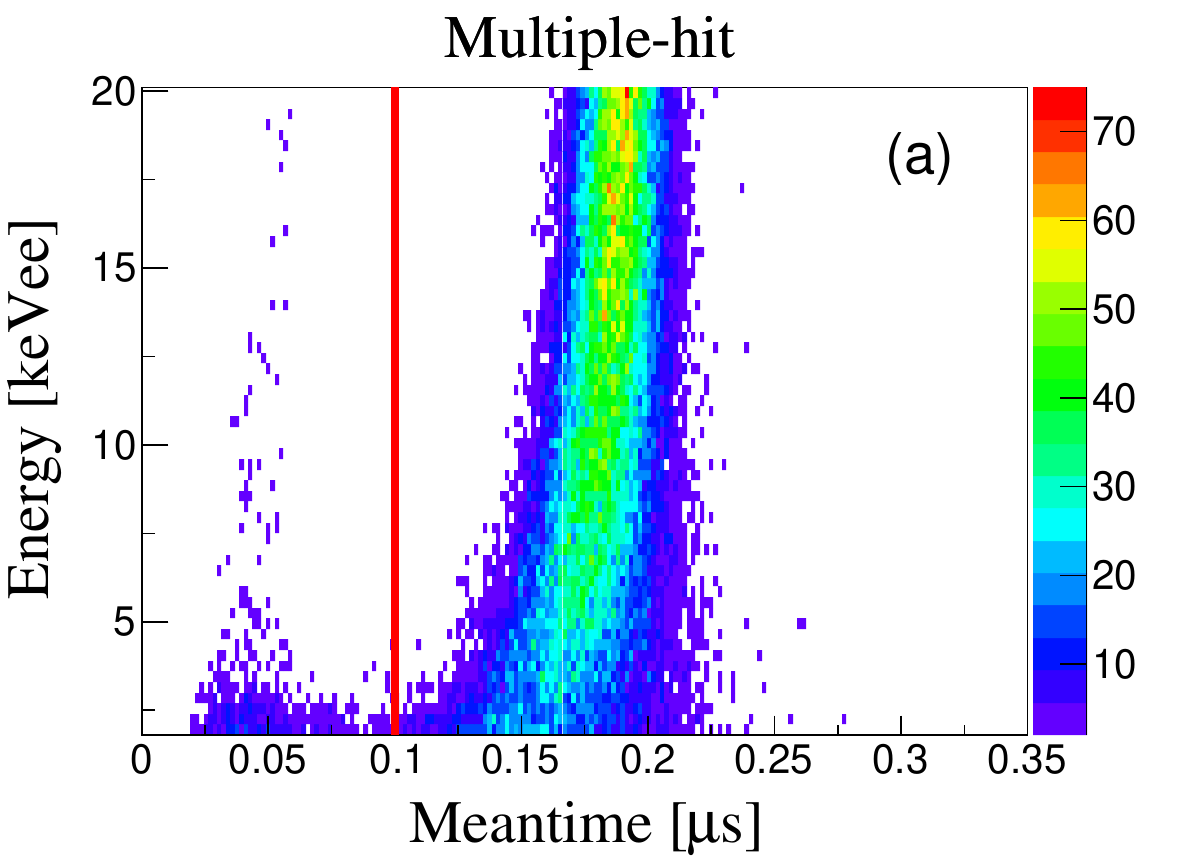} &
    \includegraphics[width=0.33\textwidth]{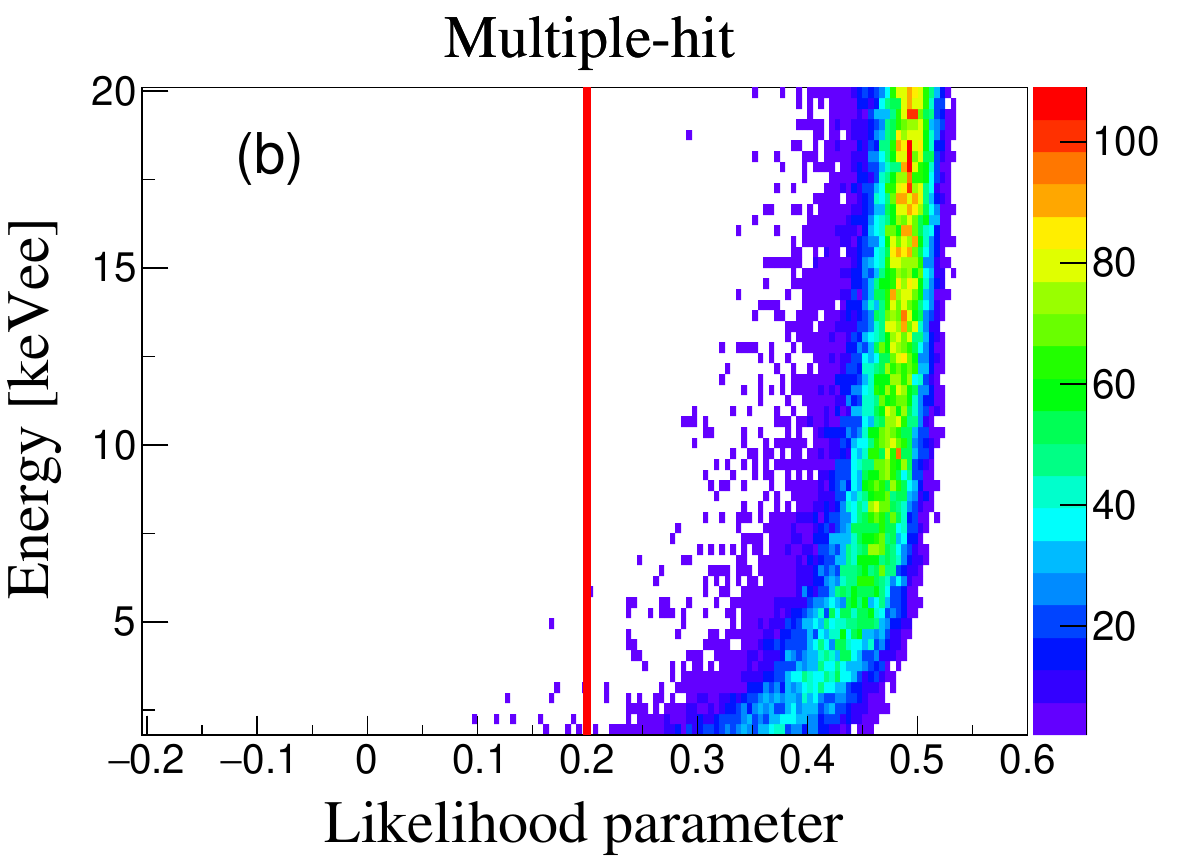} &
    \includegraphics[width=0.33\textwidth]{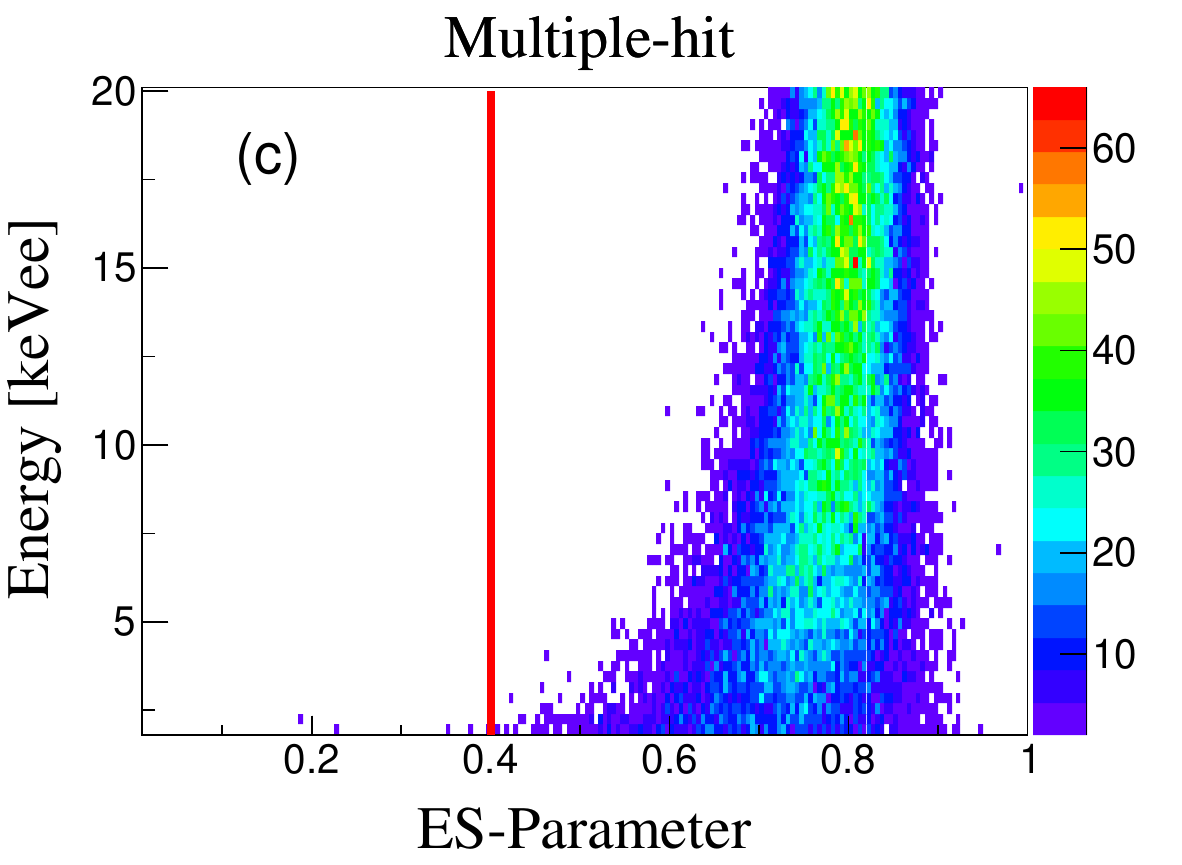} \\
    \includegraphics[width=0.33\textwidth]{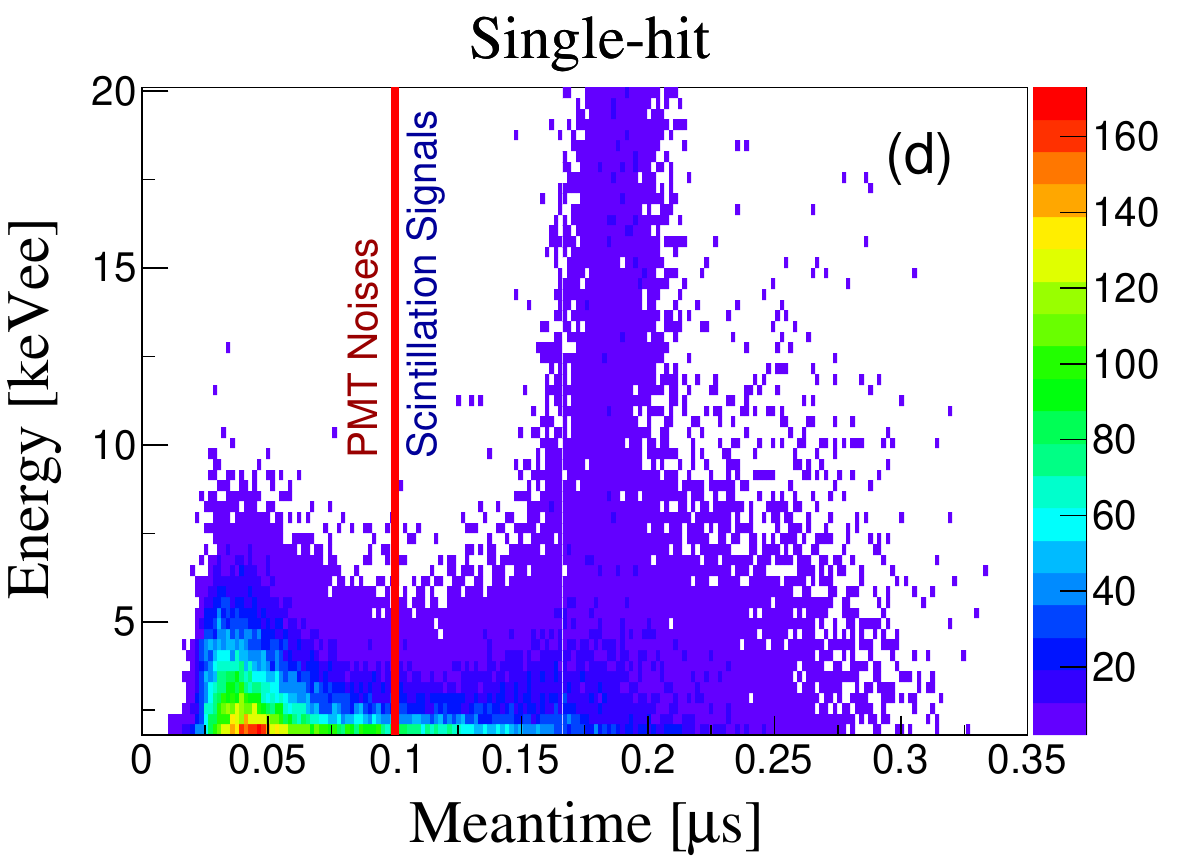} &
    \includegraphics[width=0.33\textwidth]{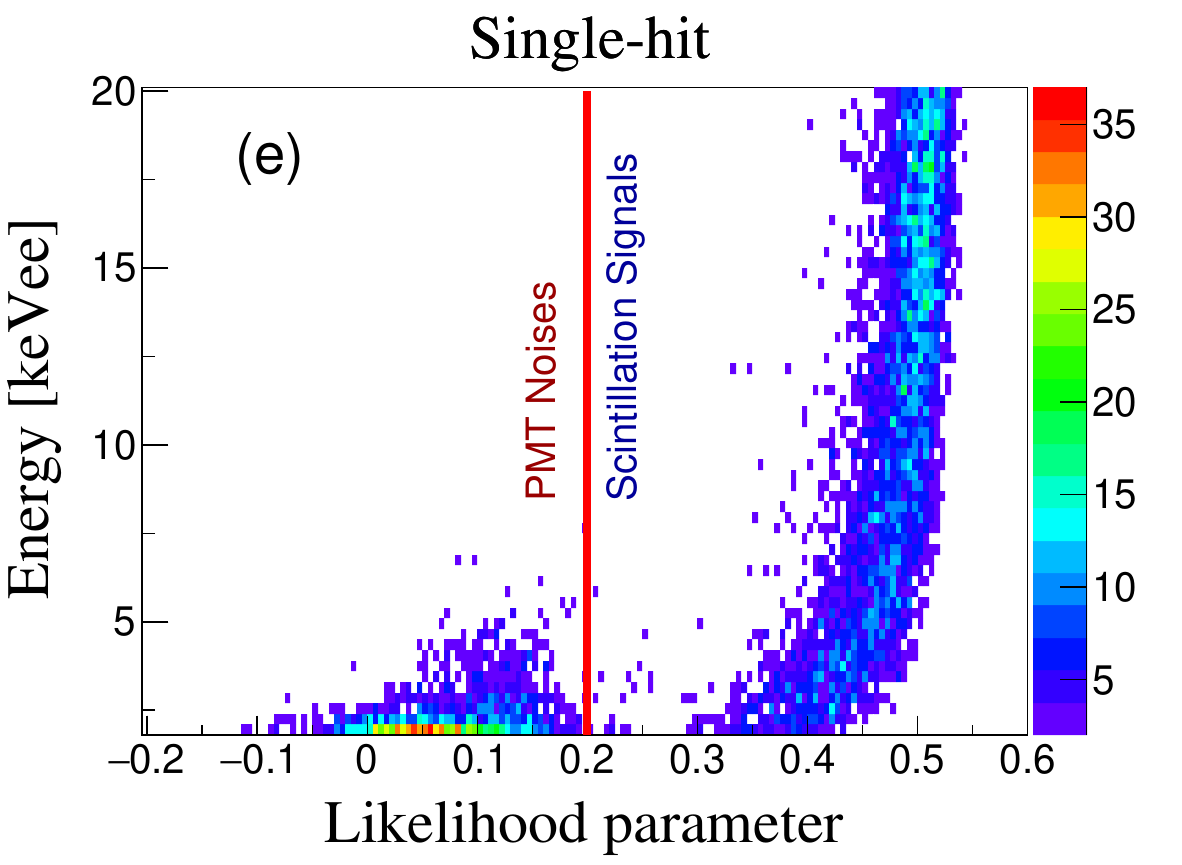} &
    \includegraphics[width=0.33\textwidth]{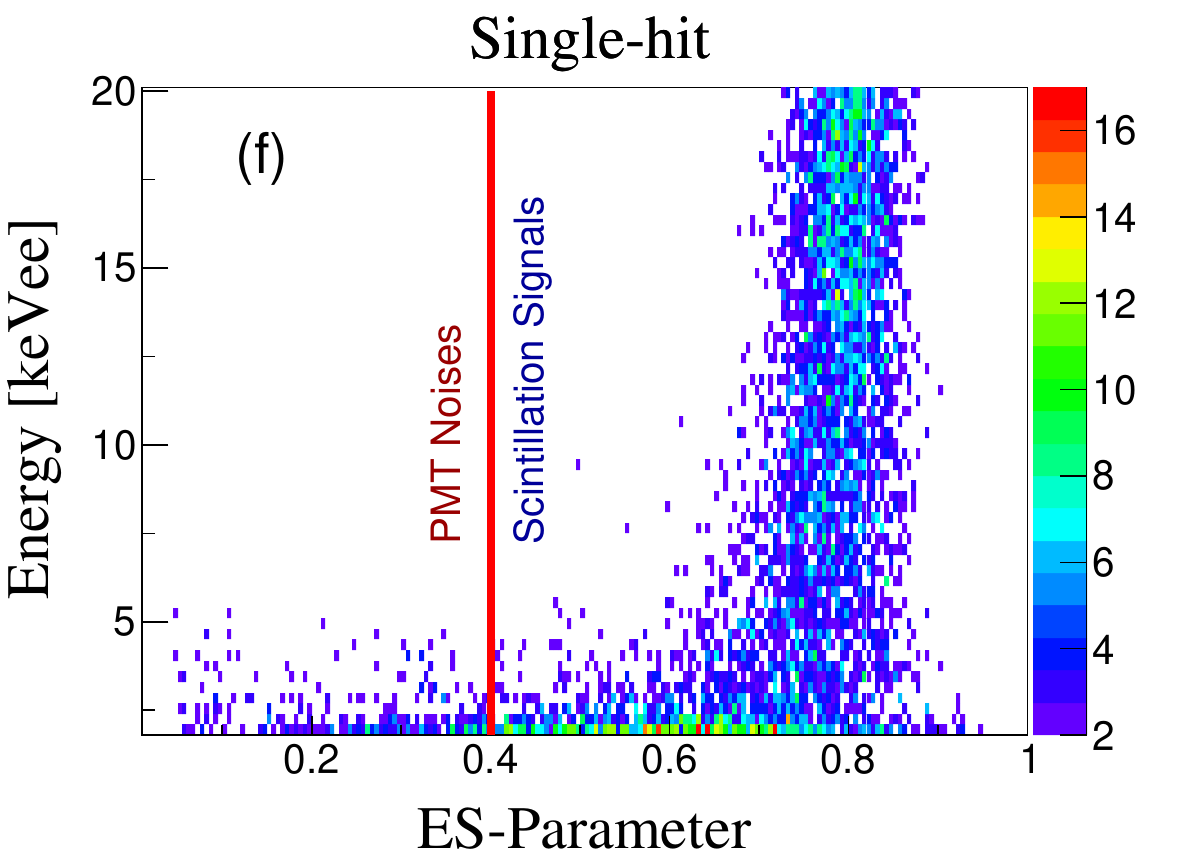} \\
    \end{tabular}
  \end{center}
	\caption{Distribution of the event selection parameters: mean time (left (a) and (d)), likelihood parameter (middle (b) and (e)), and ES parameter (right (c) and (f)), are presented separately for the multiple-hit (top (a) (b) (c)) and single-hit data (bottom (d) (e) (f)). As the multiple-hit data is less affected by the PMT-induced noise events, the selection criteria (red solid lines) were chosen so that the selection efficiency would be greater than 99\,\%. The likelihood and ES parameters are shown only for the mean time accepted events. 
	}
  \label{ref:boxcut}
\end{figure*}

Low-energy data acquired via NaI(Tl) crystals predominantly collect non-physical events that are primarily caused by PMT-induced noise. 
These noise events can be caused by the radioactive impurities in the PMTs, discharge of an accumulated space charge, PMT dark current, and large pulses. 
In the COSINE-100 experiment, an efficient noise rejection method was developed by employing a machine learning technique with multiple parameters using BDT~\cite{adhikari2020lowering}. 
As the BDT-based machine learning technique is under development for NEON data, box cuts for multiple parameters are applied. The parameters used in the event selection are the mean time, which is the amplitude-weighted average time of the events, a likelihood parameter for samples of scintillation-signal events and fast PMT-induced events~\cite{adhikari2020lowering}, and the DAMA ES parameter~\cite{Bernabei:2020mon}, which is based on the difference between the trailing-edge (100-600\,ns) and leading-edge (0-50\,ns) charge ratios to the total charge (0-600\,ns).  
Figure~\ref{ref:boxcut} shows the event selection parameters for multiple- and single-hit data. As the multiple-hit data contained fewer PMT-induced noise events, the selection criteria are developed. An event selection efficiency above 2\,keVee is maintained at more than 99\,\%. 

\begin{figure*}[!htb]
  \begin{center}
    \begin{tabular}{cc}
    \includegraphics[width=0.5\textwidth]{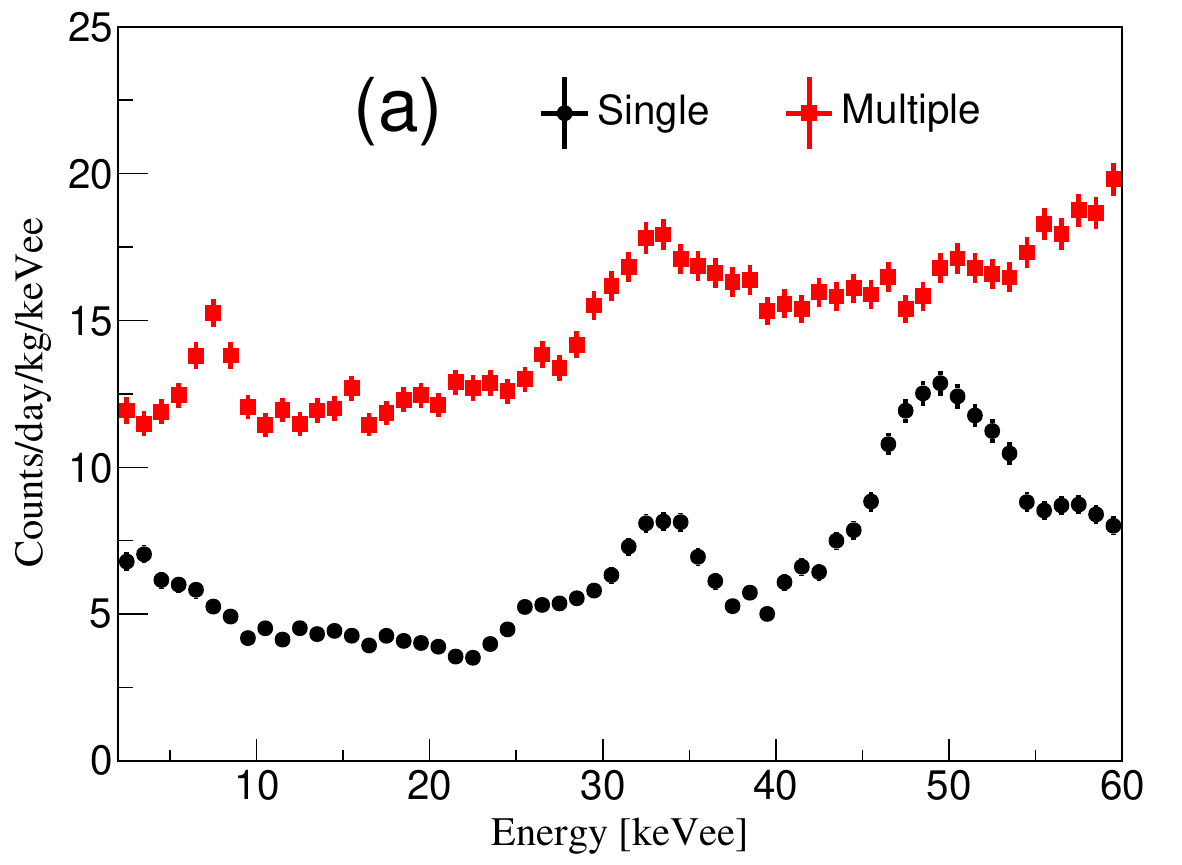} &
    \includegraphics[width=0.5\textwidth]{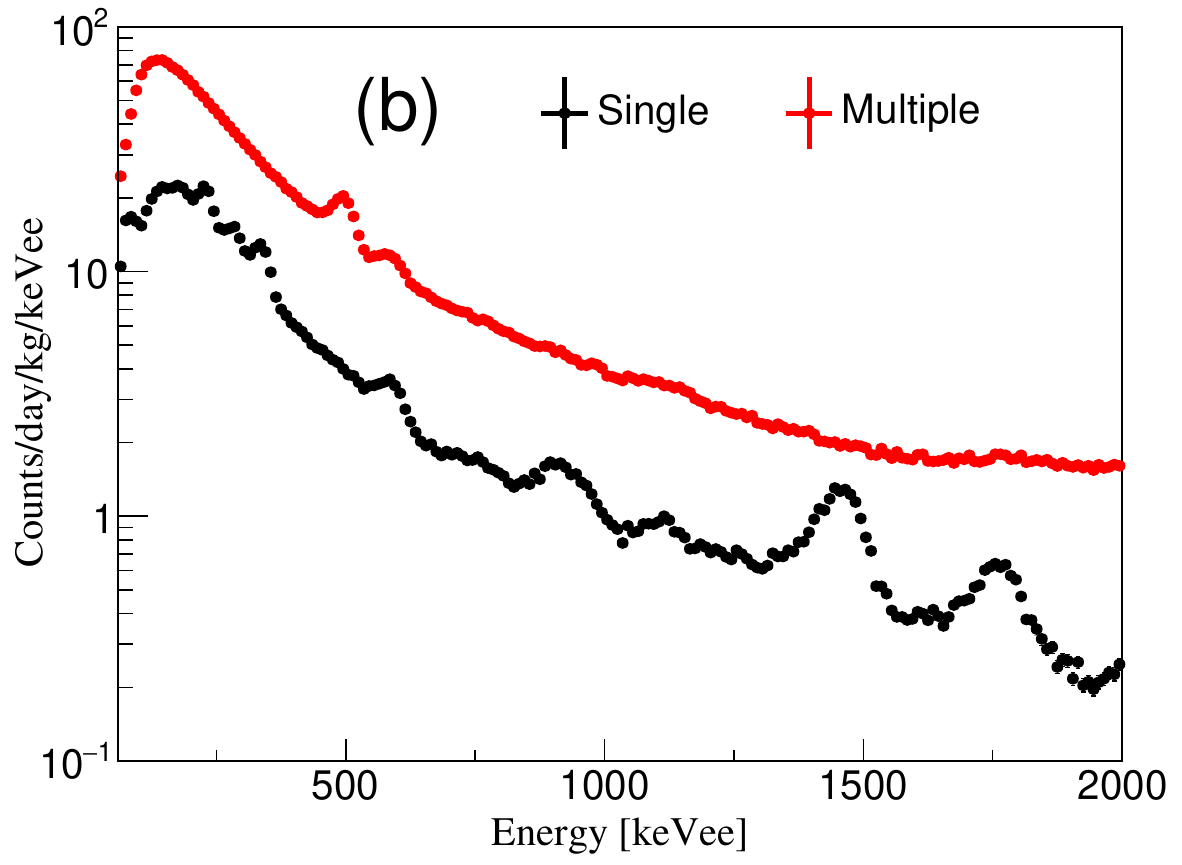} \\
		(a) 2--60\,keVee & (b) 60--2000\,keVee\\
    \end{tabular}
  \end{center}
  \caption{Background spectra of the NEO-5 crystal, which are measured using the full power of the reactor, are shown for the single-hit (black dots) and multiple-hit (red squares) events. An effective active veto using the liquid scintillator is shown as a large number of multiple-hit events. 
The single-hit 2--6\,keVee has approximately 6\,counts/kg/keV/day background level. }
  \label{ref:bkgspec}
\end{figure*}

\begin{figure}[!htb]
    \includegraphics[width=0.49\textwidth]{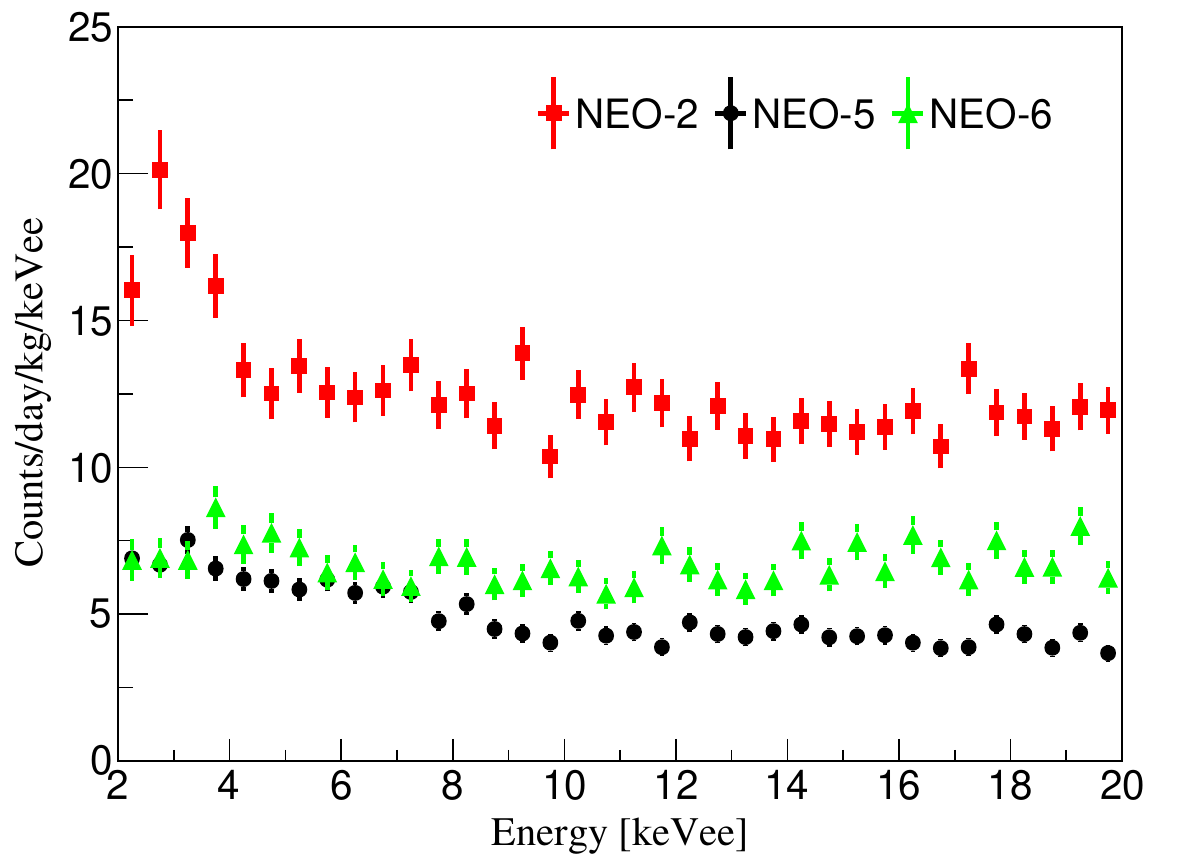} 
		\caption{Single-hit energy spectrum comparisons for three different-sized crystals (4-inch long NEO-6 and 8-inch long NEO-5) and different contamination levels (NEO-2) are shown for an energy region of 2--20\,keVee. Owing to a particularly large contamination of $^{210}$Pb and $^{40}$K, the NEO-2 crystal has approximately twice the background level. Although the 4-inch-long crystal has a higher background level above the 8\,keVee energy region owing to the larger impact of the external radiation, the low-energy spectrum around 2--6\,keVee region is similar with the 8-inch-long crystal when the internal contamination levels are similar.  }
  \label{ref:bkgspeccom}
\end{figure}

We process  data obtained between September 2021 and November 2021 at the tendon gallery of reactor unit-6 with full power in the Hanbit nuclear power complex. 
Figure~\ref{ref:bkgspec} (a) and (b) show examples of the background spectra from the NEO-5 crystal in the 2—60\,keVee and 60—2000\,keVee regions, respectively, following the application of the selection criteria. 
Here, the low-energy and high-energy spectra are calibrated with a 59.54\,keV line of $^{241}$Am and 511, 1274, and 1785\,keV lines from $^{22}$Na, assuming a linear response of NaI(Tl) scintillation.
A multiple-hit event is classified when the other crystals or LS have hits that cannot be caused by \cenns neutrino interactions. 
A single-hit event has a hit only in a single crystal. The single-hit low-energy region corresponding to 2--6\,keVee presents a background level of approximately 6\,counts/kg/keV/day, although the multiple-hit region has a background level of approximately 13\,counts/kg/keV/day. 
As the NEON shield does not have a muon tagging detector and has only a 10\,cm-thick lead, relatively large backgrounds from external radiation are observed, particularly for the multiple-hit events compared to those of the COSINE-100 crystals~\cite{Adhikari:2017esn}. 
However, the LS detector tags the dominant external background events; therefore, the single-hit physics data achieved are only twice as high as the COSINE-100 data.
Figure~\ref{ref:bkgspeccom} shows the low-energy single-hit spectra of three of the crystals: NEO-2, NEO-5, and NEO-6. The other crystals have similar background distributions with similar sizes and internal contamination crystals; for example, that of NEO-4 is similar to that of NEO-5, and those of NEO-1 and NEO-3 are similar to that of NEO-6. 
As the NEO-2 crystal contains a particularly large amount of $^{40}$K and $^{210}$Pb, as summarized in Table~\ref{table:LightYield}, the background level of the NEO-2 crystal is approximately twice larger than those of the other NEON crystals, as shown in Fig.~\ref{ref:bkgspeccom}. 
As the length of NEO-6 (4-inch long) is only half size of NEO-5 (8-inch long), the relative background contributions per unit kg weight from external radiations, such as PMT radioactivities, in the NEO-6 crystal are larger than those of the NEO-5 crystal. This results in increased rates of 4-inch-long crystals above the 8\,keVee energy regions, as shown in Fig.~\ref{ref:bkgspeccom}. However, contributions from external radiation are quickly reduced in low-energy single-hit events~\cite{cosinebg,Adhikari:2021rdm} so that the measured background levels at 2--6\,keVee are similar between 4-inch long and 8-inch-long crystals in the case of similar internal contaminations.

\begin{figure}[!htb]
  \begin{center}
    \includegraphics[width=0.3\textwidth]{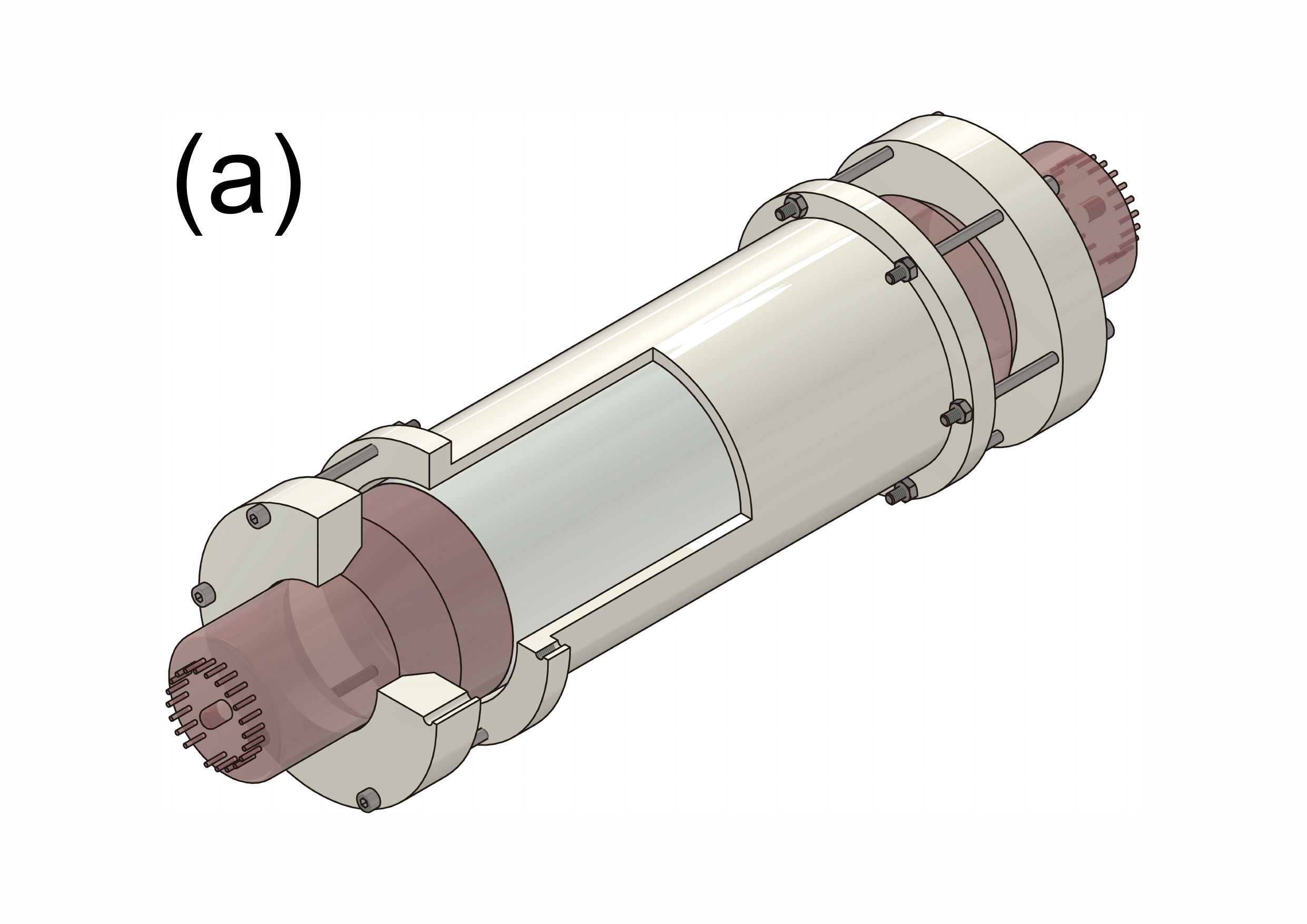}
    \includegraphics[width=0.4\textwidth]{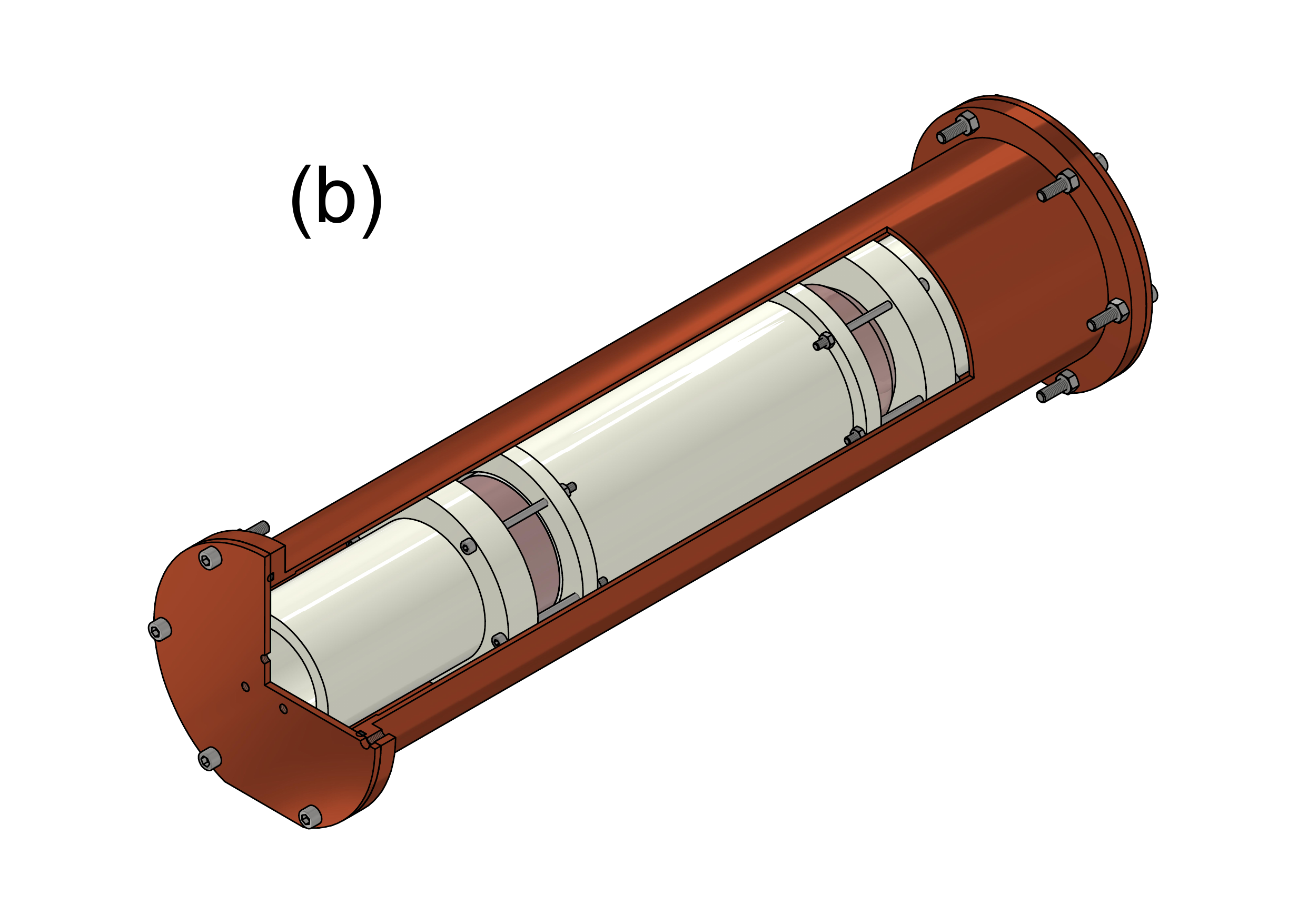}
  \end{center}
  \caption{Updated design of the NaI(Tl) detector encapsulation. (a) Inner structure for a mounting between the crystal and PMTs. (b) Encasement of the crystal-PMTs structure in the copper case.}
  \label{ref:newencap}
\end{figure}

Current NEON crystals are installed inside the inner acrylic box to avoid direct contact between the PMTs and LS. 
This design results in an increased background owing to $^{222}$Rn in the volume of the inner acrylic box and reduces tagging efficiency for the events from radioactive decay of the PMTs. 
An upgrade of the current NaI(Tl) encapsulation has been planned to immerse the detector directly into the LS, similar to the COSINE-100 design~\cite{Adhikari:2017esn}. 
Figure~\ref{ref:newencap} presents the upgraded encapsulation design for NaI(Tl) crystals that encase PMTs with air-tight O-rings in the copper housing. 
The NaI(Tl) detectors are directly immersed in the LS without the inner acrylic box, and further reduction of the background is expected. 

\section{Liquid scintillator veto system} 
The LAB-LS comprises 3\,g/L flour 2,5-diphenyloxazole(PPO), which is the primary fluorescent material, and 30 mg/L p-bis(o-methylstyryl)-benzene(bis-MSB) as a wavelength shifter~\cite{JS_Park_1,Park:2017jvs,Adhikari:2020asl}. 
To provide 800 L of LAB-LS, an 80 L concentrated master solution of PPO and bis-MSB is prepared. 
The master solution is mixed with LAB in a 1:20 ratio to obtain the final LAB-LS. The LS is produced in a surface-level laboratory and delivered to the reactor site. 

\begin{figure*}[!htb]
  \begin{center}
    \begin{tabular}{ccc}
      \includegraphics[width=0.33\textwidth]{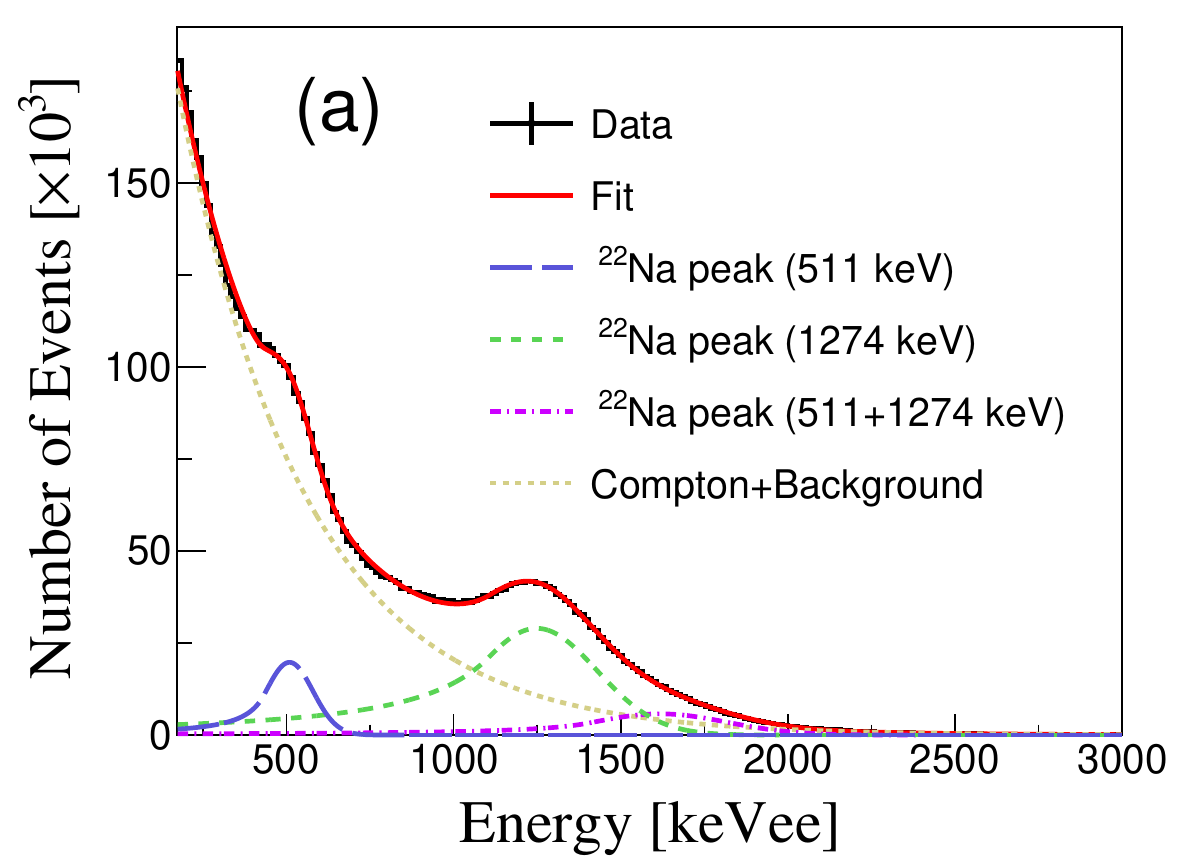} &
      \includegraphics[width=0.33\textwidth]{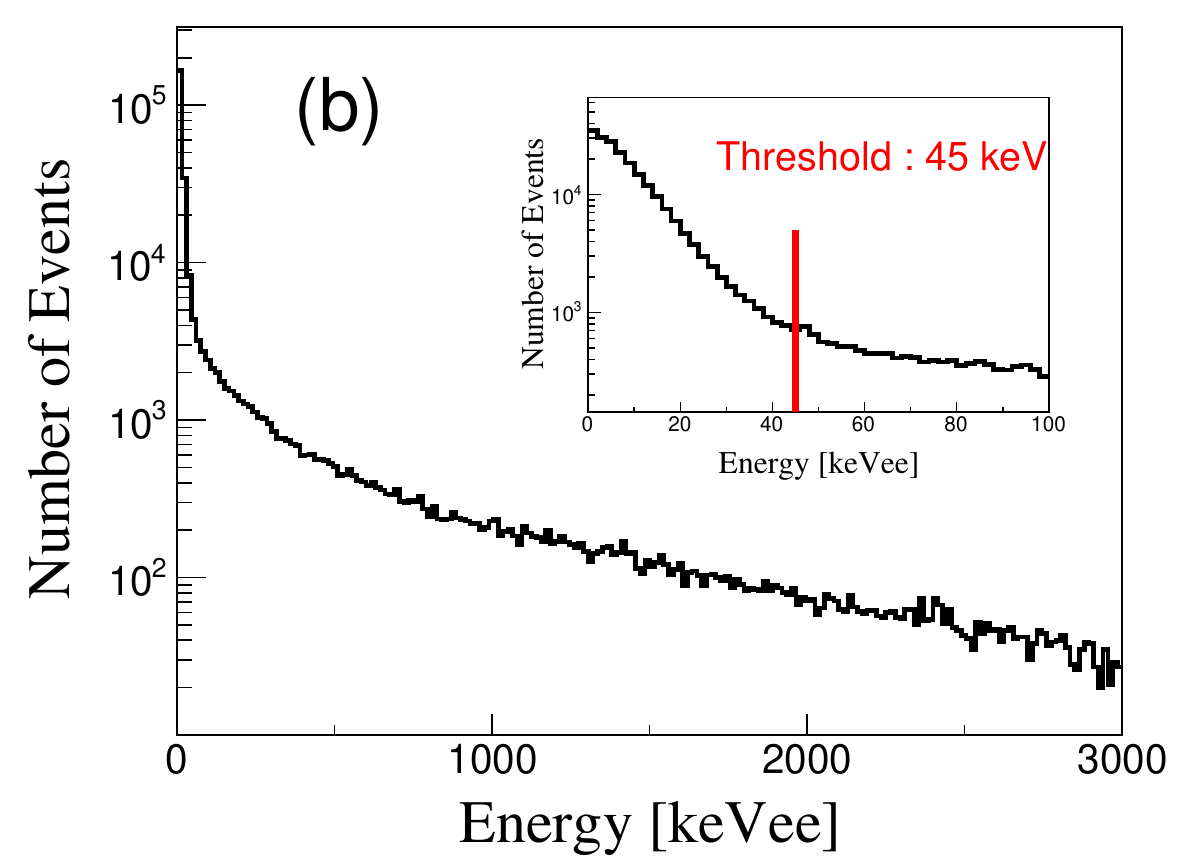} &
      \includegraphics[width=0.33\textwidth]{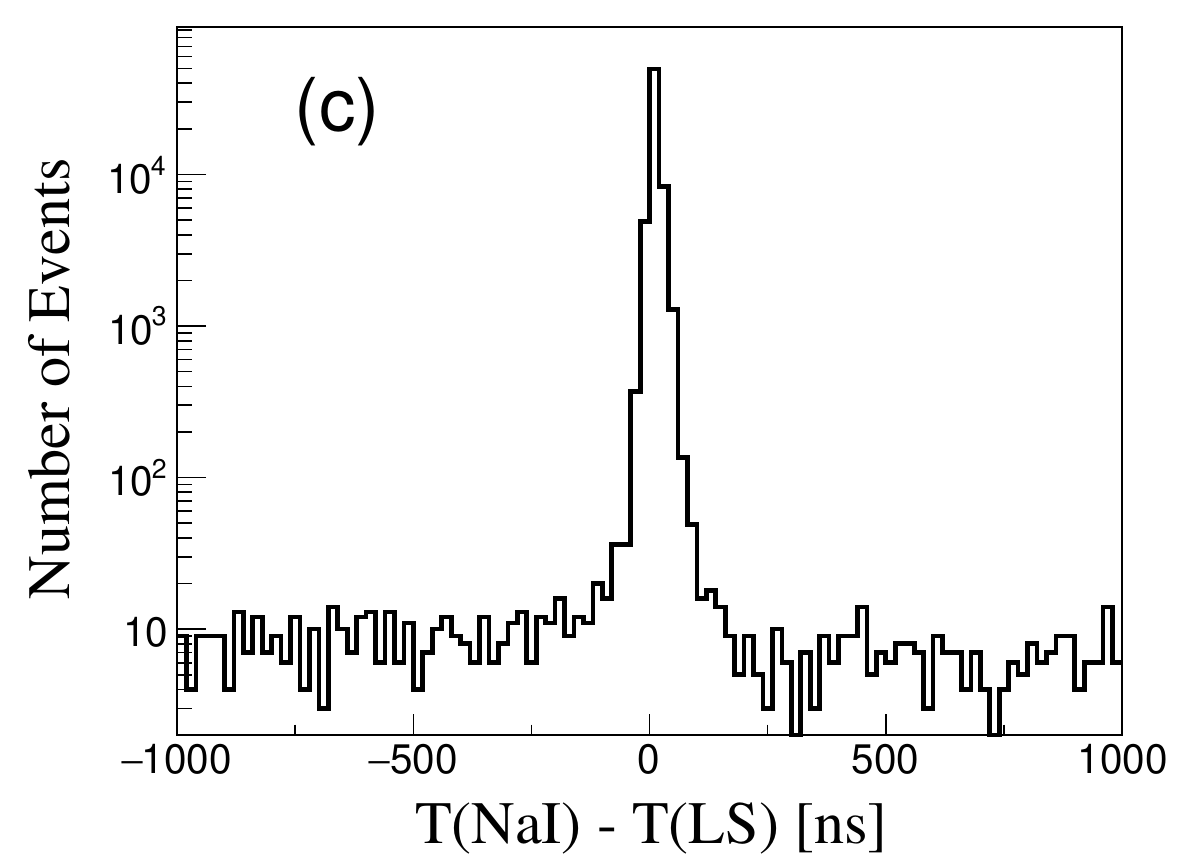} \\
    \end{tabular}
  \end{center}
  \caption{(a) \na calibration spectrum of the LS is modeled with known $\gamma$ peaks. (b) Background energy spectrum of the LS is presented. Inset shows zoomed spectrum at the low energy and presents a 45\,keV energy threshold for the multiple-hit requirement. (c) Time differences between NaI(Tl) crystals and LS are presented. Time differences between NaI and LS are required to be within $\pm$150\,ns so that the random coincidence events are less than 0.1\,\%.
  }
  \label{ref:LSplot}
\end{figure*}

Energy calibration of the LS veto system is performed with a $^{22}$Na $\gamma$-ray source that produces two 511\,keV and one 1275\,keV $\gamma$ simultaneously. 
Figure~\ref{ref:LSplot} (a) shows the energy spectra of the LS detector with $^{22}$Na calibration. 
Following the application of energy calibration to the data, the LS-deposited energy spectrum is obtained as shown in Fig.~\ref{ref:LSplot} (b). 

To avoid baseline contribution, a 45\,keV energy threshold from the LS is required for coincident multiple-hit event selection. 
Under these conditions, a clear time coincidence between the crystals and LS can be observed, as shown in Fig.~\ref{ref:LSplot} (c). For the coincident multiple-hit event, there is an additional requirement for the time difference between the NaI(Tl) crystals and LS to be within $\pm$150\,ns. 
These multiple-hit requirements provide random coincidence events of less than 0.1\,\%.

\section{Data acquisition system and electronics}
\subsection{DAQ System and Electronics}
\label{sec:DAQ}
\begin{figure}[!htb]
  \begin{center}
      \includegraphics[width=0.475\textwidth]{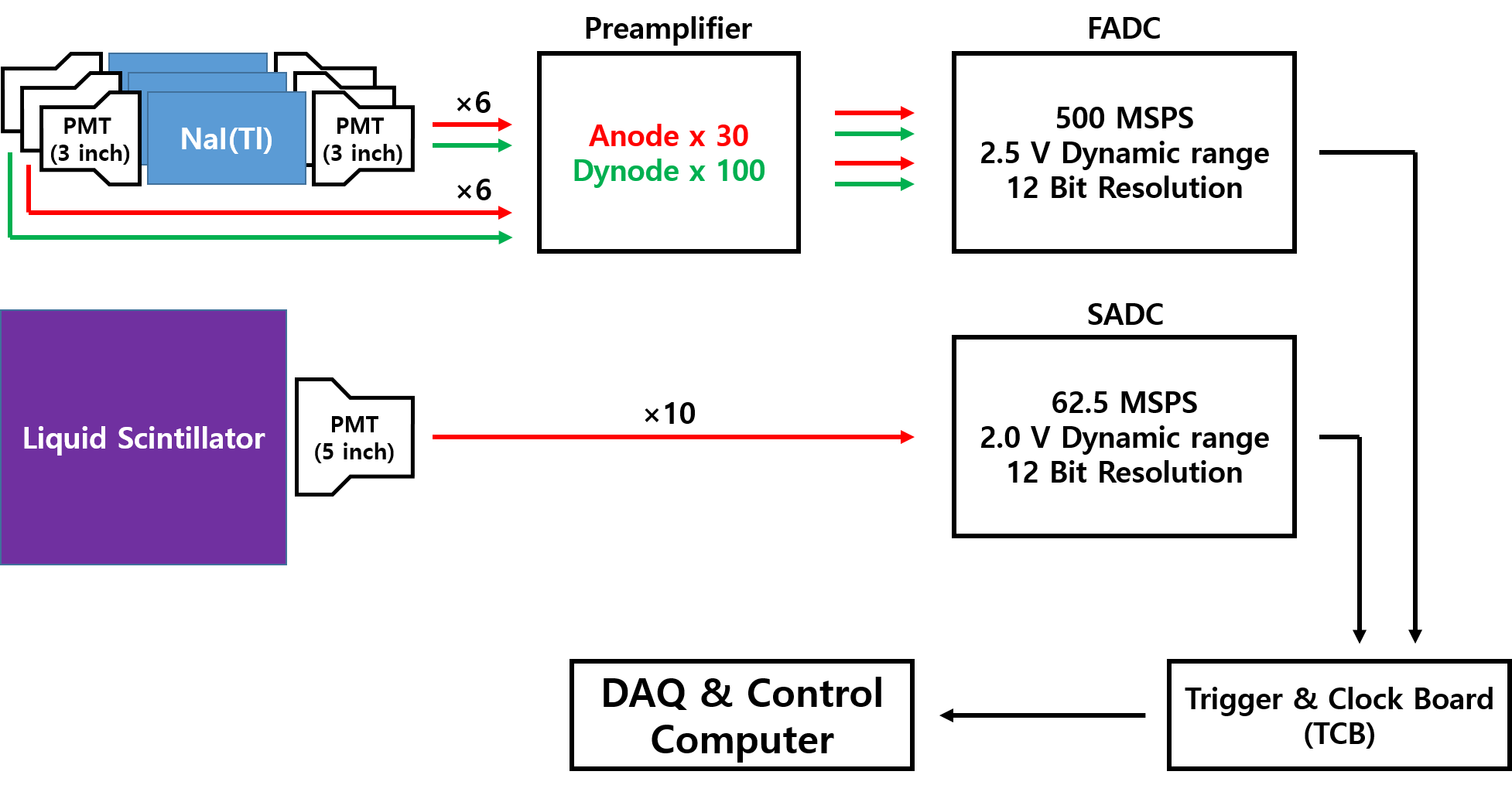} 
  \end{center}
  \caption{Data flow block diagram. The crystal signals are recorded with FADCs while the LS signals are recorded with SADC. Global triggers are formed at the TCB.}
  \label{ref:NEON_DAQ}
\end{figure}

The data acquisition (DAQ) system and electronics are installed in an electronics rack near the detector, which is placed in a temperature-controlled vinyl house. 
The system comprises DAQ modules, high-voltage suppliers, and a computer. 
The same system was used for the COSINE-100 experiment with successful long-term operation~\cite{Adhikari:2018fpo}. 
Figure~\ref{ref:NEON_DAQ} shows the overall data flow diagram of the NEON experiment. 

There are twelve 3-inch NaI(Tl) crystal-readout PMTs and ten 5-inch LS-readout PMTs. Each NaI(Tl) crystal PMT has two readout channels: a high-gain anode channel for low energy and low-gain dynode channel in the fifth stage for high energy. 
Analog signals from NaI(Tl) crystal readout PMTs are amplified using custom-made preamplifiers. The high-gain anode and low-gain dynode channel signals are amplified by 30 and 100 times, respectively. The amplified signals are converted to digital 500 mega samples per second (MSPS) using 12-bit flash analog-to-digital converters (FADCs). Further, unamplified signals from the LS PMTs are digitized using 62.5 MSPS ADC (SADC).

The events are triggered by the anode channel signals when at least one photoelectron, which is more than 20\,ADC, is observed in both PMTs in coincidence within a 200\,ns time window. A typical single photoelectron signal exhibits an average ADC count of 60. 
Triggers from individual channels are generated by field-programmable gate arrays embedded in the FADCs. The final decision for an event is made by a trigger and clock board (TCB) that synchronizes the timing of different modules. 
If one crystal satisfies the trigger condition, all FADCs (NaI(Tl) crystals) and SADC (LS) receive signals. For each FADC channel, an 8 $\mu$s waveform is recorded that starts at 2.4 $\mu$s prior to the trigger.

The triggered events are transferred to the DAQ computer through an USB3 connection in each DAQ module. 
Raw data are stored in ROOT format~\cite{BRUN199781}. 
For channels with waveforms that are only non-triggered baselines, the content is suppressed to zero. 
The data size during operation is approximately 20 GB per day. 
Owing to the security policy in the reactor complex, the DAQ system cannot have an internet connection. 
All data are transferred to CPU farms at the Institute for Basic Science every two or four weeks using portable hard disks through onsite visits by shift workers. 

\subsection{Muon phosphorus events}
Although the experimental site has an approximately 20\,m water-equivalent overburden from an approximately 10\,m concrete wall, a few muons passes through each crystal per minute.
Thus, owing to their large energy deposition, direct muon hits in the crystal generate long phosphorus photons~\cite{DM-Ice:2015aij}. 
These photons can satisfy the trigger condition continuously for up to 1\,s at a trigger rate of approximately 1,000\,Hz. 
Considering the buffer size of the DAQ modules and the speed of data transfer between the DAQ modules and computer, a maximum event rate of 500\,Hz can be maintained. 
Event veto logic is embedded for the muon phosphorus events that applies a 300\,ms dead time for energetic hit events. 
High-energy events are tagged via the requirement of above 2,500\,ADC counts for more than 300\,ns in an adjacent time bin in the dynode channels that correspond to approximately 3,000\,keV events. 
Further, trigger information regarding energetic events is stored, and {\sc in situ} estimation of dead time is provided.
Thus, approximately 10\,\% dead time for 8-inch crystals and 5\,\% dead time for 4-inch crystals are evaluated.
Furthermore, the total trigger rate in the physics run is maintained at less than 180 Hz. 

\subsection{Software trigger}
\begin{figure*}[!htb]
  \begin{center}
    \begin{tabular}{ccc}
      \includegraphics[width=0.33\textwidth]{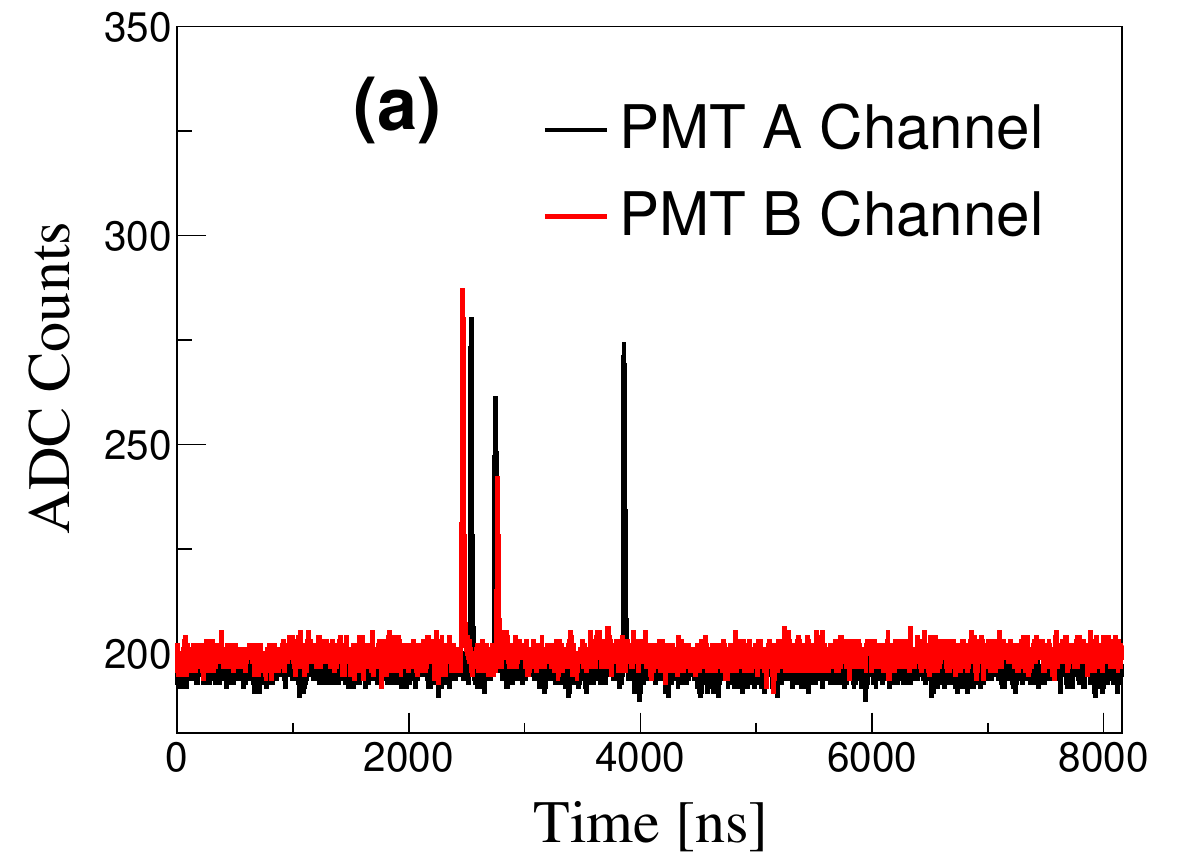} &
      \includegraphics[width=0.33\textwidth]{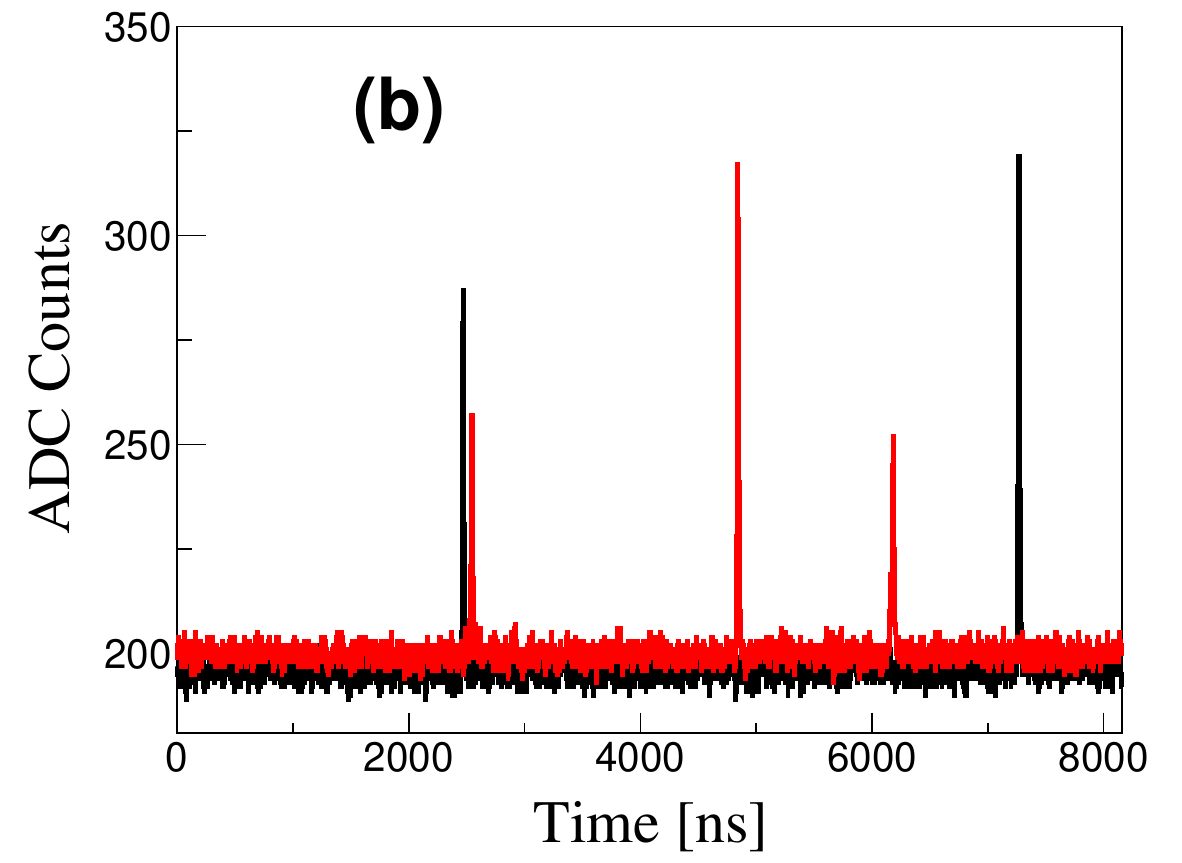} &
      \includegraphics[width=0.33\textwidth]{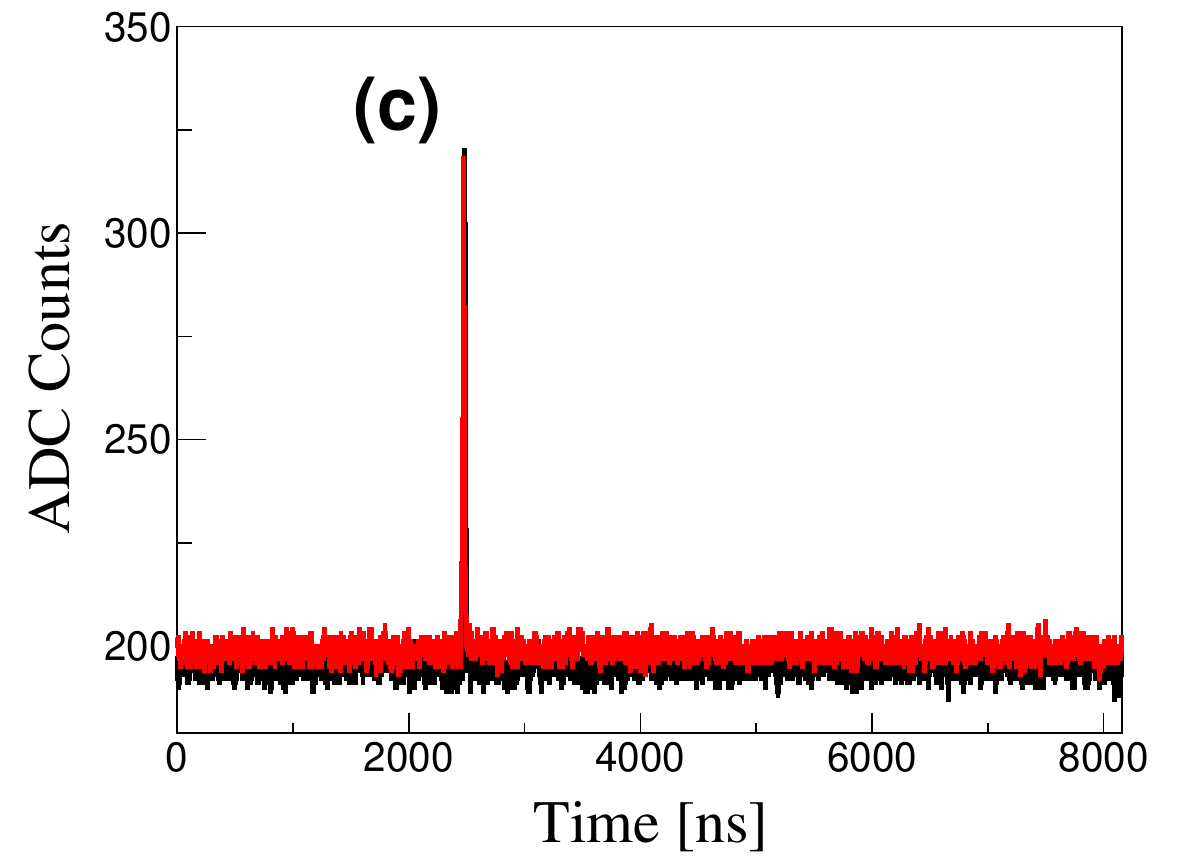} \\
    \end{tabular}
  \end{center}
  \caption{(a) Example of the scintillation pulse with five photoelectrons. (b) Example of the software trigger rejected event with five photoelectrons. (c) Example of dominant ($>$80\,\%) hardware triggered events, which have single use in each PMT, that are rejected by the software trigger.
  }
  \label{softtrg}
\end{figure*}

Although the 300\,ms hardware dead time is applied for high-energy events, most of the triggered events are due to long phosphorus events, similar to the examples shown in Fig.~\ref{softtrg}\,(b) and (c) rather than the typical scintillation candidate in (a).
To effectively use the disk space and reduce the input/output load of the DAQ computer, software trigger logic is developed and implemented in the DAQ program to select only good scintillation candidates, as shown in Fig.~\ref{softtrg}\,(a).

When the event information is transferred to the DAQ computer, the waveforms of the six crystals are quickly scanned, and the parameters are evaluated to discriminate phosphorus events. 
The number of pulses provide the most effective criterion for discriminating between the phosphorus events. The main characteristic of the phosphorus event is that the pulses are spread out, as shown in Fig.~\ref{softtrg}\,(b), and the software trigger requires at least three pulses within a 500\,ns window from the first pulse time. The coincidence time is defined as the time difference between the first pulses from the two PMTs attached to a crystal. This condition is similar to that of the PMT coincidence for the hardware trigger described in Sec.~\ref{sec:DAQ}. A software trigger applies this criterion when a hardware trigger occurs in another crystal channel. The examples of Figs.~\ref{softtrg}\,(b) and (c) are rejected owing to this requirement. 

The other parameter is the mean time, which is calculated using Eq.~\ref{meantime}. As shown in Fig.~\ref{ref:alpha_mt}, the mean time of the scintillation event is mainly greater than 0.2\,$\mu$s, as the software trigger rejects events with a mean time of less than 0.015\,$\mu$s. 

\begin{figure}[!htb]
  \begin{center}
      \includegraphics[width=0.5\textwidth]{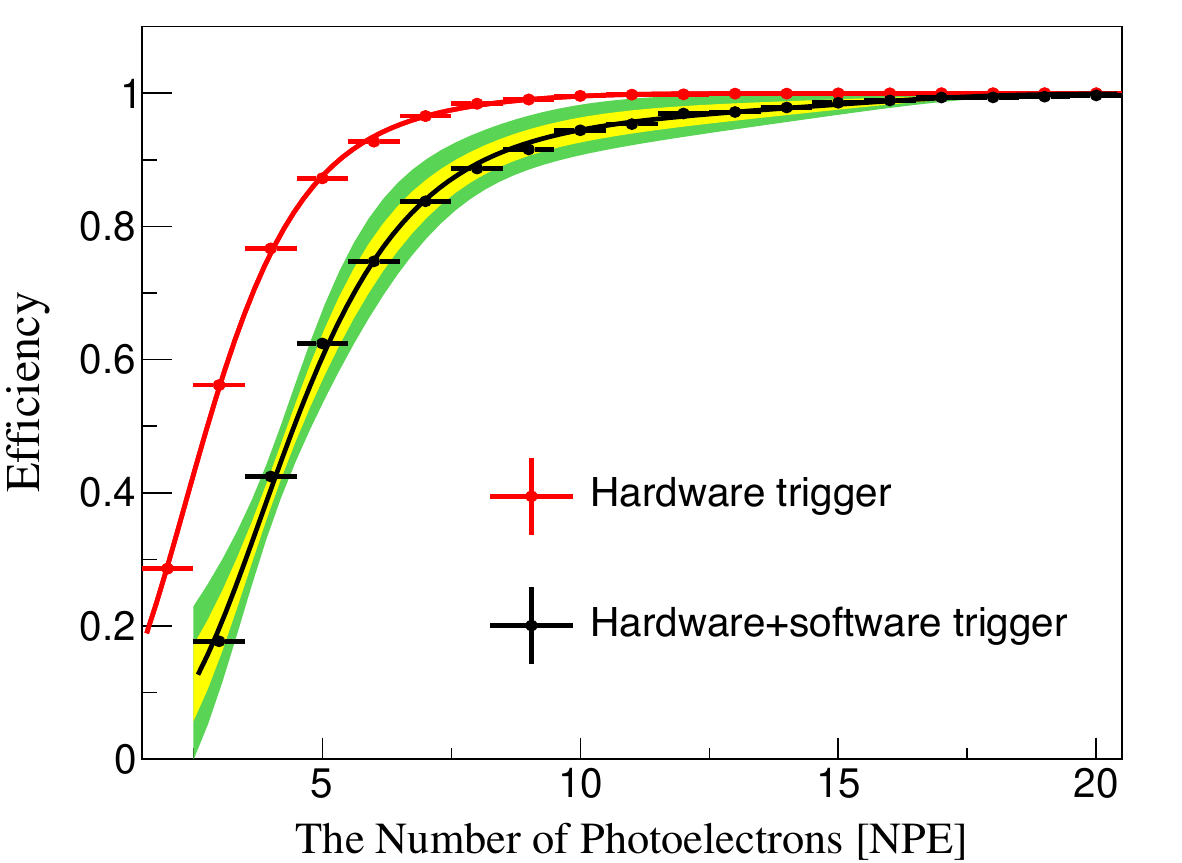}
  \end{center}
  \caption{Hardware and software trigger efficiencies from the simulated events are presented. 
  }
  \label{softtrgEff}
\end{figure}

Events filtered by the software trigger are mostly removed, with only 0.1\% being randomly written. By contrast, events that met all the criteria are tagged and fully recorded. The software trigger reduces the accepted event rates by approximately 90\% and maintains recording event rates of less than 20\,Hz. 
The efficiency of the hardware and software triggers is evaluated based on the simulated events of the scintillation photons, as shown in Fig.~\ref{softtrgEff}. 
An efficiency greater than 60\,\% from the trigger is maintained for five or more NPE events. 

\section{Monitoring system}

\begin{figure*}[!hbt]
  \begin{center}
    \includegraphics[width=1.0\textwidth]{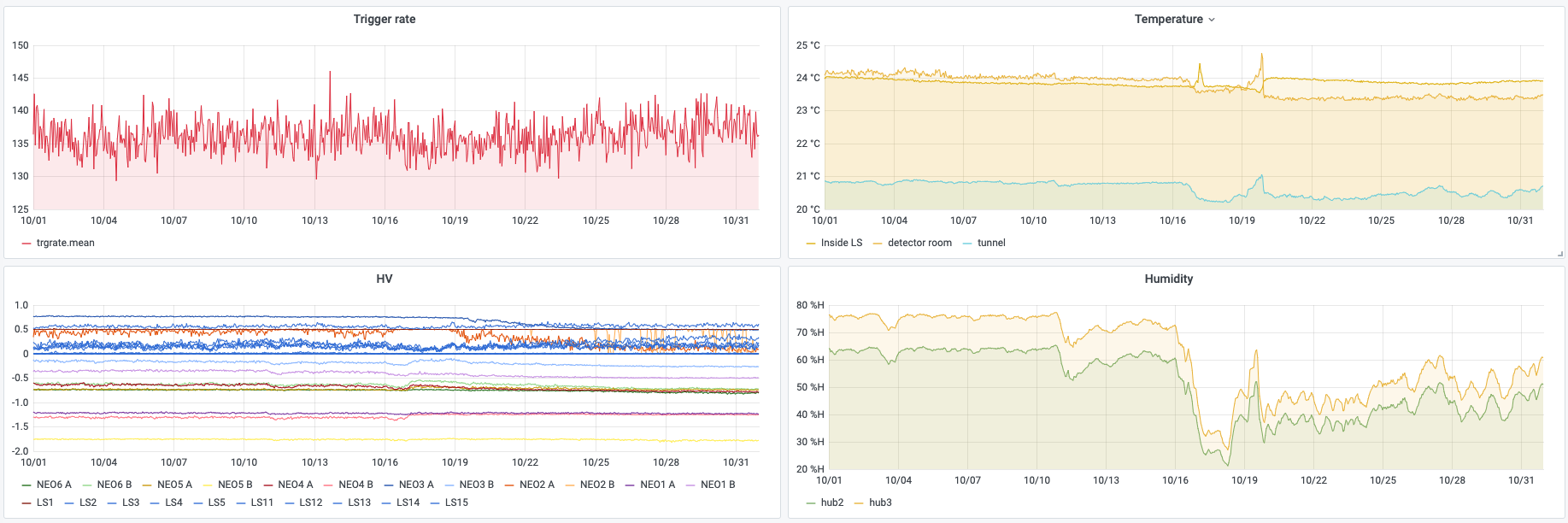}
  \end{center}
  \caption{Example of the slow monitoring panels recorded during October 2021. Trigger rate, temperatures, high voltages, and humidities are shown in this example. }
  \label{slowmonitoring}
\end{figure*}

For stable data-taking and systematic analyses of tiny signals, environmental parameters, such as the trigger rate and detector temperature, high-voltage variations, and humidity, must be controlled and  monitored. To achieve this, a variety of sensors for specific monitoring tasks are employed. These devices are controlled and read out using a common database server and a visualization program. 
This section briefly discusses the environmental monitoring system used in the NEON experiment.

Temperature and humidity are monitored using two thermo-hygrometers, and a thermocouple sensor. The thermocouple sensor is placed in contact with an LS inside an acrylic box. The other two thermo-hygrometer sensors monitor the temperature and humidity in the detector room and tunnel, respectively. The temperature and humidity status are monitored every minute. 
Furthermore, high voltages are provided and controlled by a CAEN high-voltage crate that is monitored by software provided by the company. In addition, the voltages of the preamp supply units are monitored using the Labjack U6 digital-to-analog converter module. 
All supplied high voltages, currents, preamp voltages, and PMT statuses are monitored once every 30\,s.
Moreover, the CPU, memory, and disk usage of the DAQ computer are monitored every 10\,s. 
Here, InfluxDB\,7 was used to store the monitoring data, and Grafana\,8 is used for visualization. Fig.~\ref{slowmonitoring} shows an example of environmental monitoring performed in October 2021. 
A similar system has already been employed in the COSINE-100 experiment~\cite{COSINE-100:2021mlj}. 

Owing to the security policy in the reactor complex, no online connection for the monitoring system is allowed. 
Shift crews transfer the monitoring data to an internet-connected server every two or four weeks through onsite visits. 

\section{Sensitivity}

\begin{figure}[!htb]
  \centering
  \begin{center}
    \includegraphics[width=0.45\textwidth]{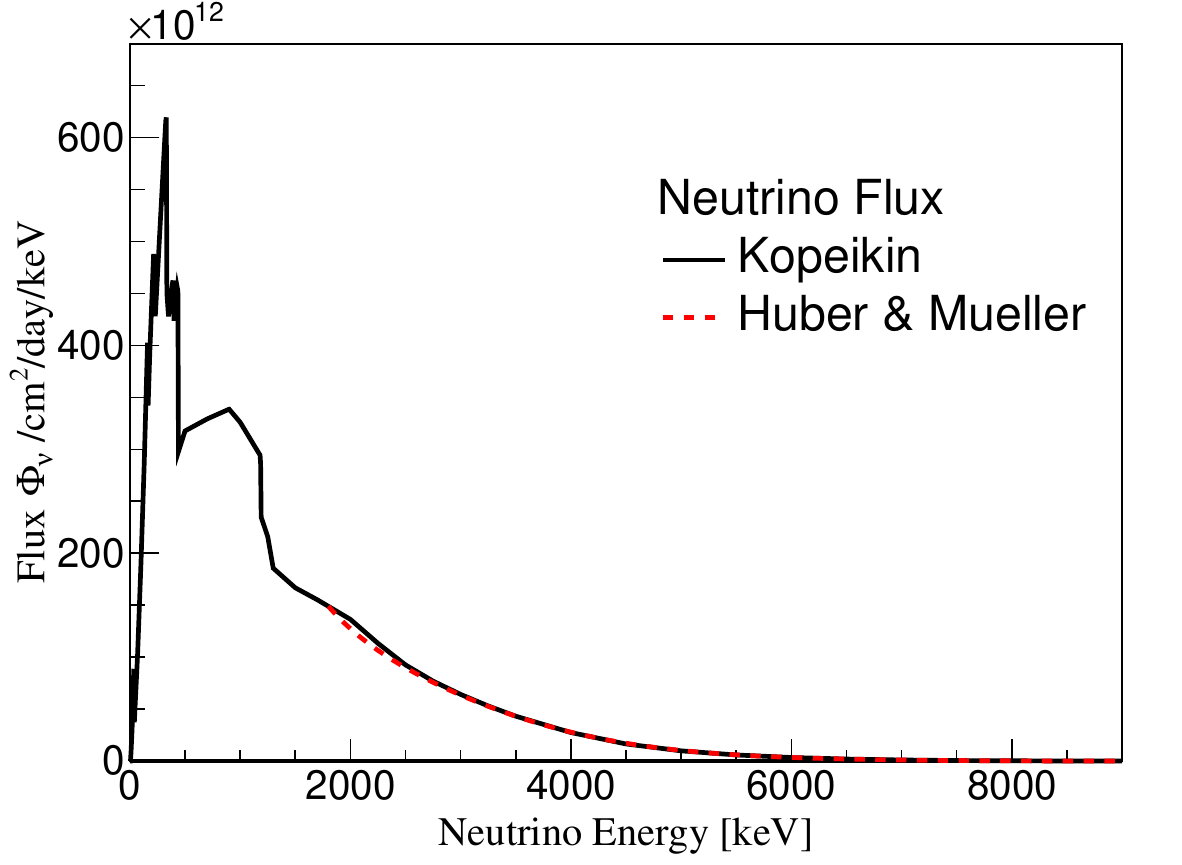}
  \end{center}
  \caption{Expected neutrino fluxes for the NEON experiment are presented for the Kopeikin~\cite{Kopeikin:2012zz} and Huber~\cite{PhysRevC.84.024617} \& Mueller~\cite{Mueller:2011nm} models. 
In the Kopeikin model, the fission fractions of 0.56, 0.31, 0.07, and 0.06 for $^{235}$U,  $^{239}$Pu,  $^{238}$U, and  $^{241}$Pu, respectively, are used. As the Huber \& Mueller model provided the neutrino flux from each isotope, we used the fission fractions of 0.577, 0.296, 0.072, and 0.055 for $^{235}$U,  $^{239}$Pu,  $^{238}$U, and  $^{241}$Pu, respectively.}
  \label{fig:nu_flux}
\end{figure}

In the reactor core, several neutrinos are emitted from fission fragments through $\beta$ decay. The fission fraction $f_i$ for isotope $i$ is 57.7\,\% of $^{235}$U, 29.6\,\% of $^{239}$Pu, 7.2\,\% of $^{238}$U, and 5.5\,\% of $^{241}$Pu, from cycle-9 of unit-5 of the Hanbit reactor complex, which has the same design as unit-6, and will be updated after completion of the current operation cycle. The fission rate $R_f$ can be described as follows: 
\begin{eqnarray}
R_f=P_\mathrm{th}/\left<E_r\right>\approx8.6\times10^{19}~\mathrm{fission}/\mathrm{s}, 
\end{eqnarray}
where $P_\mathrm{th}$ denotes the thermal power of Hanbit reactor unit-6, $P_\mathrm{th} = 2.815~\mathrm{GW_{th}}$, and $\left<E_r\right>$ denotes the average released energy $\left<E_r\right> = \sum_if_iE_i\approx 205~\mathrm{MeV}/\mathrm{fission}$~\cite{Edgar:1969,Kopeikin:2004com}. The neutrino flux at a distance $L$ from the reactor core is
\begin{eqnarray}
\Phi(E_\nu) = \frac{1}{4\pi L^2}\sum_i f_i \phi_i(E_\nu) \frac{P_\mathrm{th}}{\left<E_r\right>},
\label{eq:nu_flux}
\end{eqnarray}
where $\phi_i$ denotes the neutrino flux for isotope $i$. The neutrino flux per fission of major isotopes, such as $^{235}$U, $^{238}$U, $^{239}$Pu, and $^{241}$Pu, contributing to fission has been continuously reported~\cite{PhysRevC.84.024617,Mueller:2011nm}, and we use the flux $\sum_if_i\phi_i$ from Ref.~\cite{Kopeikin:2012zz}, which describes neutrino flux below an inverted $\beta$-decay threshold of 1.8\,MeV. The fission fractions used in Refs.~\cite{Kopeikin:2012zz} were similar to those of the Hanbit reactor; therefore, we use the neutrino flux model. Fig.~\ref{fig:nu_flux} shows the expected neutrino flux at a NEON detector 23.7\,m away from the reactor core. 
Different models by Huber~\cite{PhysRevC.84.024617} \& Mueller~\cite{Mueller:2011nm} for neutrino energies above 1.8\,MeV and the fission fraction of cycle-9 of the Hanbit reactor unit-5 show consistent neutrino flux.  The total flux at this site is $8.09\times10^{12}~/\mathrm{cm}^2/\mathrm{sec}$.

\begin{figure*}[!htb]
  \centering
  \begin{center}
    \begin{tabular}{ccc}
      \includegraphics[width=0.33\textwidth]{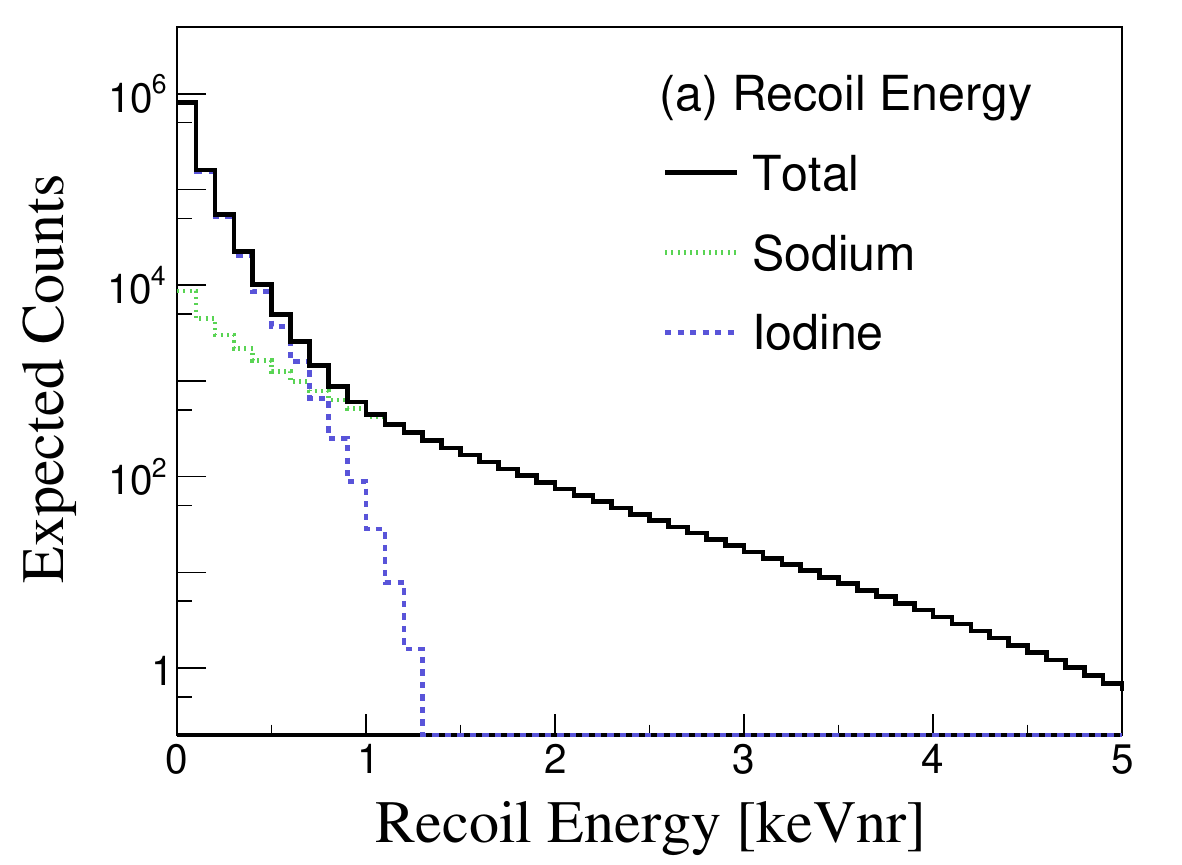} &
      \includegraphics[width=0.33\textwidth]{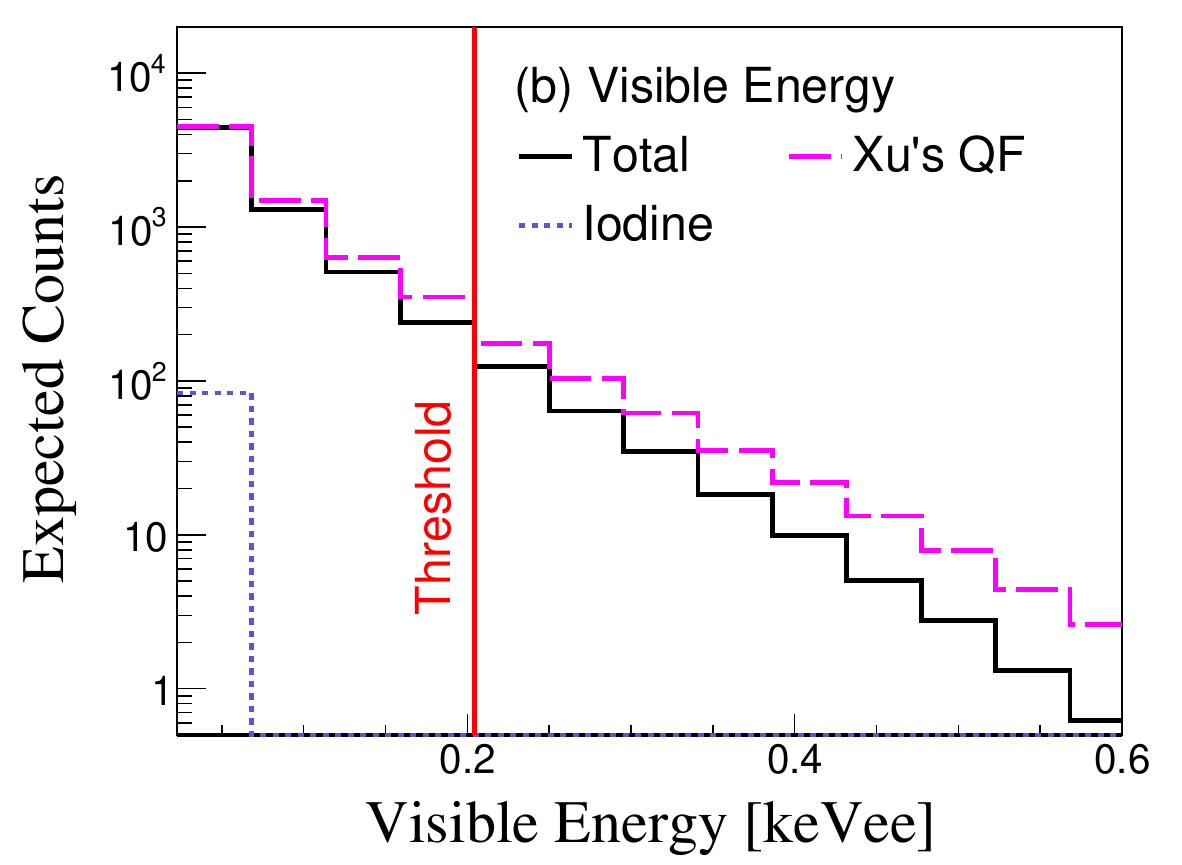} &
      \includegraphics[width=0.33\textwidth]{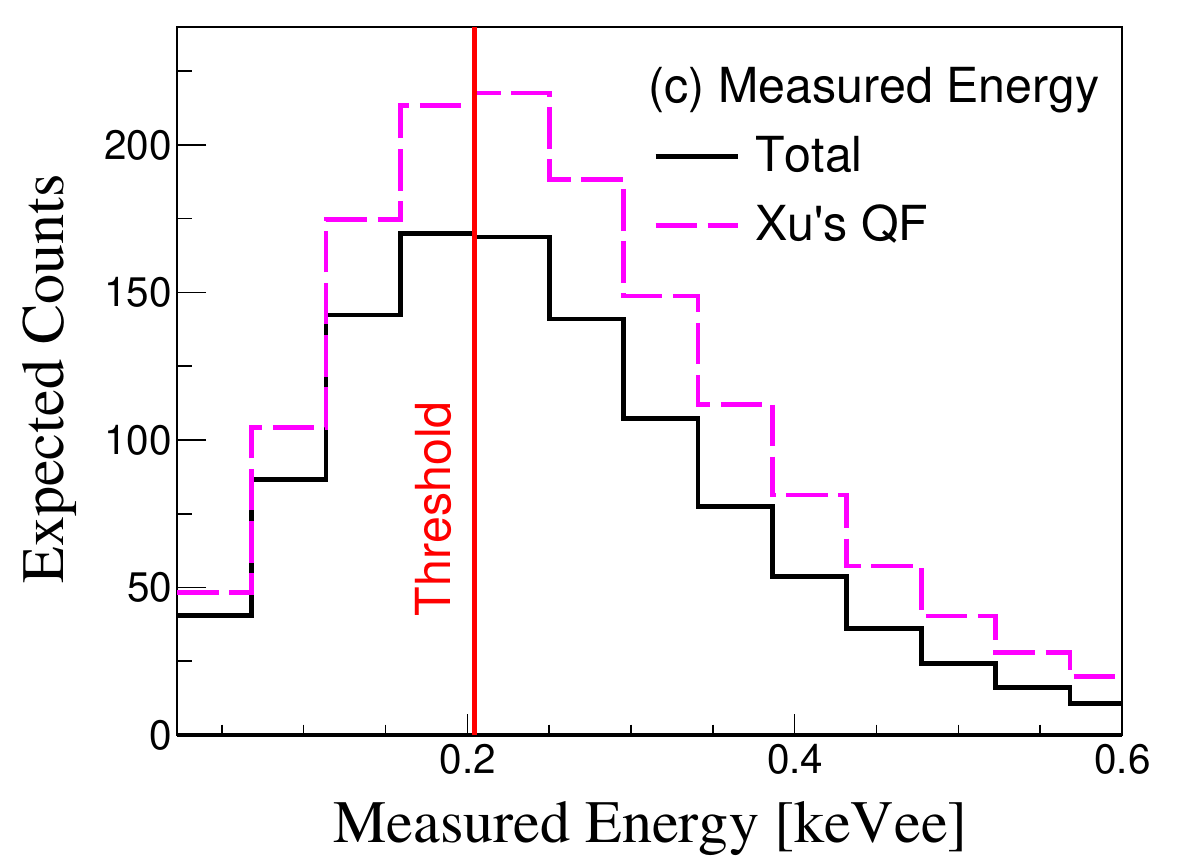} \\
    \end{tabular}
  \end{center}
	\caption{Expected \cenns signals in the NEON experiments for 1\,year of data taking are presented. (a) Expected \cenns signal rates in recoil energy are presented for sodium (green solid line), iodine (blue solid line), and combined NaI (black solid line). (b) Visible energy spectra applying the QF values for two measurements by Joo~\cite{Joo:2018hom} (solid line) and Xu~\cite{Xu:2015wha} (dashed line).  (c) Expected measured event rates after smearing resolution based on Poisson statistics and selecting only accepted events from the hardware and software triggers. }
  \label{fig:cevns_signal}
\end{figure*}

The differential cross-section of \cenns has a standard model prediction~\cite{Freedman:1977xn},
\begin{eqnarray*}
		\frac{d\sigma}{dE_\mathrm{rec}} = & \frac{G_F^2m_A}{2\pi} \big[(G_V + G_A)^2 + (G_V-G_A)^2 \left(1-\frac{E_\mathrm{rec}}{E_\nu}\right)^2 \\
		& - (G_V^2-G_A^2)\frac{m_AE_\mathrm{rec}}{2E_\nu^2}  \big], 
\label{eq:xcfull}
\end{eqnarray*} 
where $G_V$ and $G_A$ denote coefficients related to vector and axial-vector coupling, respectively, $G_F$ denotes the Fermi coupling constant, $m_A$ denotes the nuclear mass of the target, $E_\mathrm{rec}$ denotes the nuclear recoil energy, and $E_\nu$ denotes the neutrino energy. 
Considering the small contribution of the axial term and low moment transfer in the \cenns process~\cite{Hoferichter:2020osn,PhysRevD.98.053004}, the differential cross-section is approximately expressed as 
\begin{eqnarray}
\frac{d\sigma}{dE_\mathrm{rec}} = \frac{G_F^2m_A}{4\pi}\left(1-\frac{m_AE_\mathrm{rec}}{2E_\nu^2}\right)Q_\mathrm{w}^2,
\label{eq:xc}
\end{eqnarray}
where $Q_\mathrm{w}$ denotes a weak charge expressed as follows: 
\begin{eqnarray}
Q_\mathrm{w} = Z(1-4\sin^2\theta_\mathrm{W}) - N,
\end{eqnarray}
where $\theta_\mathrm{W}$ denotes the weak mixing angle and $Z$ ($N$) denotes the proton (neutron) number. The differential rate can be expressed by combining the neutrino flux in Eq.~\ref{eq:nu_flux} and the differential cross-section in Eq.~\ref{eq:xc}, 
\begin{eqnarray}
\frac{dR}{dE_\mathrm{rec}} = n_t\int_{E_\mathrm{thr}}^{\infty}dE_\nu\Phi(E_\nu) \frac{d\sigma}{dE_\mathrm{rec}}(E_\nu),
\label{eq:cevns_rate}
\end{eqnarray}
where $n_t$ denotes the number of target nuclei and $E_\mathrm{thr}=\sqrt{E_\mathrm{rec}m_A/2}$ denotes the threshold of the neutrino energy.
Figure~\ref{fig:cevns_signal}~(a) shows the expected signal rates in the recoil energy assuming 1\,year of data from the NEON experiment. 
Owing to the low atomic mass number of sodium, it generates events up to 5\,keVnr, while iodine interactions are less than 1\,keVnr, where keVnr is the unit keV nuclear recoil energy.

\begin{figure*}[!htb]
  \centering
  \begin{center}
      \includegraphics[width=0.9\textwidth]{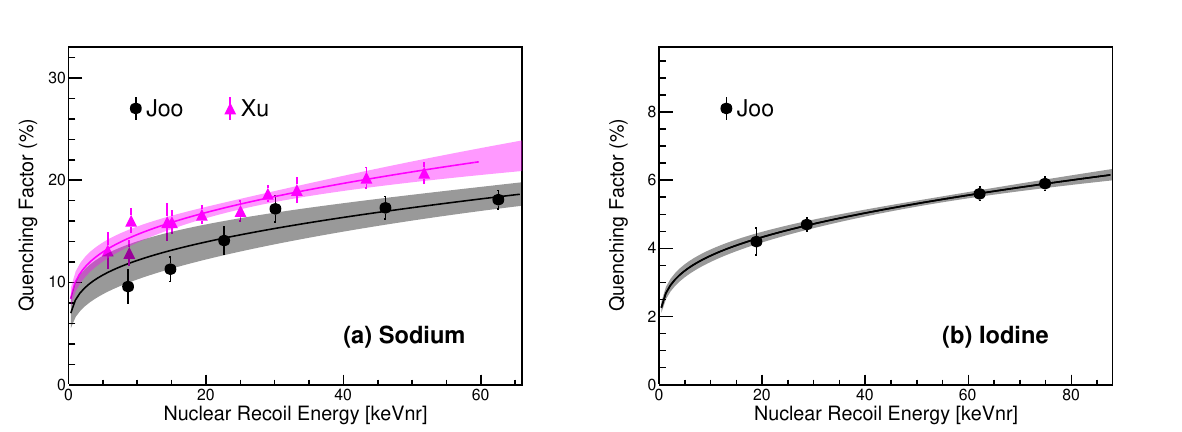} 
  \end{center}
	\caption{Measured quenching factors by Joo {\it et al.}~\cite{Joo:2018hom} (black circles) and Xu {\it et al.}~\cite{Xu:2015wha} (purple triangles) for sodium (a) and iodine (b) are fitted with the modified Lindhard model.}
  \label{ref:quench}
\end{figure*}

The scintillation light yields for nuclear recoils are quenched to those of $\gamma$/electron-induced radiation of the same energy~\cite{Birks:1951boa,osti_4701226,Tretyak:2009sr}. 
To express the \cenns rate in terms of the electron-equivalent visible energy ($ E_\mathrm{vis}$), the nuclear recoil quenching factor (QF) for each nucleus should be considered. 
\begin{eqnarray}
		\mathrm{QF}(E_\mathrm{rec}) \equiv \frac{E_\mathrm{vis}}{E_\mathrm{rec}}.
		\label{eq:qf}
\end{eqnarray}
Recent measurements of the nuclear recoil QFs of sodium and iodine in NaI(Tl) crystals have been reported as approximately 10\,\% and 5\,\%, respectively, with strong energy dependence~\cite{Collar:2013gu,Xu:2015wha,Joo:2018hom}. As there are no measurements of $E_\mathrm{rec}$ below  5\,keVnr, extrapolation of the measurements to the low-energy region is required. Here, we use parameterizations of QFs using a modified Lindhard model~\cite{osti_4701226} that was used for the COSINE-100 data interpretation~\cite{COSINE-100:2019brm},
\begin{eqnarray}
		\mathrm{QF}(E_\mathrm{rec}) = \frac{p_0\,f(\epsilon)}{1+p_0\,f(\epsilon)},
\label{eq:qf_model}
\end{eqnarray}
where $\epsilon = p_1\,E_\mathrm{rec}$, and the function $g(\epsilon)$ is~\cite{Lewin:1995rx}
\begin{eqnarray}
g(\epsilon) = 3\epsilon^{0.15} + 0.7\epsilon^{0.6} + \epsilon.
\end{eqnarray}
Here, $p_0$ and $p_1$ denote the fit parameters for describing the QF measurements. This  model describes the recent measurements well, as shown in Fig.~\ref{ref:quench}. 
Owing to the large statistical uncertainties in Ref.~\cite{Collar:2013gu}, only two measurements by Joo {\it et al..}~\cite{Joo:2018hom} and Xu {\it et al. }.~\cite{Xu:2015wha} are considered.
Measurements by Joo {\it et al..} used a crystal from Alpha Spectra, which provided all the NEON crystals. 
These two measurements used a similar energy calibration method that assumed a linear response of energy for the 59.54\,keVee line of $^{241}$Am~\cite{Joo:2018hom} or the 57.6\,keVee line from the first excited state of $^{127}$I~\cite{Xu:2015wha}. 
Owing to the nonproportionality~\cite{nonprop} of the NaI(Tl) crystal, different calibration methods lead to different QF results~\cite{Cintas:2021fvd}. 
In this study, we use the same calibration method as that used for the 59.54\,keVee line. 

As one can see in Fig.~\ref{ref:quench}, the two measurements for the sodium nuclei exhibited approximately 20\,\% different results.
Although Joo's measurement used the Alpha Spectra crystal, preliminary results~\cite{Cintas:2021fvd} using  five different Alpha Spectra crystals were consistent with Xu's measurement. 
For this reason, we evaluat the sensitivities using both the QF results. However, we do not consider the uncertainties from the model fits in this study. 
Fig.~\ref{fig:cevns_signal}(b) shows the visible energy spectra obtained by applying the two QF models. Owing to the 20\,\% higher QF values from Xu {\it et al.}, the expected events above the 0.2\,keVee threshold are approximately 30\,\% higher than those of Joo {\it et al.}. Above the 0.2\,keVee threshold, no iodine contributions are expected. 
Because of large uncertainties of the low energy calibration for \cenns measurements, we currently study  the low energy QF of sodium below 5\,keVnr using the deuteron-deuteron fusion generator~\cite{Joo:2018hom} by locating the neutron tagging detector collinear to the neutron beam that is approximately 10$^{\circ}$C with the neutron beam direction. We also consider to install deuterium-loaded neutron reflector for the low energy neutron beam as suggested in Ref.~\cite{Verbus:2016sgw}.

A fast simulation toolkit for event generation is developed to account for the energy resolution and trigger efficiencies. 
When a \cenns interaction deposits recoil energy $E_\mathrm{rec}$ in the crystal, the quenched visible energy ($E_\mathrm{vis}$) produces scintillation photons based on Poisson statistics. 
\begin{eqnarray}
N_{pe} = \mathrm{Pois}(E_\mathrm{vis}\times LY),
\end{eqnarray}
where $LY$ denotes the light yield of the NaI(Tl) crystals corresponding to approximately 22\,NPE/keVee and $N_{pe}$ denotes the number of photoelectrons after a Poisson random variation. 
The total $N_{pe}$ is distributed in the two PMTs, assuming a binomial distribution and the scintillation decay time of the NaI(Tl) crystal. 
The charge dispersion of a single photoelectron owing to PMT amplification is described by the single photoelectron charge distributions shown in Fig.~\ref{ref:LYspe}. The shape of the generated scintillation event is convoluted with the measured pedestal fluctuations. 
Simulated signal events are recorded in a format that matches that of the NEON DAQ system. 
The hardware and software triggers are simulated to use only trigger accepted events. The expected \cenns signals, considering the aforementioned process, are shown in Fig.~\ref{fig:cevns_signal} (c).

\begin{figure*}[!htb]
  \centering
  \begin{center}
    \begin{tabular}{ccc}
      \includegraphics[width=0.33\textwidth]{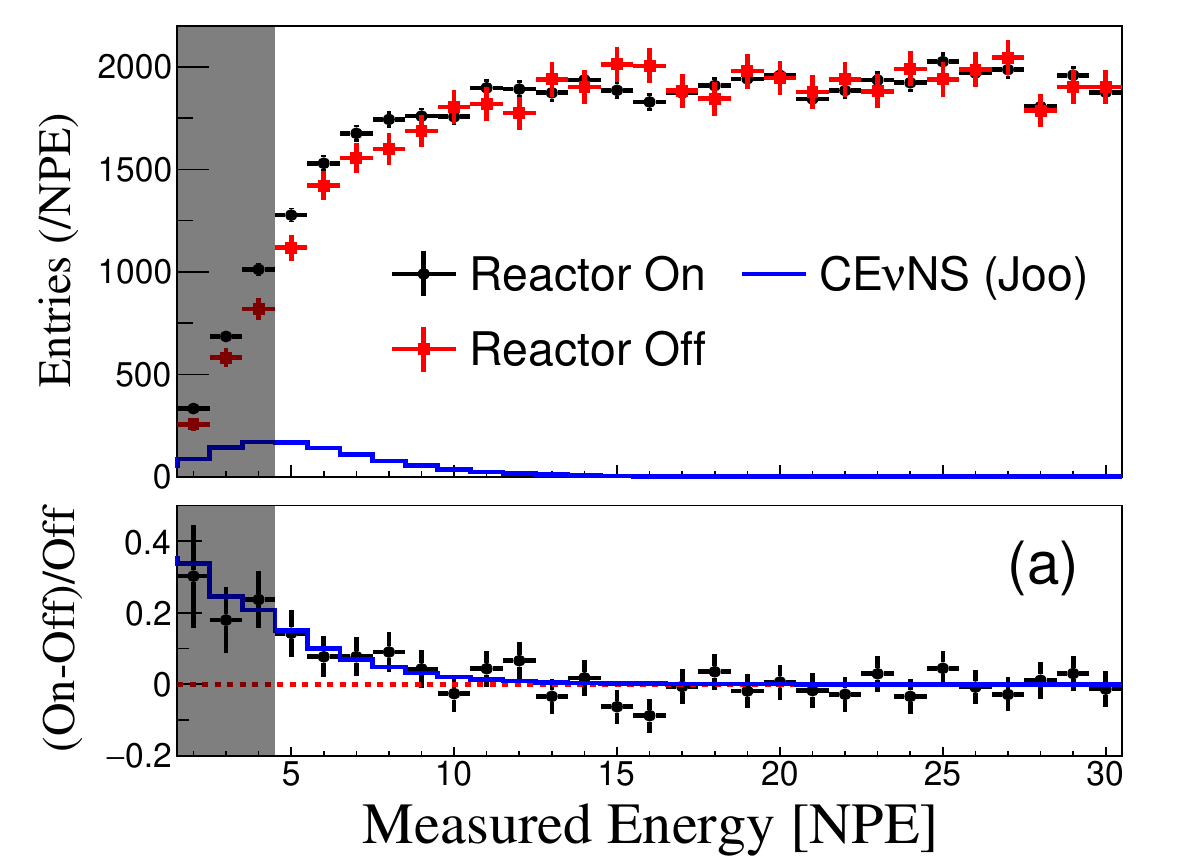} &
      \includegraphics[width=0.33\textwidth]{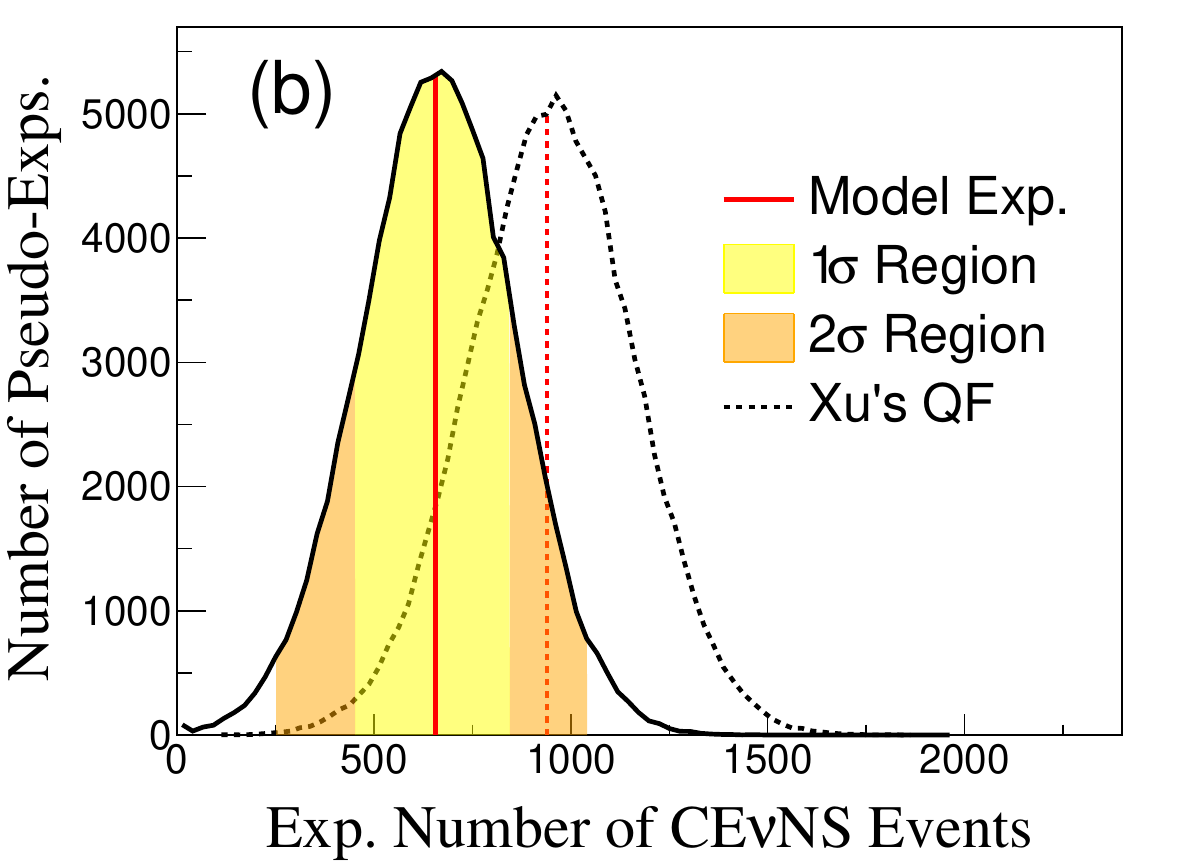} &
      \includegraphics[width=0.33\textwidth]{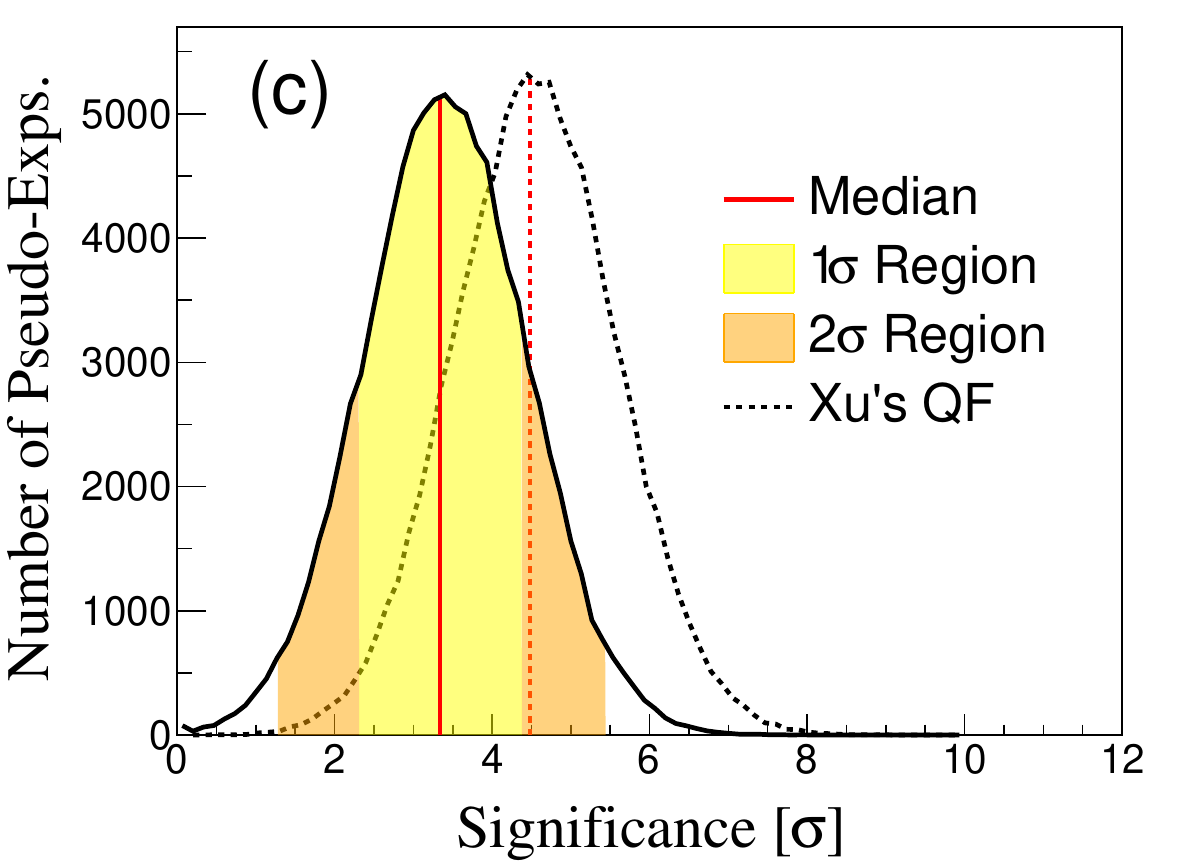} \\
    \end{tabular}
  \end{center}
  \caption{(a) Example of the simulated experiments assuming 7\,counts/day/kg/keV flat background, 22 PEs/keV light yield, one-year reactor-on data (black points), and 100-days reactor-off data (red points) are presented. 
The reactor-off data is scaled with the ratio of time exposure between the reactor-on and reactor-off periods. The reactor-on data includes the expected \cenns events (blue line) obtained in Fig.~\ref{fig:cevns_signal} (c) for the QF values from Joo. In this pseudo experiment, the $\chi^2$ fit obtains $649\pm193$ \cenns signal events.
  (b) Results of 100,000 independent simulated experiments are presented. Here a median expected number of signals, 650 events, was obtained with 1$\sigma$ variation of 197 events with the input signal 
for the Joo's QF values (solid lines). Vertical line represents the input number expected by the standard model of 656. If we use Xu's QF values (dotted lines), the median expected signal is 941$\pm$209 with an input number of 939 (vertical lines). 
  (c) Significances of the \cenns observation from 100,000 independent simulated experiments are estimated using the significance tests based on $\chi^2$ differences. A median significance of $3.34\pm1.03 \sigma$ and $4.48\pm1.04$ were obtained for the Joo and Xu QF values, respectively.
  }
  \label{ref:sensiti}
\end{figure*}

To estimate the sensitivity, we assum a few physics parameters: 7\,counts/kg/keV/day flat background, the QF models in Fig.~\ref{ref:quench} based on Joo’s and  Xu’s measurements, and a one-year reactor-on period and 100-days reactor-off period with 100\,\% live time and 100\,\% event selection efficiency. 100,000 pseudo-data sets are prepared, and each set consists of reactor-on and reactor-off data based on the aforementioned assumptions, together with the Poisson random variation in each energy bin. Black circles and red squares in Fig.~\ref{ref:sensiti}~(a) are examples of reactor-on and reactor-off simulated datasets. Blue lines present the expected \cenns signals. The \cenns signals are extracted by $\chi^2$ minimization from  NPE=$NPE_{thr}$ to NPE=30,
\begin{eqnarray}
		\chi^2(\psi) = \sum_{i = NPE_{thr}} ^{30} \frac{\left(N_{\mathrm{on},i} - \alpha_tN_{\mathrm{off},i} - \psi E_i\right)^2}{N_{\mathrm{on},i} + \alpha_t^2N_{\mathrm{off},i}},
\end{eqnarray}
where $N_{\mathrm{on},i}$ and $N_{\mathrm{off},i}$ denote the number of events in $i^\mathrm{th}$ energy ($N_{pe}$) bin for reactor-on and reactor-off data, respectively, $E_i$ denotes the expected \cenns events in the $i^{mathrm{th}}$ energy bin,  and $\alpha_t$ denotes the ratio of reactor-on to reactor-off exposure time. We assume that the NPE threshold $NPE_{thr}$=5 corresponds to 0.2\,keVee energy threshold. 
$\chi^2$ is minimized with variation in $\psi$, and the minimum chi-square $\chi^2_\mathrm{min}=\chi^2(\hat{\psi})$ is obtained with the best-fit value of $\psi$ where $\psi$=1 indicates the standard model expectation. In addition, the chi-square value $\chi^2(0)$ when $\psi=0$ is calculated as a null hypothesis. The $\chi^2$ difference between \cenns and the null hypothesis $\Delta\chi^2 = \chi^2(0) - \chi^2(\hat{\psi})$ is evaluated to estimate the strength of the \cenns hypothesis. The same procedures for 100,000 independent pseudo experiments are performed to obtain a distribution of the observed signal events and signal significance, as shown in Figs.~\ref{ref:sensiti} (b) and (c), respectively. The medians of the expected signal events are $650\pm197$ and $941\pm209$ for the QF values from Joo and Xu, respectively. The corresponding signal significances are $3.34\pm1.03$\,$\sigma$ and $4.48\pm1.04$\,$\sigma$, respectively. 
For both the QF hypotheses, we expect more than 3\,$\sigma$ significance.

\begin{figure*}[!htb]
  \centering
  \begin{center}
    \begin{tabular}{ccc}
      \includegraphics[width=0.33\textwidth]{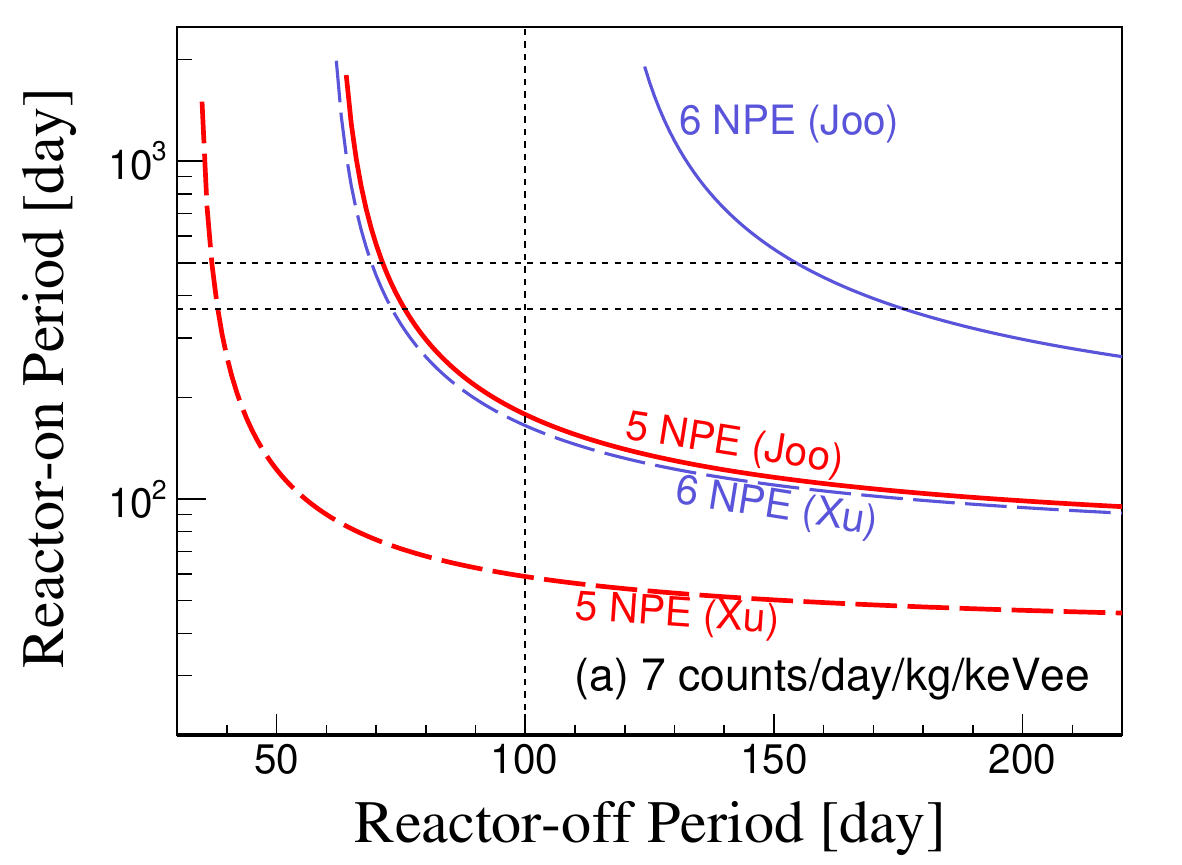} &
      \includegraphics[width=0.33\textwidth]{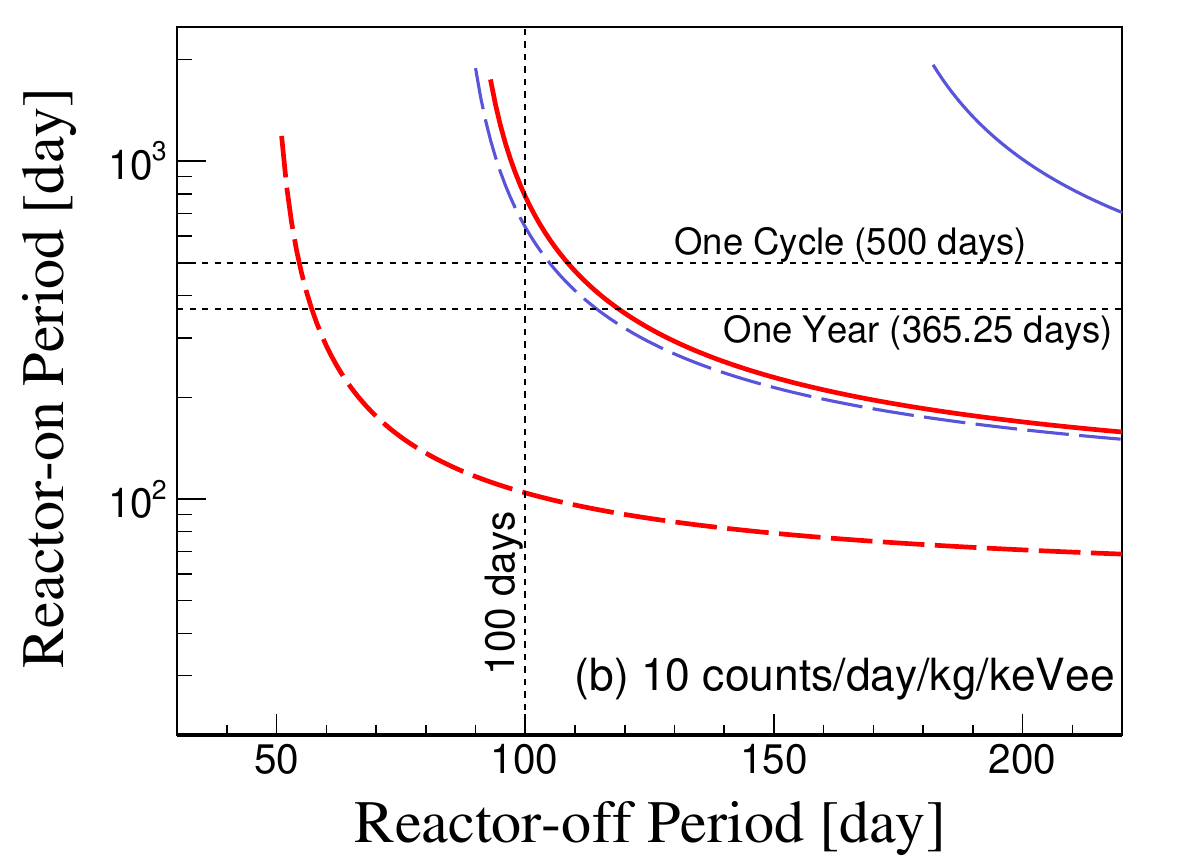} &
      \includegraphics[width=0.33\textwidth]{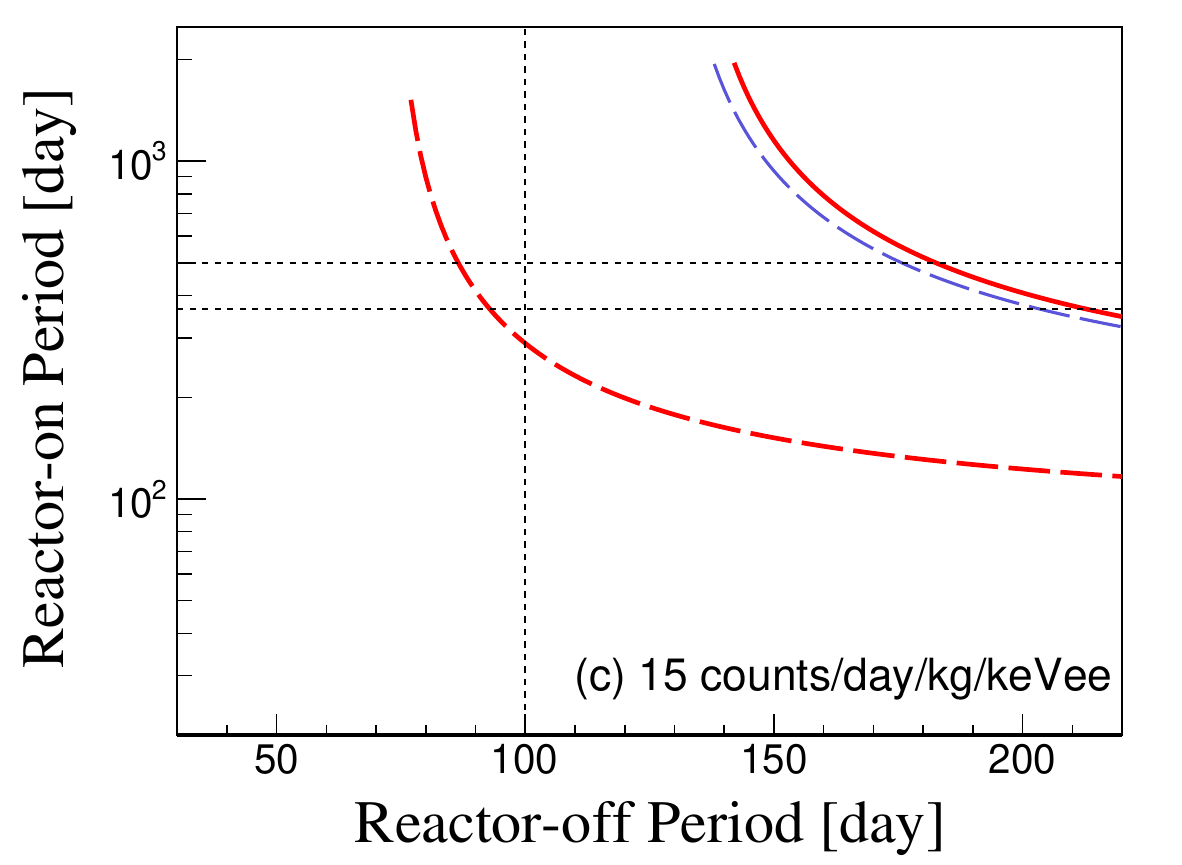} \\
    \end{tabular}
  \end{center}
  \caption{Discovery sensitivities at 3$\sigma$ confidence levels are presented in data exposures of the reactor-on (Y-axis) and reactor-off (X-axis). We vary NPE thresholds from 5 to 6 for two different QF models assuming 7\,counts/kg/keV/day (a), 10\,counts/kg/keV/day (b), and 15\,counts/kg/keV/day (c) background levels. 
	}
  \label{ref:sensi_3s}
\end{figure*}

Owing to uncertain parameters in the region of interest (0.2-- 0.5\,keVee), we consider different cases of poorer detector responses, such as higher energy thresholds from 5 to 6\,NPE, and higher background levels of 7, 10, and 15\,counts/kg/keV/day. Figure~\ref{ref:sensi_3s} presents the data exposures necessary to achieve a 3$\sigma$ significance for the \cenns observation for various detector performances. 
In most cases, we can achieve 3$\sigma$ observation significance if we take one-year reactor-on and 200-days reactor-off data. 
In the case of a higher threshold of 6\,NPE and higher backgrounds of 10 or 15\,counts/kg/keV/day, it is difficult to achieve 3$\sigma$ significance assuming Joo's QF model. 

\section{Summary}
The NEON experiment aims to observe \cenns using reactor electron antineutrinos. 
The detector comprises six NaI(Tl) crystals with several layers of shield and is installed in the tendon gallery of a nuclear reactor with a thermal power of 2.8\,GW that is approximately 23.7\,m away from the reactor core.
The detector has been acquiring data at full reactor power since May 2021. The initial data reveal an excellent performance of the detector with an acceptable background level of 6\,counts/keV/kg/day in the 2--6\,keVee energy region. 
Assuming a one-year reactor-on and 100\,days reactor-off data, 0.2\,keVee energy threshold with 22\,NPE/keVee light output, and 7\,counts/keV/kg/day background in the region of interest (0.2--0.5\,keVee), \cenns observation sensitivity of the NEON experiment is evaluated as more than 3\,$\sigma$.

\acknowledgments
This work was supported by the Institute for Basic Science (IBS) under project code IBS-R016-A1, Republic of Korea, and the National Research Foundation of Korea (NRF) grant funded by the Korean government (MSIT) (NRF-2021R1A2C1013761 and NRF-2021R1A2C3010989).
We thank the Korea Hydro and Nuclear Power (KHNP) company and acknowledge the help and support provided by the staff members of the Safety and Engineering Support Team of Hanbit Nuclear Power Plant 3. 
\providecommand{\href}[2]{#2}\begingroup\raggedright\endgroup
\bibliographystyle{JHEP} 

\begin{thebibliography}{10}

\bibitem{PhysRevD.9.1389}
D.~Z. Freedman, {\it ``Coherent effects of a weak neutral current.''} Phys.
  Rev. D {\bf 9} (Mar, 1974) 1389.

\bibitem{Kopeliovich:1974mv}
V.~B. Kopeliovich and L.~L. Frankfurt, {\it ``{Isotopic and chiral structure of
  neutral current}.''} JETP Lett. {\bf 19} (1974) 145--147.

\bibitem{Drukier19842295}
A.~Drukier and L.~Stodolsky, {\it ``Principles and applications of a
  neutral-current detector for neutrino physics and astronomy.''} Phys. Rev. D
  {\bf 30} (1984) 2295.

\bibitem{Krauss:1991ba}
L.~M. Krauss, {\it ``{Low-energy neutrino detection and precision tests of the
  standard model}.''} Phys. Lett. B {\bf 269} (1991) 407.

\bibitem{Patton:2012jr}
K.~Patton, J.~Engel, G.~C. McLaughlin, and N.~Schunck, {\it ``{Neutrino-nucleus
  coherent scattering as a probe of neutron density distributions}.''} Phys.
  Rev. C {\bf 86} (2012) 024612.

\bibitem{Dutta:2015nlo}
B.~Dutta, Y.~Gao, R.~Mahapatra, N.~Mirabolfathi, L.~E. Strigari, and J.~W.
  Walker, {\it ``{Sensitivity to oscillation with a sterile fourth generation
  neutrino from ultra-low threshold neutrino-nucleus coherent scattering}.''}
  Phys. Rev. D {\bf 94} (2016) 093002.

\bibitem{Kosmas:2017zbh}
T.~S. Kosmas, D.~K. Papoulias, M.~Tortola, and J.~W.~F. Valle, {\it ``{Probing
  light sterile neutrino signatures at reactor and Spallation Neutron Source
  neutrino experiments}.''} Phys. Rev. D {\bf 96} (2017) 063013.

\bibitem{deNiverville:2015mwa}
P.~deNiverville, M.~Pospelov, and A.~Ritz, {\it ``{Light new physics in
  coherent neutrino-nucleus scattering experiments}.''} Phys. Rev. D {\bf 92}
  (2015) 095005.

\bibitem{Cabrera:1984rr}
B.~Cabrera, L.~M. Krauss, and F.~Wilczek, {\it ``{Bolometric Detection of
  Neutrinos}.''} Phys. Rev. Lett. {\bf 55} (1985) 25.

\bibitem{Formaggio:2011jt}
J.~A. Formaggio, E.~Figueroa-Feliciano, and A.~J. Anderson, {\it ``{Sterile
  Neutrinos, Coherent Scattering and Oscillometry Measurements with
  Low-temperature Bolometers}.''} Phys. Rev. D {\bf 85} (2012) 013009.

\bibitem{Akimov:2012aya}
D.~{\relax Yu}. Akimov et~al., ({RED} Collaboration), {\it ``{Prospects for
  observation of neutrino-nuclear neutral current coherent scattering with
  two-phase Xenon emission detector}.''} JINST {\bf 8} (2013) P10023.

\bibitem{Belov:2015ufh}
V.~Belov et~al., {\it ``{The $\nu$GeN experiment at the Kalinin Nuclear Power
  Plant}.''} JINST {\bf 10} (2015) P12011.

\bibitem{Agnolet:2016zir}
G.~Agnolet et~al., ({MINER} Collaboration), {\it ``{Background Studies for the
  MINER Coherent Neutrino Scattering Reactor Experiment}.''} Nucl. Instrum.
  Meth. A {\bf 853} (2017) 53.

\bibitem{Aguilar-Arevalo:2016khx}
A.~Aguilar-Arevalo et~al., ({CONNIE} Collaboration), {\it ``{The CONNIE
  experiment}.''} J. Phys. Conf. Ser. {\bf 761} (2016) 012057.

\bibitem{Kerman:2016jqp}
S.~Kerman, V.~Sharma, M.~Deniz, H.~T. Wong, J.~W. Chen, H.~B. Li, S.~T. Lin,
  C.~P. Liu, and Q.~Yue, ({TEXONO} Collaboration), {\it ``{Coherency in
  Neutrino-Nucleus Elastic Scattering}.''} Phys. Rev. D {\bf 93} (2016) 113006.

\bibitem{Billard:2016giu}
J.~Billard et~al., {\it ``{Coherent Neutrino Scattering with Low Temperature
  Bolometers at Chooz Reactor Complex}.''} J. Phys. G {\bf 44} (2017) 105101.

\bibitem{Hakenmuller:2019ecb}
J.~Hakenmüller et~al., {\it ``{Neutron-induced background in the CONUS
  experiment}.''} Eur. Phys. J. C {\bf 79} (2019) 699.

\bibitem{Angloher:2019flc}
G.~Angloher et~al., ({NUCLEUS} Collaboration), {\it ``{Exploring $\hbox {CE}\nu
  \hbox {NS}$ with NUCLEUS at the Chooz nuclear power plant}.''} Eur. Phys. J.
  C {\bf 79} (2019) 1018.

\bibitem{CONUS:2020skt}
H.~Bonet et~al., ({CONUS} Collaboration), {\it ``{Constraints on Elastic
  Neutrino Nucleus Scattering in the Fully Coherent Regime from the CONUS
  Experiment}.''} Phys. Rev. Lett. {\bf 126} (2021) 041804.

\bibitem{Akimov:2017ade}
D.~Akimov et~al., ({COHERENT} Collaboration), {\it ``{Observation of Coherent
  Elastic Neutrino-Nucleus Scattering}.''} Science {\bf 357} (2017) 1123.

\bibitem{COHERENT:2020iec}
D.~Akimov et~al., ({COHERENT} Collaboration), {\it ``{First Measurement of
  Coherent Elastic Neutrino-Nucleus Scattering on Argon}.''} Phys. Rev. Lett.
  {\bf 126} (2021) 012002.

\bibitem{PhysRevD.73.033005}
K.~Scholberg, {\it ``Prospects for measuring coherent neutrino-nucleus elastic
  scattering at a stopped-pion neutrino source.''} Phys. Rev. D {\bf 73} (2006)
  033005.

\bibitem{COLLAR201556}
J.~Collar et~al., {\it ``{Coherent neutrino-nucleus scattering detection with a
  CsI[Na] scintillator at the SNS spallation source}.''} Nucl. Instrum. Meth. A
  {\bf 773} (2015) 56.

\bibitem{Colaresi:2021kus}
J.~Colaresi, J.~I. Collar, T.~W. Hossbach, A.~R.~L. Kavner, C.~M. Lewis, A.~E.
  Robinson, and K.~M. Yocum, {\it ``{First results from a search for coherent
  elastic neutrino-nucleus scattering at a reactor site}.''} Phys. Rev. D {\bf
  104} (2021) 072003.

\bibitem{CONNIE:2021ggh}
A.~Aguilar-Arevalo et~al., ({CONNIE} Collaboration), {\it ``{Search for
  coherent elastic neutrino-nucleus scattering at a nuclear reactor with CONNIE
  2019 data}.''} JHEP {\bf 05} (2022) 017.

\bibitem{nGeN:2022uje}
I.~Alekseev et~al., ({\ensuremath{\nu}GeN} Collaboration), {\it ``{First
  results of the \ensuremath{\nu}GeN experiment on coherent elastic
  neutrino-nucleus scattering}.''} Phys. Rev. D {\bf 106} (2022) L051101.

\bibitem{XENON:2020gfr}
E.~Aprile et~al., ({XENON} Collaboration), {\it ``{Search for Coherent Elastic
  Scattering of Solar $^8$B Neutrinos in the XENON1T Dark Matter
  Experiment}.''} Phys. Rev. Lett. {\bf 126} (2021) 091301.

\bibitem{Janka:2017vlw}
H.~T. Janka, {\it ``{Neutrino Emission from Supernovae}.''}
  \href{http://arxiv.org/abs/1702.08713}{{\tt arXiv:1702.08713}}.

\bibitem{Cogswell2016}
B.~K. Cogswell and P.~Huber, {\it ``{Detection of Breeding Blankets Using
  Antineutrinos}.''} Science \& Global Security {\bf 24} (2016) 114--130.

\bibitem{RevModPhys.92.011003}
A.~Bernstein, N.~Bowden, B.~L. Goldblum, P.~Huber, I.~Jovanovic, and
  J.~Mattingly, {\it ``Colloquium: Neutrino detectors as tools for nuclear
  security.''} Rev. Mod. Phys. {\bf 92} (2020) 011003.

\bibitem{Liao:2017uzy}
J.~Liao and D.~Marfatia, {\it ``{COHERENT constraints on nonstandard neutrino
  interactions}.''} Phys. Lett. B {\bf 775} (2017) 54.

\bibitem{Dev:2019anc}
P.~Bhupal~Dev et~al., {\it ``{Neutrino Non-Standard Interactions: A Status
  Report}.''} \href{http://arxiv.org/abs/1907.00991}{{\tt arXiv:1907.00991}}.

\bibitem{Aguilar-Arevalo:2019jlr}
A.~Aguilar-Arevalo et~al., ({CONNIE} Collaboration), {\it ``{Exploring
  low-energy neutrino physics with the Coherent Neutrino Nucleus Interaction
  Experiment}.''} Phys. Rev. D {\bf 100} (2019) 092005.

\bibitem{YJKO:2017NEOS}
Y.~J. Ko et~al., {\it ``{Sterile Neutirno Search at the NEOS Experiment}.''}
  Phys. Rev. Lett. {\bf 118} (2017) 121802.

\bibitem{Adhikari:2018ljm}
G.~Adhikari et~al., ({COSINE-100} Collaboration), {\it ``{An experiment to
  search for dark-matter interactions using sodium iodide detectors}.''} Nature
  {\bf 564} (2018) 83.

\bibitem{COSINE-100:2019lgn}
G.~Adhikari et~al., ({COSINE-100} Collaboration), {\it ``{Search for a Dark
  Matter-Induced Annual Modulation Signal in NaI(Tl) with the COSINE-100
  Experiment}.''} Phys. Rev. Lett. {\bf 123} (2019) 031302.

\bibitem{Choi:2020qcj}
J.~Choi, B.~Park, C.~Ha, K.~Kim, S.~Kim, Y.~Kim, Y.~Ko, H.~Lee, S.~Lee, and
  S.~Olsen, {\it ``{Improving the light collection using a new NaI(Tl)crystal
  encapsulation}.''} Nucl. Instrum. Meth. A {\bf 981} (2020) 164556.

\bibitem{Kim_2010}
S.~K. Kim, H.~J. Kim, and Y.~D. Kim, {\it ``{Scintillator-based detectors for
  dark matter searches I}.''} New J. Phy. {\bf 12} (2010) 075003.

\bibitem{Pandey_2018}
I.~R. Pandey, H.~J. Kim, H.~S. Lee, Y.~D. Kim, M.~H. Lee, V.~D. Grigorieva, and
  V.~N. Shlegel, {\it ``{The Na2W2O7crystal: a crystal scintillator for dark
  matter search experiment}.''} Eur. Phys. Journal C {\bf 78} (2018) 973.

\bibitem{Bernabei:2014jba}
R.~Bernabei et~al., ({DAMA/LIBRA} Collaboration), {\it ``{The Annual Modulation
  Signature for Dark Matter: DAMA/LIBRA-Phase1 Results and Perspectives}.''}
  Adv. High Energy Phys. {\bf 2014} (2014) 605659.

\bibitem{Bernabei:2018jrt}
R.~Bernabei et~al., {\it ``{First model independent results from
  DAMA/LIBRA-phase2}.''} Nucl. Phys. Atom. Energy {\bf 19} (2018) 307.

\bibitem{Kim:2014toa}
K.~W. Kim et~al., {\it ``{Tests on NaI(Tl) crystals for WIMP search at the
  Yangyang Underground Laboratory}.''} Astropart. Phys. {\bf 62} (2015) 249.

\bibitem{sabre}
J.~Xu, F.~Calaprice, F.~Froborg, E.~Shields, and B.~Suerfu, ({SABRE}
  Collaboration), {\it ``{SABRE – A test of DAMA with high-purity NaI(Tl)
  crystals}.''} AIP Conf. Proc. {\bf 1672} (2015) 040001.

\bibitem{Adhikari:2017esn}
G.~Adhikari et~al., ({COSINE-100} Collaboration), {\it ``{Initial Performance
  of the COSINE-100 Experiment}.''} Eur. Phys. J. C {\bf 78} (2018) 107.

\bibitem{Fushimi:2018qzk}
K.-I. Fushimi, {\it ``{Low Background Measurement by Means of NaI(Tl)
  Scintillator: Improvement of Sensitivity for Cosmic Dark Matter}.''}
  RADIOISOTOPES {\bf 67} (2018) 101.

\bibitem{Coarasa:2018qzs}
I.~Coarasa et~al., {\it ``{ANAIS-112 sensitivity in the search for dark matter
  annual modulation}.''} Eur. Phys. J. C {\bf 79} (2019) 233.

\bibitem{Amare:2018sxx}
J.~Amare et~al., ({ANAIS-112} Collaboration), {\it ``{Performance of ANAIS-112
  experiment after the first year of data taking}.''} Eur. Phys. J. C {\bf 79}
  (2019) 228.

\bibitem{Olivan:2017akd}
M.~Olivan et~al., {\it ``{Light yield determination in large sodium iodide
  detectors applied in the search for dark matter}.''} Astropart. Phys. {\bf
  93} (2017) 86--95.

\bibitem{Suerfu:2019snq}
B.~Suerfu, M.~Wada, W.~Peloso, M.~Souza, F.~Calaprice, J.~Tower, and G.~Ciampi,
  {\it ``{Growth of Ultra-high Purity NaI(Tl) Crystal for Dark Matter
  Searches}.''} Phys. Rev. Research {\bf 2} (2020) 013223.

\bibitem{Park:2020fsq}
B.~Park et~al., ({COSINE} Collaboration), {\it ``{Development of ultra-pure
  NaI(Tl) detectors for the COSINE-200 experiment}.''} Eur. Phys. J. C {\bf 80}
  (2020) 814.

\bibitem{Adhikari:2021rdm}
G.~Adhikari et~al., ({COSINE-100} Collaboration), {\it ``{Background modeling
  for dark matter search with 1.7 years of COSINE-100 data}.''} Eur. Phys. J. C
  {\bf 81} (2021) 837.

\bibitem{Fushimi:2021mez}
K.~Fushimi et~al., {\it ``{Development of highly radiopure NaI(Tl) scintillator
  for PICOLON dark matter search project}.''} PTEP {\bf 2021} (2021) 043F01.

\bibitem{cosinebg}
P.~Adhikari et~al., ({COSINE-100} Collaboration), {\it ``{Background model for
  the NaI(Tl) crystals in COSINE-100}.''} Eur. Phys. J. C {\bf 78} (2018) 490.

\bibitem{Abdullah:2022zue}
M.~Abdullah et~al., {\it {Coherent elastic neutrino-nucleus scattering:
  Terrestrial and astrophysical applications}},  in {\em {2022 Snowmass Summer
  Study}}, 3, 2022.
\newblock \href{http://arxiv.org/abs/2203.07361}{{\tt arXiv:2203.07361}}.

\bibitem{Amare:2019jul}
J.~Amare et~al., ({ANAIS-112} Collaboration), {\it ``{First Results on Dark
  Matter Annual Modulation from the ANAIS-112 Experiment}.''} Phys. Rev. Lett.
  {\bf 123} (2019) 031301.

\bibitem{Amare:2021yyu}
J.~Amare et~al., ({ANAIS-112} Collaboration), {\it ``{Annual Modulation Results
  from Three Years Exposure of ANAIS-112}.''} Phys. Rev. D {\bf 103} (2021)
  102005.

\bibitem{BDT}
J.~H. Friedman, {\it ``{Greedy function approximation: A gradient boosting
  machine}.''} Ann. Stat. {\bf 29} (2001) 1189.

\bibitem{adhikari2020lowering}
G.~Adhikari et~al., ({COSINE-100} Collaboration), {\it ``{Lowering the energy
  threshold in COSINE-100 dark matter searches}.''} Astropart. Phys. {\bf 130}
  (2021) 102581.

\bibitem{COSINE-100:2021xqn}
G.~Adhikari et~al., ({COSINE-100} Collaboration), {\it ``{Strong constraints
  from COSINE-100 on the DAMA dark matter results using the same sodium iodide
  target}.''} Sci. Adv. {\bf 7} (2021) abk2699.

\bibitem{COSINE-100:2021poy}
G.~Adhikari et~al., ({COSINE-100} Collaboration), {\it ``{Searching for
  low-mass dark matter via Migdal effect in COSINE-100}.''} Phys. Rev. D {\bf
  105} (2022) 042006.

\bibitem{RENO:2019otc}
C.~D. Shin et~al., ({RENO} Collaboration), {\it ``{Observation of reactor
  antineutrino disappearance using delayed neutron capture on hydrogen at
  RENO}.''} JHEP {\bf 04} (2020) 029.

\bibitem{Kopeikin:2012zz}
V.~I. Kopeikin, {\it ``{Flux and spectrum of reactor antineutrinos}.''} Phys.
  Atom. Nucl. {\bf 75} (2012) 143--152.

\bibitem{Ko:2019cip}
Y.~Ko et~al., {\it ``{NEOS Experiment}.''} J. Phys. Conf. Ser. {\bf 1216}
  (2019) 012004.

\bibitem{Ko:2016neu}
Y.~J. Ko et~al., {\it ``Comparison of fast neutron rates for the neos
  experiment.''} Journal of the Korean Physical Society {\bf 69} (2016)
  1651--1655.

\bibitem{KIM2016285}
Y.~Kim, {\it ``Detection of antineutrinos for reactor monitoring.''} Nuclear
  Engineering and Technology {\bf 48} (2016) 285--292.

\bibitem{Agostinelli:2002hh}
S.~Agostinelli et~al., ({GEANT4} Collaboration), {\it ``{GEANT4: A Simulation
  toolkit}.''} Nucl. Instrum. Meth. A {\bf 506} (2003) 250.

\bibitem{Note1}
procured from Eckert and Ziegler Isotope Products.

\bibitem{adhikari16}
P.~Adhikari et~al., ({KIMS} Collaboration), {\it ``{Understanding internal
  backgrounds in NaI(Tl) crystals toward a 200 kg array for the KIMS-NaI
  experiment}.''} Eur. Phys. J. C {\bf 76} (2016) 185.

\bibitem{Bernabei:2008yh}
R.~Bernabei et~al., ({DAMA} Collaboration), {\it ``{The DAMA/LIBRA
  apparatus}.''} Nucl. Instrum. Meth. A {\bf 592} (2008) 297--315.

\bibitem{Adhikari:2017gbj}
G.~Adhikari et~al., ({KIMS} Collaboration), {\it ``{Understanding NaI(Tl)
  crystal background for dark matter searches}.''} Eur. Phys. J. C {\bf 77}
  (2017) 437.

\bibitem{Amare:2018ndh}
J.~Amar\'e et~al., {\it ``{Analysis of backgrounds for the ANAIS-112 dark
  matter experiment}.''} Eur. Phys. J. C {\bf 79} (2019) 412.

\bibitem{Antonello2021}
M.~Antonello and M.~Zurowski, {\it ``Characterization of sabre crystal nai-33
  with direct underground counting.''} Eur. Phys. J. C {\bf 81} (2021) 299.

\bibitem{Bernabei:2020mon}
R.~Bernabei et~al., {\it ``{The DAMA project: Achievements, implications and
  perspectives}.''} Prog. Part. Nucl. Phys. {\bf 114} (2020) 103810.

\bibitem{JS_Park_1}
J.~Park et~al., ({RENO} Collaboration), {\it ``Production and optical
  properties of gd-loaded liquid scintillator for the reno neutrino
  detector.''} Nucl. Instrum. Meth. A {\bf 707} (2013) 1016.

\bibitem{Park:2017jvs}
J.~S. Park et~al., ({KIMS} Collaboration), {\it ``{Performance of a prototype
  active veto system using liquid scintillator for a dark matter search
  experiment}.''} Nucl. Instrum. Meth. A {\bf 851} (2017) 103.

\bibitem{Adhikari:2020asl}
G.~Adhikari et~al., ({COSINE-100} Collaboration), {\it ``{The COSINE-100 liquid
  scintillator veto system}.''} Nucl. Instrum. Meth. A {\bf 1006} (2021)
  165431.

\bibitem{Adhikari:2018fpo}
G.~Adhikari et~al., ({COSINE-100} Collaboration), {\it ``{The COSINE-100 Data
  Acquisition System}.''} JINST {\bf 13} (2018) P09006.

\bibitem{BRUN199781}
R.~Brun and F.~Rademakers, {\it ``{ROOT — An object oriented data analysis
  framework}.''} Nucl. Instrum. Meth. A {\bf 389} (1997) 81--86.

\bibitem{DM-Ice:2015aij}
J.~Cherwinka et~al., ({DM-Ice} Collaboration), {\it ``{Measurement of Muon
  Annual Modulation and Muon-Induced Phosphorescence in NaI(Tl) Crystals with
  DM-Ice17}.''} Phys. Rev. D {\bf 93} (2016) 042001.

\bibitem{COSINE-100:2021mlj}
H.~Kim et~al., ({COSINE-100} Collaboration), {\it ``{The environmental
  monitoring system at the COSINE-100 experiment}.''} JINST {\bf 17} (2022)
  T01001.

\bibitem{Edgar:1969}
P.~Edgar and M.~B. James, {\it ``{Stored Energy in Irradiated Sodium
  Chloride}.''} Phys. Rev. {\bf 181} (1969) 1290.

\bibitem{Kopeikin:2004com}
V.~I. Kopeikin, L.~Mikaelyan, and V.~V. Sinev, {\it ``{Components of
  Antineutrino Emission in Nuclear Reactor}.''} Physics of Atomic Nuclei {\bf
  67} (2004) 1963.

\bibitem{PhysRevC.84.024617}
P.~Huber, {\it ``Determination of antineutrino spectra from nuclear
  reactors.''} Phys. Rev. C {\bf 84} (Aug, 2011) 024617.

\bibitem{Mueller:2011nm}
T.~A. Mueller et~al., {\it ``{Improved Predictions of Reactor Antineutrino
  Spectra}.''} Phys. Rev. C {\bf 83} (2011) 054615.

\bibitem{Joo:2018hom}
H.~W. Joo, H.~S. Park, J.~H. Kim, S.~K. Kim, Y.~D. Kim, H.~S. Lee, and S.~H.
  Kim, {\it ``{Quenching factor measurement for NaI(Tl) scintillation
  crystal}.''} Astropart. Phys. {\bf 108} (2019) 50.

\bibitem{Xu:2015wha}
J.~Xu et~al., {\it ``{Scintillation Efficiency Measurement of Na Recoils in
  NaI(Tl) Below the DAMA/LIBRA Energy Threshold}.''} Phys. Rev. C {\bf 92}
  (2015) 015807.

\bibitem{Freedman:1977xn}
D.~Z. Freedman, D.~N. Schramm, and D.~L. Tubbs, {\it ``{The Weak Neutral
  Current and Its Effects in Stellar Collapse}.''} Ann. Rev. Nucl. Part. Sci.
  {\bf 27} (1977) 167--207.

\bibitem{Hoferichter:2020osn}
M.~Hoferichter, J.~Men\'endez, and A.~Schwenk, {\it ``{Coherent elastic
  neutrino-nucleus scattering: EFT analysis and nuclear responses}.''} Phys.
  Rev. D {\bf 102} (2020) 074018.

\bibitem{PhysRevD.98.053004}
V.~A. Bednyakov and D.~V. Naumov, {\it ``Coherency and incoherency in
  neutrino-nucleus elastic and inelastic scattering.''} Phys. Rev. D {\bf 98}
  (Sep, 2018) 053004.

\bibitem{Birks:1951boa}
J.~B. Birks, {\it ``{Scintillations from Organic Crystals: Specific
  Fluorescence and Relative Response to Different Radiations}.''} Proc. Phys.
  Soc. A {\bf 64} (1951) 874--877.

\bibitem{osti_4701226}
J.~Lindhard, V.~Nielsen, M.~Scharff, and P.~Thomsen, {\it ``{Integral equations
  governing radiation effects. (Notes on Atomic collisions, III)}.''} Mat. Fys.
  Medd. Dan. Vid. Selsk. {\bf 33} (1, 1963).

\bibitem{Tretyak:2009sr}
V.~I. Tretyak, {\it ``{Semi-empirical calculation of quenching factors for ions
  in scintillators}.''} Astropart. Phys. {\bf 33} (2010) 40--53.

\bibitem{Collar:2013gu}
J.~I. Collar, {\it ``{Quenching and channeling of nuclear recoils in NaI(Tl):
  Implications for dark-matter searches}.''} Phys. Rev. C {\bf 88} (2013)
  035806.

\bibitem{COSINE-100:2019brm}
Y.~J. Ko et~al., ({COSINE-100} Collaboration), {\it ``{Comparison between
  DAMA/LIBRA and COSINE-100 in the light of Quenching Factors}.''} JCAP {\bf
  11} (2019) 008.

\bibitem{Lewin:1995rx}
J.~Lewin and P.~Smith, {\it ``{Review of mathematics, numerical factors, and
  corrections for dark matter experiments based on elastic nuclear recoil}.''}
  Astropart. Phys. {\bf 6} (1996) 87.

\bibitem{nonprop}
L.~Swiderski, {\it ``{Response of doped alkali iodides measured with gamma-ray
  absorption and Compton electrons}.''} Nucl. Instrum. Meth. A {\bf 705} (2013)
  42.

\bibitem{Cintas:2021fvd}
D.~Cintas et~al., {\it ``{Quenching Factor consistency across several NaI(Tl)
  crystals}.''} J. Phys. Conf. Ser. {\bf 2156} (2021) 012065.

\bibitem{Verbus:2016sgw}
J.~R. Verbus et~al., {\it ``{Proposed low-energy absolute calibration of
  nuclear recoils in a dual-phase noble element TPC using D-D neutron
  scattering kinematics}.''} Nucl. Instrum. Meth. A {\bf 851} (2017) 68--81.

\end{thebibliography}
\end{document}